\documentclass[a4paper,10pt]{article}

\usepackage{latexsym}
\usepackage[dvips]{graphics,lscape,rotating}
\usepackage{euscript}
\usepackage{bbm}
\usepackage{amsfonts,amsmath,amssymb}
\usepackage{mathrsfs}
\usepackage{array,longtable}
\usepackage{pstricks,psfrag}

\usepackage{makeidx}
\makeindex

\renewcommand{\arraystretch}{1.5}

\numberwithin{equation}{section}

\newcommand{\lu}[1]{_{#1}\!\!}

\newcommand{\cmt}[2]{\begin{minipage}[t]{#1cm} #2 \end{minipage}}
\newcommand{\mb}[1]{\mathbbm{#1}} 
 
\newcommand{\Muserfunction}[1]{A}               

\newcommand{\ul}[1]{\underline{#1}}  
\newcommand{\ol}[1]{\overline{#1}}

\newcommand{\NN}{\nonumber\\}

\newcommand{\barr}[1]{\begin{eqnarray}\begin{array}{#1}}
\newcommand{\earr}{\end{array}\end{eqnarray}}

\newcommand{\WT}[1]{\widetilde{#1}}

\newcommand{\MC}[1]{\mathcal{#1}}

\newcommand{\supp}{\text{supp}}
 
\newcommand{\bs}{\backslash}

\def\M{\MC{M}}
\def\C{\MC{C}}

\newcommand{\jmax}{j_{\mathrm{max}}}

\def\ba{\begin{eqnarray}}
\def\ea{\end{eqnarray}}
\def\be{\begin{equation}}
\def\ee{\end{equation}}

\newtheorem{Theorem}{Theorem}[section]                             
\newtheorem{Definition}{Definition}[section]
\newtheorem{Lemma}{Lemma}[section]

\newcommand{\Ho}{\mathcal{H}_{0}}
\newcommand{\Hgauss}{\mathcal{H}_{\text{Gauss}}}
\newcommand{\Hdiff}{\mathcal{H}_{\text{diff}}}
\newcommand{\Hphys}{\mathcal{H}_{\text{phys}}}

\DeclareMathOperator{\sgn}{sgn}

\begin{document}

\title{Oriented Matroids --\\Combinatorial Structures Underlying \\Loop Quantum Gravity}

\author{Johannes Brunnemann$^1$\thanks{johannes.brunnemann@math.upb.de}~,
       ~David Rideout$^2$\thanks{drideout@perimeterinstitute.ca}\\
$^1$Department of Mathematics, University of Paderborn,  33098 Paderborn, Germany\\
$^2$Perimeter Institute for Theoretical Physics, N2L 2Y5, Waterloo, ON, Canada 
}

\maketitle

\begin{abstract}

We analyze combinatorial structures which play a central role in determining spectral properties of the volume operator \cite{Ashtekar1998} in loop quantum gravity (LQG). These structures encode geometrical information of the embedding of arbitrary valence vertices of a graph in 3-dimensional Riemannian space,
 and can be represented by sign strings containing relative orientations of embedded edges. 
We demonstrate that these signature factors are 
a special representation of the general mathematical concept of an oriented matroid \cite{Ziegler1998,Bjorner1999}. Moreover, we show that oriented matroids can also be used 
to describe the topology (connectedness) of directed graphs. Hence the mathematical methods developed for oriented matroids can be applied 
to 
the difficult combinatorics of embedded graphs underlying the construction of LQG.
As a first application we revisit the analysis of  \cite{BrunnemannRideout2008,BrunnemannRideout2008a}, and find that
enumeration of all possible sign configurations used there is equivalent to enumerating all realizable oriented matroids of rank 3 \cite{Ziegler1998,Bjorner1999}, and thus can be greatly simplified. We find that for 7-valent vertices having no coplanar triples of edge tangents, the smallest non-zero eigenvalue of the volume spectrum does not grow as one increases the maximum spin $\jmax$ at the vertex, for \emph{any} orientation of the edge tangents. This indicates that, in contrast to the area operator, considering large $\jmax$ does not necessarily imply large volume eigenvalues.
In addition we give an outlook to
possible starting points for rewriting the combinatorics of LQG in terms of oriented matroids.  

\end{abstract}
\tableofcontents

\section{\label{Introduction}Introduction}
Any approach to quantum gravity is expected to exhibit a discrete nature of spacetime at distances close to the Planck scale ($\sim 10^{-35}$\,m\,). Taking this perspective, our traditional picture of continuous spacetime geometry can only arise as a kind of (semi-)classical limit from a more fundamental combinatorial theory, linking matter and gravity. 
Loop quantum gravity (LQG) is supposed to give a canonical quantization of gravity, based on the initial value formulation of general relativity (GR), in a manner which is independent of any background geometry. LQG provides a mathematically rigorous way to non-perturbatively quantize geometry itself by promoting classical geometric quantities such as length, area or volume into operators in quantum theory. Remarkably these operators indeed turn out to have discrete spectra\footnote{Thus far this is a kinematical statement. See \cite{Dittrich2009,Rovelli2007} for a discussion on how this property can be transferred to the physical sector of LQG.  Additionally in \cite{Giesel2010, Giesel2007b} a setup is given where these properties can be regarded as physical.}.
During the last 20 years LQG has already produced results which are regarded as physically relevant:  the absence of a cosmological Big-Bang singularity\footnote{In a symmetry reduced setup.} \cite{Ashtekar:2006}, as well as the reproduction of the entropy-area law for black holes from first principles \cite{Barbero:2008}, just to mention a couple.
Moreover for the first time there exists a proposal 
for implementing the Hamilton constraint, which encodes the dynamics, as an operator in quantum theory \cite{Thiemann1996}. Even better, if geometry is kept dynamical, matter can be quantized without the occurrence of anomalies or UV-divergences, which plague  quantum field theories constructed on a fixed background \cite{Thiemann1997}. 

Nevertheless there are important questions regarding the contruction of the physical sector of LQG which remain unanswered so far.   
One obstacle is the construction of the theory in terms of a projective limit over a poset of arbitrarily complicated directed graphs embedded into three dimensional Riemannian space, the Cauchy surfaces which arise in the initial value formulation of GR. Without reference to a background geometry, the characterization of such embeddings by coordinates, edge lengths or angles is meaningless. Rather one is referred to diffeomorphism invariant characterizations such as closed circuits\footnote{Also called cycles in graph theory.} in graphs or the local intersection behavior of edges at graph vertices. 
While the mathematical construction of the projective limit over the graph poset is well understood, the complicated combinatorics of generic elements of the poset still lack an effective characterization method which makes a study of the full theory viable.

In practical computations, the combinatorial difficulty is often circumvented by imposing simplifications to LQG.
One possible simplification is based on the assumption that the underlying classical theory is not full GR, but a symmetry reduced, so-called cosmological model. This is equivalent to partially fixing a background geometry. The resulting model is quantized using techniques similar to LQG using a restricted set of graphs, adapted to the symmetry reduction. In the resulting loop quantum cosmology (LQC) (see \cite{Bojowald2006,Ashtekar2003} and references therein) it is then possible to address questions such as the Big-Bang singularity, which is far beyond the possibilities one has in full LQG at present. 
Another ansatz (called Algebraic Quantum Gravity, AQG) is to keep GR, that is full background independence, but to modify the quantization procedure by restricting only to certain types of graphs and embeddings in the poset \cite{Giesel2007}. This makes it possible to analyze the classical limit of the resulting model  \cite{Giesel2007a} as well as to construct a physical sector of LQG in a reduced phase space quantization \cite{Giesel2007b}.  
  
Given the promising results of the outlined simplified models, 
it is highly desireable to get a better understanding of the physical implications of the underlying assumptions, as simplifications may modify dynamical properties of the analyzed model.
For example, in the context of LQC it is found that the operator corresponding to the classical inverse scale factor of the universe is bounded in LQC \cite{Bojowald2006}, but unbounded in full LQG \cite{Brunnemann2006}. Moreover the Hilbert space of LQC cannot be continuously embedded into the Hilbert space obtained in LQG \cite{Brunnemann2007}.
In AQG, on the other hand, one special graph and its embedding is fixed, in order to compute the volume spectrum.  
It has been suggested that only this particular embedding gives a consistent semiclassical limit of the model \cite{Flori2008}.  The mathematical tools presented here may allow us to consider this question for arbitrary embedded vertices.

In full LQG, it turns out that the global circuit structure of generic graphs and their local embedding properties are relevant for the construction of the Hamilton constraint operator of \cite{Thiemann1996}, which encodes the dynamics of the theory. 
This arises because a crucial ingredient of the Hamilton constraint operator consists in the quantum analogue to the classical volume integral of a region in three dimensional Riemannian space, called the volume operator. In  \cite{BrunnemannRideout2008,BrunnemannRideout2008a} the spectral properties of this operator due to \cite{Ashtekar1998} were analyzed in full LQG, and it was found that they are highly sensitive to the local graph embedding. In particular the presence of a finite smallest non-zero eigenvalue is characterized by the chosen embedding.

In order to bring the results of the outlined simplified approaches in a closer context to the full theory, it is 
important to continue work on full LQG, and to develop techniques 
to better understand its combinatorics.
This paper is intended as a first step toward developing such a method. It is based on the observation that local embedded graph geometry and global graph topology have a common home in the mathematical field of (oriented) matroid theory \cite{Bjorner1999,Ziegler1998}, which provides a precise combinatorial abstraction of simplices and directed graphs. 

~\\
Oriented matroids have already been used in the physics literature. One example is quantum information theory \cite{Raussendorf2009}. Another is a considerable amount of work done by the author of \cite{Nieto1998}--\cite{Nieto2005} and collaborators, to investigate connections between matroid theory and supergravity \cite{Nieto1998, Nieto2005a}, Chern-Simons theory \cite{Nieto2000}, string / M- theory  \cite{Nieto2003a,Nieto2006} and 2D-gravity \cite{Nieto2009}. In the context of supergravity, possible implications on the Ashtekar formalism in higher dimesions have been outlined \cite{Nieto2005}.  
Their approach is mainly based on the observation that certain antisymmetry properties of tensors and forms in the according contexts can be captured by oriented matroids. 

~\\ 
We suggest a different approach here. We are going to treat the combinatorics inherent in the construction of LQG by means of oriented matroids. As we will demonstrate, the occurring combinatorial structures match in a very natural way. 
This opens up a new mathematical arena to study LQG, and to overcome certain technical and conceptual obstacles which, at present, make concrete analytical and numerical investigations of the full theory very difficult. 
In particular it becomes feasible to develop a systematic and unified treatment of global (connectedness) and local (embedding) properties of the graph poset underlying LQG.

~\\
The plan of this paper is as follows. In the next section we will describe the occurrence of geometrical and topological properties of graphs in the present formulation of LQG in more detail and characterize their combinatorics. 
We then introduce the concept of a matroid and an oriented matroid and show how the latter can be used in order to encode the information on global and local properties of directed graphs. 
Furthermore we will comment on the issue of realizability of oriented matroids, which is directly related to computing the number of possible diffeomorphic local graph embeddings, as well the number of certain graph topologies. 
We will finally show how this can be applied to the results of \cite{BrunnemannRideout2008,BrunnemannRideout2008a}, which will be revisited and simplified. 
Our findings will be discussed in section \ref{Discussion}. There we will also comment on the notion of a semiclassical limit of the volume operator in light of the revisited numerical computation.
In the summary and outlook section we outline upcoming work, and steps necessary 
to fully establish the oriented matroid formalism within LQG, as well as its potential impact on further studies in full LQG.    

\section{\label{LQG Combinatorics}LQG Combinatorics}
In this section we will only outline the current construction 
of LQG as necessary for our work. For a detailed introduction we refer to \cite{Ashtekar2004,Thiemann2007}.

LQG is based on the initial value formulation of Einstein's equations for GR,
which is possible for any globally hyperbolic spacetime. 
According to a theorem of Geroch \cite{Geroch1970}, the spacetime is then homeomorphic\footnote{In fact it is even diffeomorphic to $\mb{R}\times \Sigma$, as proved in \cite{Sanchez2003}.} to the Cartesian product of the real numbers $\mb{R}$ (`time axis') times an orientable spacelike three dimensional Cauchy surface $\Sigma$. An according $3+1$ decomposition procedure introduced by Arnowitt, Deser and Misner (ADM) \cite{ADM1962} then rewrites GR in terms of canonically conjugated variables. These are components of the induced spatial metric tensor on $\Sigma$ and its first order (``time-") derivatives in the $\mb{R}$-direction given by components of extrinsic curvature.  In this setup the dynamics of Einstein's equations is encoded completely in constraints (three spatial diffeomorphism and a scalar constraint) which restrict admissible initial data on a chosen Cauchy surface $\Sigma$, and assure background independence\footnote{The physical content of the theory must not depend on the choice of Cauchy surfaces.}. The resulting theory can then be treated using Dirac's approach to constrained Hamiltonian systems \cite{Dirac2001}, which proposes to first construct the canonical theory and second to implement the constraints formulated in terms of the canonical variables on phase space. This results  in the  well known Dirac algebra of Poisson brackets among the constraints.    

Inspired by the work of Sen, Ashtekar realized \cite{Ashtekar1986} that the canonical ADM-variables of GR can be extended  in terms of Lie-algebra valued densitized triads $E$ (local orthonormal frames) and according connections $A$ on a principal fibre bundle  with basis $\Sigma$ and compact structure group $G$. In practice $G$ is taken to be $SU(2)$, however the construction described here works to a certain extent for general compact Lie groups $G$. As different triads can encode identical information on the spatial metric, another constraint, called the Gauss contraint, is added to the Dirac algebra. 

This opens a way to formulate the theory in terms of holonomies and fluxes, very similar to lattice gauge theory. Consider a one dimensional embedded directed path\footnote{A path is an equivalence class of semianalytic curves $c$ with respect to re-parametrization and finite semianalytic retracings. A curve is a map $c:~\mb{R}\supset[0,1]\ni t\rightarrow c(t)\subset\Sigma$. Here $c(0)=:b(c)$ is called the beginning point, $c(1)=:f(c)$ the end point of $c$. Semianalyticity is needed to ensure that two paths can only have a finite number of intersections.}
$p\subset\Sigma$. By the holonomy map $h$, a connection $A\in\MC{A}$ can be understood as a mapping of $p$ to the gauge group $G$, $A:~p \stackrel{h}{\rightarrow} A(p)\in G$, where $\MC{A}$ denotes the space of smooth connections\footnote{In the context of LQG a holonomy is understood as a parallel transport of the connection $A$ along $p$, where $p$ is not necessarily a closed loop.}. In order to make this construction more explicit one often writes $h_p(A)$ instead of $A(p)$.  The set $\MC P$ of all paths carries a groupoid structure with respect to composition\footnote{Two paths can only be composed if they have a common point.}.
As can be shown, the set $\MC{A}$ is contained in the set $\ol{\MC{A}}=Hom(\MC{P},G)$ of all homomorphisms (no continuity assumption)
of the path groupoid $\MC{P}$ to the gauge group $G$. The set $\ol{\MC{A}}$ is consequently called the space of generalized connections. $\ol{\MC{A}}$ is equipped with a topology as follows. One considers finite collections $E(\gamma):=\{e_1,\ldots,e_N\}$ of non self-intersecting paths $e_k$, called edges\footnote{Modulo finite retracings. Strictly speaking an edge corresponds to a path which contains a representative curve which has no self intersections.}, which mutually intersect at most at their beginning $b(e_k)$ and end (final) points $f(e_k)$, called vertices $V(\gamma):=\{b(e_k),f(e_k)\}_{e_k\in E(\gamma)}$. Such a collection is called a graph $\gamma$, the set of all finite semianalytic graphs is denoted by $\Gamma$. Now one considers the subgroupoid $l(\gamma)\subset\MC{P}$ generated by all elements contained in $E(\gamma)$ together with their inverses and finite composition\footnote{It follows that $l(\gamma)$ is in fact an equivalence class of graphs: $l(\gamma)=l(\gamma')$ if their edge / ground-sets $E(\gamma')$ and $E(\gamma)$ differ only by a re-labelling of elements and/or reorientation of edge directions plus finite compositions. One only considers graphs over the full $n$-element ground set, called maximal representatives.}. 
The characteristic function of $E(\gamma)$ is also called the {\it support} $\supp(l(\gamma))$.

Certainly the label set $\MC{L}=\{l(\gamma)\}_{\gamma\in\Gamma}$ carries a 
partial order, that is $l(\gamma)\prec l'(\gamma')$ if $l(\gamma)$ is a subgroupoid of $l'(\gamma')$. By construction $X_{l(\gamma)}:=Hom(l(\gamma),G)$, is understood as the set of all homomorphisms from $E(\gamma)$ to $G^{N}$. That is, one copy of $G$ is associated to each edge via the holonomy map.

Using the partial order of the set $\MC{L}$ one then constructs the projective limit $\ol{X}$ among the  $X_{l(\gamma)}$. It can be shown that there is bijection between $\ol{\MC{A}}$ and $\ol{X}$.
Therefore $\ol{\MC{A}}$ can be equipped with a topology inherited by $\ol{X}$ and turns out to be compact Hausdorff by the compactness of $G$ and Tychonoff's theorem. This makes it possible to construct a Hilbert space $\MC{H}_{0}=L^2(\ol{\MC{A}},d\mu_0)$ as an inductive limit of Hilbert spaces $\MC{H}_{l(\gamma)}= L^2(X_{l(\gamma)},d\mu_\gamma)=L^2(G^{ N}_{(l(\gamma))},{d\mu_H}^{ N})$

In each $\MC{H}_{l(\gamma)}$ one considers functions $f_\gamma:G^{N}_{(l(\gamma))}\rightarrow \mb{C}$ which are called cylindrical, because their support consists of those $N$ copies of $G$ which are labelled by  $\supp(l(\gamma))$. These functions are square integrable continuous with respect to the Haar measure $d\mu_H$ on each copy of $G$. An orthonormal basis on $\MC{H}_0$ is given by so called spin network functions $T_{\gamma\vec{j}\vec{m}\vec{n}}$, labelled 
by a graph $\gamma$
and irreducible representations $\vec{j}:=(j_1,\ldots,j_N)$ of $G$ (``spins'') plus a pair of matrix indices $\vec{m}:=(m_1,\ldots,m_N)$, $\vec{n}:=(n_1,\ldots,n_N)$ one for each edge contained in $E(\gamma)$. This is derived from the Peter\&Weyl theorem and rests on the compactness of $G$.  
Holonomies act as multiplication operators, fluxes as derivations on $\MC{H}_0$. 

Note that the construction of the measure $\mu_0$ depends only on the topology of $\Sigma$ via $\supp(l(\gamma))$. It does not refer to a choice of coordinates on $\Sigma$. 
In this sense $\mu_0$ is defined background independently. 
In particular it can be shown that finite (semi-) analytic diffeomorphisms on $\Sigma$ can be unitarily implemented on $\MC{H}_0$. Even more,  $\mu_0$ is uniquely determined if unitarity of diffeomorphisms is imposed \cite{LOST,Fleischhack2009}.
\\
 
So far all these constructions are on the kinematical level. By the quantum analogue to Dirac's procedure\footnote{Also referred to as refined algebraic quantization.}  the classical constraint functions have to be implemented as constraint operators on $\Ho$. Their common kernel gives the physical Hilbert space $\Hphys$. 
Indeed it is straightforward to obtain the set of solutions to the Gauss constraint $\Hgauss$ as a subset of $\Ho$.

However, one cannot in general expect $\Hphys\subseteq \Ho$, rather\footnote{Consider for example eigen``functions'' of the position operator in quantum mechanics, given by Dirac's $\delta$-distribution.} elements of $\Hphys$ are likely to be distributions contained in the algebraic dual $\MC{D}^*$ of $\Ho$.
   
This is what happens when one constructs the space $\Hdiff$ of solutions to the spatial diffeomorphism constraint. Nevertheless it is possible to rigorously construct $\Hdiff$ via a rigging map construction \cite{Ashtekar1995}. Elements of $\Hdiff$ are then labelled by equivalence classes of $\supp(l(\gamma))$ with respect to finite semianalytic diffeomorphisms. As described in the introduction, each such equivalence class carries global (topological) information about the connectedness of the representatives $\gamma$, that is their circuit structure, as well as local (geometric) information about the intersection behavior of their edges at the vertices of $\gamma$.
\\

Despite the successful solution of Gauss and diffeomorphism constraints, the scalar or Hamilton constraint, encoding the dynamics, has not been solved so far. However there is a proposal to implement it in quantum theory on $\Ho$ \cite{Thiemann1996} as well as a composite operator called the master constraint on $\Hdiff$ \cite{Thiemann2006}. Moreover in the context of \cite{Thiemann2006,Thiemann2005} a rigorous program for constructing an inner product on $\Hphys$ was proposed for the first time. Completing this  step is crucial in order to construct the physical sector of LQG and to make physical predictions. 

A major difficulty in the treatment of the Hamilton- respectively master constraint operator in the full theory comes from the fact that both operators are graph changing, due to the presence of holonomies in the classical constraint expression which are promoted to multiplication operators as mentioned above\footnote{When a cylindrical function $f_\gamma$ is multiplied by a holonomy $h_e$ supported on an edge $e$ not contained in $E(\gamma)$ then the support of  $h_e\cdot f_\gamma$ is given by $l(\gamma\cup \{e\})$.}. Illustratively speaking, constructing $\Hphys$ is equivalent to finding eigenstates to graph changing operators at a projective limit. 

Nevertheless operators only containing fluxes, such as the volume operator, can in principle be discussed at the level of $\Ho$, just by evaluating their action on a function $f_\gamma$ cylindrical over a graph $\gamma$. If cylindrical consistency\footnote{That is, one has to show that the action of this operator on a cylindrical function $f_{\WT{\gamma}}\in \MC{H}_{l(\WT{\gamma})}$ is equal to its action on a cylindrical function $f_{\gamma}\in \MC{H}_{l(\gamma)}$ with $\gamma\subset\WT{\gamma}$, if $\MC{H}_{l(\WT{\gamma})}$ is restricted to its subspace $\MC{H}_{l(\gamma)}$ supported on $\gamma$.}  can be shown, then the according operator is automatically defined on all of $\Ho$.   
\\~\\
In the next section we will introduce the concept of matroids and oriented matroids and show how the connectedness of graphs as well as their local geometric embedding properties can be described within this framework.

~\\

\section{\label{Matroids and Oriented Matroids}Matroids and Oriented Matroids}
In this section we will give a brief introduction to the subject of matroids and oriented matroids. The presentation is mostly based on \cite{White1986} and \cite{Bjorner1999}. Partly we have also used \cite{Oxley1992,Goodman2004}. We have tried to stay close to the notation used there. 
Many definitions and proofs have been taken 
directly from these books. At some places we 
give page numbers. 
However we would like to emphasize that, in order to keep this introduction short, we have changed the manner of presentation of the subject.
In the literature one starts from different axiom systems, e.g.\ for chirotopes, (oriented) matroids in terms of (signed) circuits, (signed) bases, etc.\ and proves that these are equivalent. We take this equivalence for granted and use these objects equivalently in our presentation.  We have extended the explanation where we felt it would be for the benefit of the reader not familiar with (oriented) matroids.

In general matroids \cite{White1986} and oriented matroids \cite{Bjorner1999} can be thought of as a combinatorial abstraction of linear dependencies in a real vector space. In order to make the presentation more accessible to the reader, we will give some simple vector examples in section \ref{Matroids} and \ref{Oriented Matroids}.
However, as we will see later, the vector point of view is only one out of several possible representations for (oriented) matroids. Nevertheless for the beginning and in the subsequent definitions it is instructive to keep the vector picture in mind. This will be made more explicit in section \ref{Oriented Matroids from Vector Configurations}.

First we would like to fix some notation following \cite{White1986,Bjorner1999}. 
Let $E$ be a given finite\footnote{We will limit our presentation here to the finite case. However, there exists work on the infinite case e.g.\ \cite{Bruhn2009}.} set of $n$ elements.  We will also write $|E|=n$. 
We will denote by $2^E$ the set of all subsets of  $E$. It has cardinality $|2^E|=2^{|E|}=2^n$, as $\sum_{k=0}^n \left(n \atop k\right)=2^n$. The subsets of $2^E$ are called $families$ and will be denoted by script capital letters, i.e. $\MC{B}$.
The families can be regarded as elements of the set of subsets of $2^E$, $2^{2^E}$. We will call the subsets of $2^{2^E}$ $collections$ and denote them by boldface capital letters, i.e.\ $\bf B$.

An $antichain$ (also called \emph{incomparable family}) in $E$ is a family $\MC{B}$ of subsets of $E$ such that, for all $X,Y\in\MC{B}$, from $X\subseteq Y$ it follows that $X=Y$.

We will also need the notion of a {\it minimal} and {\it maximal} subset $B$ of $E$ with respect to a certain property $prop$ that holds for $B$.
This is achieved by ordering the subsets by set inclusion, and regarding a subset $B$ to be maximal (minimal) in $\MC{B}$ iff $\nexists B' \in \MC{B}$ such that $B \subset B'$ ($B \supset B'$) and also $prop$ holds for $B'$.

\subsection{\label{Matroids}Matroids} 
In this section we introduce the concept of a matroid. The reason for using underlined quantities as in \cite{Bjorner1999} will become clear in the next section, where we introduce signed subsets of a set. 

We start with the definition of a matroid in terms of its circuits. In the vector picture a circuit can be thought of as a minimal linearly dependent set of vectors. 

\begin{Definition}[Matroid from Circuits \cite{White1986}.]
   A family $\ul{\MC C}$ of subsets of a set $E$ is called the set of circuits of a matroid $\ul{\MC{M}}=(E,\ul{\MC{C}})$ on $E$ if
   
   \begin{tabular}{llp{0.7\linewidth}}
      (C0)&Non-emptiness: & $\emptyset\notin {\ul{\MC{C}}}$
      \\
      (C2)&Incomparability: & if $ {\ul{C}_1}\subseteq  {\ul{C}_2}$ then $ {\ul{C}_1}= {\ul{C}_2}$  $~~\forall~  {\ul{C}_1}, {\ul{C}_2}\in {\ul{\MC{C}}}$  ~~~($ {\ul{\MC{C}}}$ is an antichain)

      \\
      (C3)&Elimination: &For all $ {\ul{C}_1}, {\ul{C}_2}\in {\ul{\MC{C}}}$ with $ {\ul{C}_1}\!\ne\! {\ul{C}_2}$, and $\forall$  $e\in E$, $\exists~ {\ul{C}_3}\!\in\! {\ul{\MC{C}}}$ such that\linebreak  $ {\ul{C}_3}\subseteq ( {\ul{C}_1}\cup  {\ul{C}_2})\backslash\{e\}$.
   \end{tabular}
\end{Definition} 

Moreover we have the following definitions:
\begin{Definition}[Bases, Circuits, Hyperplanes and Cocircuits of a matroid $\ul{\M}$ \cite{White1986}.] \label{Circuits, Hyperplanes and Cocircuits of a matroid in terms of its bases intro}~\\

 \begin{tabular}{llll}

      (i)&Bases& \multicolumn{2}{l}{$\MC{B}:=\{B\subseteq E:~ B \text{~is maximal~}, ~\nexists\, \ul{C}\in\ul{\C},~B\supseteq \ul{C}~\}$}
      \\
      &&&\cmt{12}{$\MC{B}$ is a family of $maximal$ subsets $B$ of $E$ which contain no circuit $\ul{C}$ of $\ul{\MC M}$ as a subset. $\MC{B}$ is called a family of bases or basic family of $\ul{\MC M}$ and its sets $B$ are called bases of $\ul{\MC M}$. }
      \\\\
      (ii)&Circuits&\multicolumn{2}{l}{$\ul{\C}:=\{~\ul{C}\subseteq E:~ \ul{C} \text{~is minimal~}, ~\nexists\, B\in\mathcal{B},~\ul{C}\subseteq B~\}~~$}
      \\
      &&&\cmt{12}{Conversely to the definition of bases, the  circuits  $\ul{\C}$ of $\ul{\M}$ can be seen as a family of minimal subsets $\ul{C}$ of $E$ which are not contained in a basis $B$. }
      \\\\
      (iii)&Hyperplanes&\multicolumn{2}{l}{$\ul{\MC{H}}:=\{~\ul{H}\subseteq E:~ \ul{H} \text{~is maximal~}, ~\nexists\, B\in\mathcal{B},~\ul{H}\supseteq B\}$}
      \\
      &&&\cmt{12}{ The family $~\ul{\MC{H}}$ of hyperplanes of $\ul{\MC{M}}$ is given by the maximal subsets $\ul{H}$ of $E$ which contain no basis of   $\ul{\MC{M}}$ as a subset.}
      \\
      \\
      (iv)&Cocircuits&\multicolumn{2}{l}{$\ul{\C}^*=\{\ul{C}^*\subseteq E:~E\backslash \ul{C}^*\in\ul{\MC{H}} \}$ }
      \\
      &&&\cmt{12}{The family $~\ul{\MC{C}}^*$ consists of all subsets $\ul{C}^*$ of $E$ whose complement is a hyperplane. $~\ul{\MC{C}}^*$ is called the family of cocircuits (also called bonds in the literature) of $\ul{\MC{M}}$ .}
      
   \end{tabular}  
\end{Definition}

We can equivalently give the definition of a matroid in terms of its bases. 

\begin{Definition}[Matroid from Bases \cite{White1986}.] \label{Matroid from Bases of a Set}
   Let $E$ be a finite set. For a basic family $\MC{B}$ of subsets of $E$ the following axioms hold:

\begin{tabular}{rp{0.22\linewidth}p{0.64\linewidth}}
   (b1)&Nontriviality:& $\MC{B}\ne \emptyset$.
   \\
   (b2)&  Incomparability:&
           $\MC{B}$ is an antichain of subsets in $E$
   \\    
   (b3)& Basis-exchange axiom:&
    $\forall B_1,B_2\in\MC{B}$ and for every $b_1\in B_1\backslash B_2$, $\exists$  $b_2\in B_2\backslash B_1$ such that \linebreak $\big(B_1\backslash\{b_1\}\big)\cup \{b_2\}\in\MC{B}$. Note that this implies  $\big(B_2\backslash\{b_2\}\big)\cup \{b_1\}\in\MC{B}$ as well.      
\end{tabular}
 \\\\
 A pair $\ul{\MC{M}}=(E,\MC{B})$ is called a finite matroid on the ground set $E$.    
\end{Definition}
According to \cite{White1986} we will denote the collection of all basic families of $E$ by ${\bf B}(E)$. 
In this notion of a basis one can introduce a map $\ul{\chi}: 2^E\rightarrow \{0,1\}$ indicating whether a given subset $B\subseteq E$ is a basis (a member of the basic family $\MC{B}(\ul{\MC{M}})\in {\bf B}(E)$) or not:
\[
  \ul{\chi}_{\MC{B}}(B)=\left\{\begin{array}{ccl}
                        1 &~&\text{iff~} B\in\MC{B}
                        \\
                        0 &&\text{else}
                      \end{array}\right.
\]
Moreover as a consequence of {\it (b3)} any two bases $B_1,B_2\in \MC{B}$ of a matroid $\ul{\MC{M}}=(E,\MC{B})$ have the same cardinality, $|B_1|=|B_2|=:r$. The cardinality $r$ is often called the {\it rank}  of the  matroid $\ul{\MC{M}}$.
Finally we would like to introduce the notion of basic (co-)circuits of a matroid.
\begin{Definition}[Basic (fundamental) (co-)circuits of an element with respect to a basis \cite{Bjorner1999}.]\label{Basic (fundamental) (co-)circuits}
   Let $\ul{\MC{M}}=(E,\MC{B})$ be a matroid on $E$ and $B\in\MC{B}$ be a basis
 of $\ul{\MC{M}}$. Let $e\in E\backslash B$. Then there is a unique circuit $\ul{C}(e,B)$ of $\ul{\MC{M}}$ contained in $\{e\}\cup B$. The circuit $\ul{C}(e,B)$ is called the basic (or fundamental) circuit of $e$ with respect to $B$.
   \\
   Dually if $e\in B$ then there is a unique cocircuit $\ul{C}^*(e,B)$ of $\ul{\MC{M}}$ which is disjoint from the hyperplane $(B\backslash e)\in\ul{\MC{H}}$, that is $\ul{C}^*\cap (B\backslash e)=\emptyset$ or equivalently $E\backslash\ul{C}^* = (B\backslash e)\in\ul{\MC{H}}$. The cocircuit $\ul{C}^*(e,B)$ is called the basic (or fundamental) cocircuit of $e$ with respect to $B$.
\end{Definition}

Notice that the opposite does not hold in general: Given a (co-)circuit, there can be several bases from which this (co-)circuit can be obtained as a fundamental (co-)circuit by adding an additional element.

\begin{Lemma}[Intersection of fundamental circuits and cocircuits \cite{Bjorner1999}.]\label{Intersection of fundamental circuits and cocircuits} 
   Let $\ul{\MC{M}}$ be a matroid. Given any circuit $\ul{C}\in \ul{\MC{C}}(\ul{\MC{M}})$ and $e,f\in  \ul{C}$, $e\ne f$, then there is a cocircuit $\ul{C}^*\in \ul{\MC{C}}^*(\ul{\MC{M}})$, such that $\ul{C}\cap\ul{C}^*=\{e,f\}$. 
\end{Lemma}
To see this, let $B$ be a basis of  $\ul{\MC{M}}$, containing $\ul{C}\backslash f$. Clearly $f\notin B$, as otherwise $\ul{C}$ would be contained in $B$. By construction the basic cocircuit $\ul{C}^*(e,B)$ is such that $\ul{C}\cap\ul{C}^*=\{e,f\}$.
~\\\\

\subsubsection{Dual Matroid}
Finally we have a natural notion of duality between matroids \cite{White1986}. 
\begin{Definition}[Dual Matroid.]\label{Def dual matroid}
  Let  $\ul{\MC{\M}}=(E,\MC{B})$ be a matroid. A family $\MC{B}^\star\in{\bf B}(E)$ with
  \[
    \MC{B}^\star:=\{B^\star\subseteq E:~\exists B\in\MC{B},~ B^\star=E\backslash B\}
  \]
is called a dual basic family and gives rise to the matroid $\ul{\MC{M}}^\star=(E,\MC{B}^\star)$ which is called the {\it dual} matroid to $\ul{\MC{\M}}=(E,\MC{B})$.
\end{Definition}

It follows that the circuits $\ul{\C}$ of $\ul{\MC{\M}}=(E,\MC{B})$ are the cocircuits $\ul{C}^\star$ of $\ul{\MC{M}}^\star=(E,\MC{B}^\star)$ and vice versa.
To see this, notice that by definition \ref{Def dual matroid} the map ``$\star$'' is bijective. That is,  $\forall B\in \MC{B}$ there is a unique $B^\star\in\MC{B}^\star$ and vice versa, $\forall B^\star\in\MC{B}^\star$ there is a unique $ B\in \MC{B}$ such that $(B^\star)^\star=B$. Then, recalling the definition of $\ul{\C}$ in terms of $\MC{B}$ from definition \ref{Circuits, Hyperplanes and Cocircuits of a matroid in terms of its bases intro}  {\it (ii)}
\[
   \ul{\C}:=\{~\ul{C}\subseteq E:~ \ul{C} \text{~is minimal~}, ~\nexists\, B\in\mathcal{B},~\ul{C}\subseteq B~\}
\]
we can formally dualize it and write
\ba
   \ul{\C}^\star
   &:=&\{~\ul{C}^\star\subseteq E:~ \ul{C}^\star \text{~ minimal}, ~\nexists\, B^\star\in\mathcal{B}^\star,~\ul{C}^\star\subseteq B^\star~\}
  \NN
  &=&\{~\ul{C}^\star\subseteq E:~ \ul{C}^\star \text{~ minimal}, ~\nexists\, B\in\mathcal{B},~\ul{C}^\star\subseteq E\backslash B~\}
  \NN
  &=&\{~\ul{C}^\star\subseteq E:~ E\backslash\ul{C}^\star \text{ maximal}, ~\nexists\, B\in\mathcal{B},~E\backslash\ul{C}^\star\supseteq  B~\}
  \NN
  &=&\{~\ul{C}^\star\subseteq E:~ E\backslash\ul{C}^\star\in\ul{\MC{H}}\}
  \NN
  &=&\ul{\C}^*~~~.
\ea
Here, from the first to the second line we have used definition \ref{Def dual matroid} and bijectivity of ``$\star$'' in order to replace conditions of the form ``$\nexists B^\star\in\MC{B}^\star$" by ``$\nexists B\in\MC{B}$". From the second to the third line we have used the fact from elementary set theory that for two subsets $X,Y\subseteq E$ if $X\supset E\backslash Y$ then also $E\backslash X \subset E\backslash \{E\backslash Y\}$ which is equivalent to $E\backslash X \subset Y$. From the third to the fourth line we use definition \ref{Circuits, Hyperplanes and Cocircuits of a matroid in terms of its bases intro}  {\it (iii)} and finally from the fourth to the fifth line we use definition \ref{Circuits, Hyperplanes and Cocircuits of a matroid in terms of its bases intro}  {\it (iv)}.

Hence it is obvious that the circuits of $\ul{\M}$ are the cocircuits of $\ul{\M}^\star$ and the cocircuits of $\ul{\M}$ are the circuits of $\ul{\M}^\star$. Therefore we will drop the symbol ``$\star$'' in what follows, and write instead the common symbol ``$*$'' for dual objects, such as circuits $\ul{\C}$ and cocircuits $\ul{\C}^*$, bases $\MC{B}$ and dual bases $\MC{B}^*$, matroids $\ul{\MC{M}}$ and dual matroids $\ul{\MC{M}}^*$ etc.

\begin{figure}[hbt!]
\center
\cmt{8}
  {\center
    \includegraphics[width=3cm]{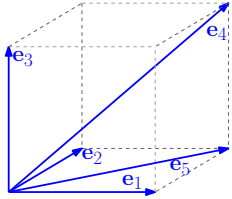}
    \caption{Vector example for a matroid.}
    \label{vector example}
   }
\end{figure}

\paragraph{Example.} Consider the set $E=\{{\bf e}_1,{\bf e}_2,{\bf e}_3,{\bf e}_4,{\bf e}_5\}$ of 5 vectors in $\mb{R}^3$ shown in figure \ref{vector example}. It is given by the column vectors of the matrix
\[
   A=\left(\begin{array}{ccccc}
      1&0&0&1&1 \\
      0&1&0&1&1 \\
      0&0&1&1&0
     \end{array}\right).     
 \]
 For every 3-element subset\footnote{This can be extended to all other subsets $B'$ of cardinality different from 3 by setting $\ul{\chi}_{\MC{B}}(B')=0$. } $B\subset E$ we consider the map
 \[   
  \ul{\chi}_{\MC{B}}(B)=\left\{\begin{array}{ccl}
                        1 &~&\text{iff~} \det{B}\ne 0
                        \\
                        0 &&\text{else}
                      \end{array}\right.  
                      ~~~.
\] 
We will abbreviate the elements ${\bf e}_k$ by its label $k$ for $k=1\,\ldots,5$, such that we can write $E=\{1,2,3,4,5\}$.
We have the non-uniform\footnote{For this example this means that we have colinear triples of vectors in $E$. See section \ref{Further Definitions} for the definition of uniform.} rank 3 vector matroid $\M=(E,\ul{\C})=(E,\MC{B})$ where the sets $\ul{\C}$ of circuits and respectively the set $\MC{B}$ of bases  are given by  
\barr{rcl}
   \ul{\C}&=&\big\{ \{1,2,3,4\},\{1,2,5\},\{3,4,5\}         
         \big\}
   \\
   \MC{B}&=&\big\{\{1,2,3\},\{1,2,4\},\{1,3,4\},\{1,3,5\},\{1,4,5\},\{2,3,4\},\{2,3,5\},\{2,4,5\}
          \big\}   
   \nonumber             
\earr
Moreover we have the set of $\ul{\MC{H}}$ of hyperplanes and the set $\ul{\C}^*$ of cocircuits and dual bases $\MC{B}^*$
\barr{rcl}
   \ul{\MC{H}}&=&\big\{ \{1,2,5\},\{3,4,5\},\{1,3\},\{1,4\},\{2,3\},\{2,4\}         
         \big\}
   \\
   \MC{\ul{\C}^*}&=&\big\{\{3,4\},\{1,2\},\{2,4,5\},\{2,3,5\},\{1,4,5\},\{1,3,5\}
          \big\} 
   \\
   \MC{B}^*&=&\big\{\{4,5\},\{3,5\},\{2,5\},\{2,4\},\{2,3\},\{1,5\},\{1,4\},\{1,3\}
          \big\}           
   \nonumber             ~~,
\earr
which can be used in order to define the rank 2 dual matroid $\MC{M}^*=(E,\ul{\C}^*)=(E,\MC{B}^*)$.
\subsection{\label{Oriented Matroids}Oriented Matroids} 
The concept of a rank $r$ matroid $\ul{\M}=(E,\MC{B})$  can be extended to the case of an oriented matroid $\M=(E,\MC{B})$. Notice however that not every matroid can be extended to an oriented matroid\footnote{For a discussion of criteria for extensibility of a matroid to an oriented matroid we refer to \cite{Bjorner1999}.}. 

For this we need to extend the definitions of the previous section.  We label
the elements of the ground set $E$ with positive integers $1, 2, \ldots
n=|E|$, and will often refer to $n$-tuples of subsets of $E$ in round
brackets $(\cdot)$ (for which the ordering of the elements is important).
The elements of a \emph{sorted} set (such as a basis) are understood to be
tuples in which the element labels are listed in increasing order.

Secondly we would like to extend the concept of circuits to signed circuits. For this we need the definition of {\it signed subsets} of $E$.

\begin{Definition}[Signed Subset.] \label{Signed Subset Def}
See \cite{Bjorner1999} p.\ 101 and following.

A signed subset $X$ of $E$ is a set $\underline{X}\subset E$ together with a partition $(X^+,X^-)$ of $\underline{X}$ into two distinguished subsets. $X^+$ is called the set of positive elements of $X$, $X^-$ is called the set of negative elements of $X$, $X^+\cap X^-=\emptyset$. The underlying set $\underline{X}=X^+\cup X^-$ is then called the support of $X$. 
Let $X,Y$ be signed subsets, $F$ be an unsigned subset of $E$. Then the following properties hold. 

\begin{tabular}{rlp{0.75\linewidth}}
   (i)&Equality:& $X=Y$ means $X^+=Y^+$ and $X^-=Y^-$.
   \\
   (ii)&  Restriction:&
           $X$ is called a restriction of $Y$ if $X^+\subseteq Y^+$ and $X^-\subseteq Y^-$
           \\&&
          $Y=X|_F$ is called the restriction of $X$ to $F$ if $Y^+=X^+\cap F$ and \linebreak  $Y^-=X^-\cap F$. 
   \\    
   (iii)& Composition:& $X\circ Y$ is the signed set defined by $(X\circ Y)^\pm=X^\pm\cup (Y^\pm\backslash X^\mp)$. Notice that this operation is associative, but not commutative in general.
   \\
   (iv)& Positivity:& $X$ is called positive/negative if $X^-=\emptyset$ / $X^+=\emptyset$. $X$ is called empty if \linebreak $X^-=X^+=\emptyset$.
   \\
   (v)& Opposite:& $-X$ with $(-X)^\pm=X^\mp$ is called the opposite of $X$.
   The signed set $\lu{-F~}X$ is obtained from $X$ by sign reversal or reorientation on $F$ by $(\lu{-F~}X)^\pm=(X^\pm \backslash F)\cup (X^\mp \cap F)$
   
\end{tabular}
\end{Definition}
For two signed sets $X,Y$ we write $e\in X$ if $e\in \ul{X}$. We denote the cardinality of $\underline{X}$ by $|\underline{X}|=|X|$. $X\backslash Y$ denotes the restriction of $X$ to $\underline{X}\backslash\underline{Y}$.

\begin{Definition}[Signed Subset. Signature.]\label{Signed Set I}
   A signed set $X$ can also be viewed as a set $\ul{X}\subset E$ together with a mapping $\sgn_X:\ul{X}\rightarrow \{-1,1\}$ 
such that $X^\pm = \big\{~e\in X~ :~\sgn_X(e)=\pm 1~\big\}$. The mapping $\sgn_X$ is called the signature of $X$.
\end{Definition}

\begin{Definition}[Signed Subset. Extended Signature.]\label{Signed Subset II}
   The signature of $X$ can be extended to all of $E$ by defining the extended signature map $\sgn_X: E \rightarrow \{-1,0,1\}^{|E|}$. Here $\sgn_X(e)=\pm 1$ if $e\in X^\pm\subset E$,  $\sgn_X(e)=0$ if $e\in E\backslash\underline{X}$.
\end{Definition}
$X$ can then be identified with an element in $\{-1,0,+1\}^{|E|}\equiv\{+,0,-\}^{|E|}$ and we will shortly write $X\in \{-1,0,+1\}^{|E|}$. 
Moreover by $\MC{S}\subseteq \{-1,0,+1\}^{|E|}$ we will denote a family $\MC{S}$ of signed subsets of $E$. 
For the $n$-element `ground tuple' 
$E=E_n=(1, 2, \ldots, n)$
we may denote $X$ as a {\it sign vector} $X\in  \{-1,0,+1\}^{n}$.
Finally we give a notion of orthogonality of signed subsets.
\begin{Definition}[Orthogonality of Signed Subsets.]\label{Orthogonality of signed sets}
    Two signed subsets $X=(X^+,X^-),Y=(Y^+,Y^-)$, $X,Y\subset E$ are said to be orthogonal $X\perp Y$ if $X\cap Y=\emptyset$ or the restrictions of $X$ and $Y$ to their intersection are neither equal nor opposite. \\That is, there are elements $e,f\in\ul{X}\cap\ul{Y}$ such that $\sgn_X(e)\sgn_Y(e)=-\sgn_X(f)\sgn_Y(f)$.
\footnote{This is motivated from the sign properties of the scalar product of two real vectors.
Consider a finite set $E=\{1,\ldots,n\}$ with cardinality $|E|=n$. 
Let ${\bf{u}}=\big(u_1,\ldots,u_n\big),{\bf v}=\big(v_1,\ldots,v_n\big)\in\mb{R}^{n}$ be orthogonal with respect to the Euclidean inner product $\big<\cdot,\cdot\big>$ of  $\mb{R}^{n}$. Define $X=(X^+,X^-)$, where $X^\pm:=\{i~:~u_i\gtrless 0\}$, and $Y=(Y^+,Y^-)$, where $Y^\pm:=\{i~:~v_i\gtrless 0\}$. From
\[
   \big<u,v\big>=\sum_{k=1}^n u_k\cdot v_k=0
\]
it follows that the non-zero terms in the sum cannot all have the same sign, that is if there are non-zero components  $u_k,v_k$, then their signs have to be different for at least one $k$. Hence there exists at least one $k$, contained in $X^+$ and $Y^-$ or $X^-$ and $Y^+$ respectively. According to definition \ref {Orthogonality of signed sets} we then say that the signed sets $X$ and $Y$ are orthogonal, $X\perp Y$.}

\end{Definition}
With these definitions we are now ready to give the definition of an oriented matroid in terms of its signed circuits \cite{Bjorner1999}:

\begin{Definition}[Oriented Matroid from Signed Circuits.]
\label{Oriented matroid circuit axioms}
   A family~ $\MC C$ of signed subsets of a set $E$ is called the set of signed circuits of an oriented matroid $\MC{M}=(E,\MC{C})$ on $E$ if
   
   \begin{tabular}{lll}
      (C0)&Non-emptiness: & $\emptyset\notin\MC{C}$
      \\
      (C1)&Symmetry: & $\MC{C}=-\MC{C}$~, that is for every $C\in\C$ also its opposite $-C\in\C$.
      \\
      (C2)&Incomparability: & if $\underline{C_1}\subseteq \underline{C_2}$ then either $C_1=C_2$ or $C_1=-C_2$ $~~\forall C_1,C_2\in\MC{C}$
      \\
      (C3)&Elimination: &\cmt{12}{For all $C_1,C_2\in\MC{C}$ with $C_1\!\ne\!-C_2$, if $e\in C_1^+\cap C_2^-$ $\exists~C_3\!\in\!\MC{C}$ such that\newline $C_3^\pm\subseteq (C_1^\pm\cup C_2^\pm)\backslash\{e\}$.}
   \end{tabular}
\end{Definition}

Now we would like to define the oriented matroid $\M$ in terms of a family $\MC{B}$ of bases of the underlying rank $r$ matroid $\ul{\M}$. An element $B\in\MC{B}$ can be written as a (sorted) $r$-tuple $B=(b_1,b_2,\ldots,b_r)$, where each element $b_k$ stands for an element of $E$. In particular there exists a permutation $\pi$, which brings all elements into lexicographic order according to $E$, such that $b_1<b_2<\ldots<b_r$. Then $B$ is called an ordered basis.

\begin{Definition}[Oriented Matroid from Oriented Bases of a Set.]
\label{Oriented Matroid from Oriented Bases of a Set}
   Let $\ul{\M}=(E,\MC{B})$ be a finite rank $r$ matroid over the ground set $E$. 
   An oriented matroid $\M=(E,\MC{B})$ on $E$ is given by the bases $\MC{B}$ of the underlying matroid $\ul{\MC{M}}$ together with a mapping 
   $\chi_{\MC{B}}:~E^r\ni B \rightarrow \{-1,0,+1\}$  with
   \[
    \chi_{\MC{B}}(B)=\left\{\begin{array}{ccl}
                        \pm1 &~&\text{iff~} B\in\MC{B}
                        \\
                        0 &&\text{else}
                      \end{array}\right.
   \]
   called the basis orientation or chirotope. The rank $r$ of $\M$ equals the rank of $\ul{\M}$ and for 
   $\chi_{\MC{B}}$ it holds that
     
   \begin{tabular}{lp{0.17\textwidth}p{0.68\textwidth}}
      (B1)& $\chi_{\MC{B}}$ is alternating: &  for $B=(b_1,\ldots,b_r)$ and $B'=(b_{\pi(1)},\ldots,b_{\pi(r)})$ consisting of the same elements as $B$ but in different order, related to the order in $B$ by a permutation $\pi$ we have 
      $\chi_{\MC{B}}(B)=\sgn(\pi)\cdot\chi_{\MC{B}}(B')$.
      \\
      (PV)&Pivoting property:& 
      for any two sorted bases $B,B'\in \MC{B}(\MC{M})$ with $B=(y,b_2,\ldots,b_r)$ and \linebreak  $B'=(z,b_2,\ldots,b_r)$, $y\ne z$ we have 
      $\chi_{\MC{B}}(B)=-\sgn_C(y)\cdot \sgn_C(z) \cdot \chi_{\MC{B}}(B')$ where $C$ is one of either of the two (opposite) signed circuits of $\MC{M}$ whose support is contained in the set $\{y,z,b_2,\ldots,b_r\}$.
      
   \end{tabular}
\end{Definition}
We would like to remark that in general $\chi_{\MC{B}}$ is only fixed up to  an overall sign: if $\chi_{\MC{B}}$ is a basis orientation for $\M(E,\MC{B})$, then also $-\chi_{\MC{B}}$ is a basis orientation. This must be kept in mind in subsequent computations. See also section \ref{Signed Basis Graphs}.

\paragraph{Signed Cocircuits.}  With definition \ref{Orthogonality of signed sets} we can define the set $\C^*$ of
{\it signed cocircuits} of the rank $r$ oriented matroid $\M$ as follows:  
\be\label{Def Cocircuit Signature}
 \C^*:=\{C^*=({C^*}^+,{C^*}^-)~:~\ul{C}^*\subseteq E~,~C^*\perp C~\forall C\in \C~\text{(in the sense of definition \ref{Orthogonality of signed sets})}\}~~.
\ee
Alternatively to this definition we can equip the set $\ul{\C}^*$ of cocircuits of the underlying matroid $\ul{\M}$ with a signature as follows \cite{Bjorner1999}.
Each support $\ul{C}^*$ is the complement of a hyperplane $\ul{H}\in\ul{\MC{H}}$ as in definition \ref{Circuits, Hyperplanes and Cocircuits of a matroid in terms of its bases intro}, that is $\ul{H}=E\backslash \ul{C}^*$. By construction $\ul{H}=\{h_1,\ldots,h_{r-1}\}$ is a maximal subset of $E$ not containing a basis $B\in\MC{B}(\M)$. Then the signature of $C^*$ can be constructed as   
$(C^*)^\pm:=\{e\in E\backslash \ul{H}~:~\chi_{\MC{B}}(h_1,\ldots,h_{r-1},e)=\pm 1\}$\footnote{
Here it makes no difference if e.g.\ $e<h_{r-1}$, since this would only affect the sign of $\pm 1$.

If $\ul{\M}$ is non-uniform, it might happen that $\ul{H}$ contains more than $r-1$ elements. In that case take any ordered subset of $\ul{H}$ which contains $r-1$ elements that span $\ul{H}$.

}.
In section \ref{Oriented Matroids from Vector Configurations} a geometric interpretation for $\C^*$ is given. 
\\

Additionally we have the following definition.
\begin{Definition}[Basic (fundamental) signed (co-)circuits of an element wrt.\ an oriented basis.]~\\
\label{Basic (fundamental) signed (co-)circuits of an oriented matroid.}
   Let $\M=(E,B)$ be a given oriented matroid of rank r with underlying matroid $\ul{\M}$. Let $\ul{C}=(e,B)$ be a fundamental circuit of $\ul{\M}$ as given in definition \ref{Basic (fundamental) (co-)circuits}. Then the signed circuit $C\in\C$ supported on $\ul{C}$ and having $\sgn_C(e)=+1$ is called the basic or fundamental circuit of $e$ with repsect to $B\in\MC{B}$. Dually, if $e\in B$ (as in definition \ref{Basic (fundamental) (co-)circuits}) let $\ul{C}^*=\ul{C}^*(e,B)$ be the unique cocircuit of $\ul{\M}$ disjoint from the hyperplane $B\bs e$. Then the cocircuit $C^*\in\C^*$ supported on $\ul{C}^*$ and positive on $e$ is called the basic or fundamental cocircuit of $e$ with respect to $B\in\MC{B}$.  
\end{Definition}

\subsubsection{Dual Oriented Matroid}

Given any $B,B'\in\MC{B}$ with $B=(y,b_2,\ldots,b_r)$, $B'=(z,b_2,\ldots,b_r)$ and $y\ne z$ we may construct the basic circuit
\[
  \ul{C}:=\ul{C}(z,B)=\ul{C}(y,B')=\{y,z,b_2,\ldots,b_r\}~ .
\]
 Moreover using $B^*,B'^*\in\MC{B}^*$ with $B^*=E\backslash B$ and $B'^*=E\backslash B'$ we may construct the basic cocircuit 
\[
      \ul{C}^*:=\ul{C}^*(z,B)
               = E\backslash \{B\backslash z\}
               =\{E\backslash B\}\cup\{z\}
               = B^*  \cup\{z\}
               = B'^* \cup\{y\}
               =\{E\backslash B'\}\cup\{y\}
               = E\backslash \{B'\backslash y\} 
               =\ul{C}^*(y,B').
\]
Here once again we see very nicely that the basic cocircuits with respect to the basic family $\MC{B}$ can be regarded as basic circuits with respect to the basic family $\MC{B}^*$.
Clearly we have $\ul{C}\cap \ul{C}^*=\{y,z\}$. Now choose one of the two oppositely signed circuits $\pm C\in \C$ and $\pm C^*\in \C^*$ such that for example $\sgn_{C}(y)=\sgn_{C^*}(y)=+1$.\footnote{This can always be done without loss of generality, because $\pm C\in\C$ and $\pm C^*\in\C^*$ by construction.}
As $C\perp C^*$ by construction we get  from definition \ref{Orthogonality of signed sets} a chirotope $\chi_{\MC{B}^*}(B^*)$ for any $B^*\in\MC{B}^*$ from 
\be\label{Orthogonality and dual chirotope}
   \chi_{\MC{B}}(B)\cdot\chi_{\MC{B}}(B')
   =-\sgn_C(y)\cdot\sgn_C(z)=\sgn_{C^*}(y)\cdot\sgn_{C^*}(z)
   =-\chi_{\MC{B}^*}(B^*)\cdot\chi_{\MC{B}^*}(B'^*)~~.
\ee
In the last step we have used the pivoting property of definition \ref{Oriented Matroid from Oriented Bases of a Set} in its dualized form\footnote{Explicitly given in definition \ref{Dual Oriented Matroid}.}. Note that this identity is independent of the choice of $C,C^*$ above, as it only uses the relative signs of $y,z$ with respect to $C,C^*$. 
Using (\ref{Orthogonality and dual chirotope}) we may define the dual matroid chirotope $\chi_{\MC{B}^*}: E^{n-r} \rightarrow \{-1,0,+1\}$ as follows. Consider the lexicographically sorted  $E=(1,2,\ldots,n)$ and an arbitrary sorted $(n-r)$-tuple $T^*=(t_1^*,\ldots,t_{n-r}^*)$ of elements in $E$. Write $T=(t_1,\ldots,t_r)$ for some permutation of $E\backslash T^*$. Clearly, there is a permutation $\pi$ mapping $(T^*,T):=(t_1^*,\ldots,t_{n-r}^*,t_1,\ldots,t_r)$ to $(1,2,\ldots,n)$ with parity $\sgn{\pi}=:\sgn\big((T^*,T)\big)$. Now we can define
\ba\label{Def dual chirotope}
     \chi_{\MC{B}^*}: E^{n-r} &\rightarrow& \{-1,0,+1\} 
     \NN
     T^*=(t_1^*,\ldots,t_{n-r}^*) &\mapsto&\chi_{\MC{B}^*}(T^*):=
     \chi_{\MC{B}}(T)\cdot\sgn\big((T^*,T)\big) \;\;.
\ea  
Note that this definition is compatible with (\ref{Orthogonality and dual chirotope}), and directly relates the values of chirotope and dual chirotope, as follows.
Choose $B=(y,b_2,\ldots,b_r)$, $B'=(z,b_2,\ldots,b_r)$ and $B^*,B'^*\in\MC{B}^*$ with $B^*=E\backslash B=(z,b_2^*,\ldots,b_{(n-r)}^*)$, 
$B'^*=E\backslash B'=(y,b_2^*,\ldots,b_{(n-r)}^*)$ and $C,C^*$as above. Clearly
\ba
   \chi_{\MC{B}^*}(B^*)
   &(\ref{Def dual chirotope})\atop =&\chi_{\MC{B}}(B)\cdot\sgn(z,b_2^*,\ldots,b_{(n-r)}^*,y,b_2,\ldots,b_r)
   \NN
   &{\it (PV)}\atop =&-\sgn_C(y)\cdot\sgn_C(z)\cdot\chi_{\MC{B}}(B')
                              \cdot\sgn(z,b_2^*,\ldots,b_{(n-r)}^*,y,b_2,\ldots,b_r)
   \NN
   &(\ref{Def Cocircuit Signature})\atop=&~~ \sgn_{C^*}(y)\cdot\sgn_{C^*}(z)\cdot\chi_{\MC{B}}(B')
                              \cdot\sgn(z,b_2^*,\ldots,b_{(n-r)}^*,y,b_2,\ldots,b_r)
   \NN
   &z\leftrightarrow y\atop =& -\sgn_{C^*}(y)\cdot\sgn_{C^*}(z)\cdot\chi_{\MC{B}}(B')
                              \cdot\sgn(y,b_2^*,\ldots,b_{(n-r)}^*,z,b_2,\ldots,b_r)
   \NN
   &(\ref{Def dual chirotope})\atop=& -\sgn_{C^*}(y)\cdot\sgn_{C^*}(z)\cdot\chi_{\MC{B^*}}(B'^*)     ~~~.                    
\ea

\begin{Definition}[Dual Oriented Matroid.]\label{Dual Oriented Matroid}

  Let $\MC{M}=(E,\MC{B})$ be an oriented rank $r$ matroid with chirotope
  $\chi_{\MC{B}}$ over the ordered ground set $E=(1,\ldots,n)$. A family
  $\MC{B}^*\in{\bf B}(E)$ with
  \[
    \MC{B}^*:=\{E\backslash B~:~B\in\MC{B}\}
  \]
together with the map $\chi_{\MC{B}^*}:E^{n-r}\ni B^*\rightarrow \{-1,0,+1\}$ for
which $B^*=E\backslash B$, $B\in E^r$
\[
  \chi_{\MC{B}^*}(B^*)=\chi_{\MC{B}}(B)\cdot\sgn((B^*,B))
                      =\left\{\begin{array}{cl}
                          \pm1 & \text{iff~} B^*\in\MC{B}^*
                                 \text{which is equivalent to~} 
                                 B\in\MC{B}
                           \\
                           0 &  \text{iff~} B^*\notin\MC{B}^*
                                 \text{which is equivalent to~} 
                                 B\notin\MC{B}                
                       \end{array}\right. 
\]
with $\sgn((B^*,B))$ defined as above gives rise to an oriented matroid
$\M^*=(E,\MC{B}^*)$, called the dual oriented matroid to $\MC{M}=(E,\MC{B})$,
of rank $(n-r)$.
In particular it holds for $\chi_{\MC{B}^*}$ that

\begin{tabular}{lp{0.17\textwidth}p{0.66\textwidth}}
      (B1)& $\chi_{\MC{B}^*}$ is alternating: & for all $B^*=(b_1^*,\ldots,b_{n-r}^*)$ and $B'^*=(b_{\pi(1)}^*,\ldots,b_{\pi(n-r)}^*)$ consisting of the same elements as $B^*$ but in different order, related to the order in $B^*$ by a permutation $\pi$, we have 
      $\chi_{\MC{B}^*}(B^*)=\sgn(\pi)\cdot\chi_{\MC{B^*}}(B'^*)$.
      \\
      (PV*)&Dual Pivoting property:& 
      for any  $B^*,B'^*\in \MC{B}^*$ with $B^*=(y,b_2^*,\ldots,b_{(n-r)}^*)$~
       and  $B'=(z,b_2^*,\ldots,b_{(n-r)}^*)$ where $y\ne z$ we have~~~~~~~~~~~\linebreak 
      $\chi_{\MC{B}^*}(B^*)=-\sgn_{C^*}(y)\cdot \sgn_{C^*}(z) \cdot \chi_{\MC{B}^*}(B'^*)$, where $C^*$ is one of the two oppositely signed cocircuits of $\MC{M}$ whose support is contained in the set $\{y,z,b_2^*,\ldots,b_{(n-r)}^*\}$.
\end{tabular}

\end{Definition}
Note that again it holds that $(\MC{M}^*)^*=\MC{M}$. The signed cocircuits of $\M$ are the signed circuits of $\M^*$ and the signed circuits of $\M$ are the signed cocircuits of $\M^*$.
\\\\
Notice that $(PV)$ and $(PV^*)$ are equivalent for a map $\chi_{\MC B}$ if $\M$ is an oriented matroid due to (\ref{Orthogonality and dual chirotope}) \cite{Bjorner1999}.
\\\\
In practice the identity in equation (\ref{Orthogonality and dual chirotope}) might often give an efficient way to compute the dual chirotope $\chi_{\MC{B}^*}$ from the chirotope $\chi_{\MC{B}}$.

\paragraph{Example continued.} For the vector example at the end of section \ref{Matroids} we have $E=\big(1,2,3,4,5\big)$ and the according set of sorted bases $\MC{B}$. The set $\C$ of signed circuits is given by $\MC{C}=\{\pm C_1,\pm C_2,\pm C_3\}$ with
\barr{c||c|c|c}
  & C_1 &C_2&C_3
   \\\hline
   C_k^+&\{ 1,2,3 \}
    & \{ 1,2\}&\{ 3,5 \}
   \\
   C_k^-&\{4\}&\{5\} & \{4\}
   \nonumber~~.
\earr
For the chirotope we consider for every 3 element subset $B$ of $E$ the map
 \[   
  \chi_{\MC{B}}(B)=\left\{\begin{array}{ccl}
                        \sgn(\det B) &~&\text{iff~} \det{B}\ne 0
                        \\
                        0 &&\text{otherwise}
                      \end{array}\right.  
                      ~~~.
\]
which leads to the following basis orientation\footnote{If $\chi_{\MC{B}}$ is a chirotope, then $-\chi_{\MC{B}}$ is also, depending of our notion of `positive' orientation.} :
\barr{c||c|c|c|c|c|c|c|c|c|c}
   B &(1,2,3) &(1,2,4)&(1,2,5)&(1,3,4)&(1,3,5)&(1,4,5)&(2,3,4)&(2,3,5)&(2,4,5)&(3,4,5)
   \\\hline
   \chi_{\MC{B}}(B)&
       +&+&0&-&-&-&+&+&+&0
   \nonumber
\earr
~\\

\subsubsection{\label{Signed Basis Graphs}Conversion from Signed Circuits to Signed Bases  via Signed Basis Graphs} 

Here we follow \cite{Bjorner1999} p.\ 132 and following.

The equivalence of defining an oriented matroid $\MC{M}$ of rank $r$ over a ground set $E$ in terms of its signed circuits $\MC{C}$, as in definition \ref{Oriented matroid circuit axioms}, or in terms of its oriented basis $\MC{B}$, as in theorem \ref{Oriented Matroid from Oriented Bases of a Set}, can be made explicit by using the so-called basis graph $BG_{\ul{\MC{M}}}$ of the underlying matroid $\ul{\MC{M}}$, which will be constructed in the sequel.
Choose a family $\MC{B}$ of bases of $\ul{\MC{M}}$. Associate to each basis $B\in\MC{B}$ a vertex of $BG_{\ul{\MC{M}}}$. Now link any two bases $B,B'\in\MC{B}$ by an edge if they differ by exactly one element,
for example $B=(b_1,b_2,\ldots,b_r)$, $B'=(b'_1,b_2,\ldots,b_r)$ and $b_1\ne b_1'$. By construction $BG_{\ul{\MC{M}}}$ is connected.
\\\\
Now let $\MC{M}=(E,\MC{C})$ be given. Fix an ordering of the ground set $E$ such that $E=(1,2,\ldots,n)$. Sort the elements of each basis $B\in\MC{B}$ according to that ordering. Now assume we have two sorted bases $B,B'\in\MC{B}$ with $B=(b_1,\ldots,b_{r})$ and  $B'=(b'_1,\ldots,b'_{r})$ such that 
$B\Delta B':=(B\cup B')\backslash (B\cap B')=\{b_i,b'_j\}$. Let $C\in\MC{C}$ be one of the two oppositely signed  circuits\footnote{That is, the fundamental (basic) signed circuit or its opposite, as given by definition \ref{Basic (fundamental) signed (co-)circuits of an oriented matroid.}.} $C(b'_j,B)$ whose support is contained\footnote{See section \ref{Oriented Matroids from Oriented Graphs} for an example in which it is a proper subset.} in the set $\{b_1,\ldots,b_{r},b'_j\}\supseteq \ul{C}(b'_j,B)$. 
\\\\
Then by {\it (B1)} of definition \ref{Oriented Matroid from Oriented Bases of a Set} we have:
\barr{rclclcl}
   \chi_{\MC{B}}(b_i,b_1,\ldots,b_{i-1},b_{i+1},\ldots,b_r)&=&(-1)^{i-1}\chi_{\MC{B}}(B)\\
   \chi_{\MC{B}}(b'_j,b'_1,\ldots,b'_{j-1},b'_{j+1},\ldots,b'_r)&=&(-1)^{j-1}\chi_{\MC{B}}(B')~~~.
\earr
Moreover by construction $(b_1,\ldots,b_{i-1},b_{i+1},\ldots,b_r)\equiv(b'_1,\ldots,b'_{j-1},b'_{j+1},\ldots,b'_r)$ and we can define
\ba\label{Sign of SBG}
  \eta(B,B'):=\chi_{\MC{B}}(B)\cdot\chi_{\MC{B}}(B')
  &=&(-1)^{i+j}\cdot \chi_{\MC{B}}(b_i,b_1,\ldots,b_{i-1},b_{i+1},\ldots,b_r)\cdot
   \chi_{\MC{B}}(b'_j,b'_1,\ldots,b'_{j-1},b'_{j+1},\ldots,b'_r)
  \NN
  &=&(-1)^{1+i+j}  \cdot\sgn_C(b_i)\cdot\sgn_C(b'_j) 
\ea
where we have used {\it (PV)} of definition \ref{Oriented Matroid from Oriented Bases of a Set} in the last line.
Hence, starting from an arbitrarily chosen $B_0\in\MC{B}$, we can construct the so-called $signed~basis~graph$ $SBG_{\MC{M}}$ by attaching $\eta(B,B')$ to every edge of $BG_{\ul{\MC{M}}}$. Clearly   $SBG_{\MC{M}}$ depends on the ordering chosen on the ground set $E$.
\\\\
A chirotope $\chi: E^r\rightarrow \{-1,0,+1\}$ can be constructed from $SBG_{\MC{M}}$ as follows. Choose an arbitrary $B_0\in\MC{B}$. Set $\chi(B_0)=1$. Then for any $B\in\MC{B}$ we have
\[
  \chi(B)=\prod_{i=1}^k \eta(B_{i-1},B_i)
\]
where $B_0,B_1,B_2,\ldots,B_k=B$ is an arbitrary path from $B_0$ to $B$ contained in $SBG_{\MC{M}}$.
It can be shown that this definition is consistent, that is $\chi(B)$ is independent of the choice of a path, see \cite{Bjorner1999} p.\ 132 and following.

Notice that this construction can easily be applied to cocircuits and the dual oriented matroid $\M^*$. Moreover having computed the chirotope $\chi_{\MC{B}}$ of $\M=(E,\C)$, one can read off the chirotope for the dual matroid by using identity (\ref{Orthogonality and dual chirotope}).

\subsection{\label{Further Definitions}Further Definitions}  
Let us complete our 
 introduction to oriented matroids by giving definitions which frequently occur in the literature \cite{Bjorner1999}. Let the oriented matroid  $\M=(E,\C)$ be given with its set $\C$ of signed circuits. 
\paragraph{Vectors and covectors.} A  composition $C_1\circ\ldots\circ C_k$ of signed circuits in the sense of {\it (iii)} of definition \ref{Signed Subset Def} is called a {\it vector} of $\M$. The set of all vectors of $\M$ is denoted by $\MC{V}$. 
Accordingly a composition of signed cocircuits of $\M$ is called a {\it covector} of $\M$. The set of all covectors of $\M$ is denoted by $\MC{L}$. Note that by construction vectors and covectors of $\M$ are signed subsets of $E$. In particular any $v\in\MC{V}$ and any $l\in\MC{L}$ determines an extended signature in the sense of definition  \ref{Signed Subset II}.
Instead of giving $\M$ in terms of its signed circuits $\C$ or cocircuits $\C^*$, $\M$ can also be determined in terms of its vectors $\MC{V}$ or covectors $\MC{L}$ \cite{Bjorner1999}.

\paragraph{Loops and coloops.} Moreover an element $e\in E$ is called a {\it loop}\footnote{In the picture of a vector configuration a loop corresponds to a zero vector.} of $\M$ if the 
signed set $(\{e\},\emptyset)\in\C$. The subset of loops contained in $E$ is denoted by $E_o$.
If $e\notin C$ $\forall C\in\C$ then $e$ is called a {\it coloop} of $\M$. 

\paragraph{Parallel elements.} Two elements $e,f\in E\backslash E_o$ are called {\it parallel} $e\| f$ if $\sgn_l(e)=\sgn_l(f)$ or  $\sgn_l(e)=-\sgn_l(f)$ for all covectors $l\in\MC{L}$.

\paragraph{Simple oriented matroid.}An oriented matroid $\M$ is called {\it simple} if it does not contain loops or parallel elements.

\paragraph{Uniform oriented matroid.}An oriented matroid $\M=(E,\MC{B})$ of rank $r$ is called {\it uniform} if {\bf every} $r$-element subset $X\subseteq E$ is contained  in the family of bases $\MC{B}$, that is $\chi_{\MC{B}}(X)\ne 0$ $\;\forall X\subset E$ such that $|X|=r$.

\subsection{\label{Oriented Matroids from Oriented Graphs}Oriented Matroids from Directed Graphs}
In this section we will show how the abstract framework of oriented matroids naturally arises from directed graphs. This will be done by reviewing an example from \cite{Bjorner1999} in great detail.

\begin{figure}[hbt!]
\center
\cmt{8}
  {\center
    \psfrag{e1}{$e_1$}
    \psfrag{e2}{$e_2$}
    \psfrag{e3}{$e_3$}
    \psfrag{e4}{$e_4$}
    \psfrag{e5}{$e_5$}
    \psfrag{e6}{$e_6$}
    
    \psfrag{v1}{$v_1$}
    \psfrag{v2}{$v_2$}
    \psfrag{v3}{$v_3$}
    \psfrag{v4}{$v_4$}
       
    \includegraphics[width=5cm]{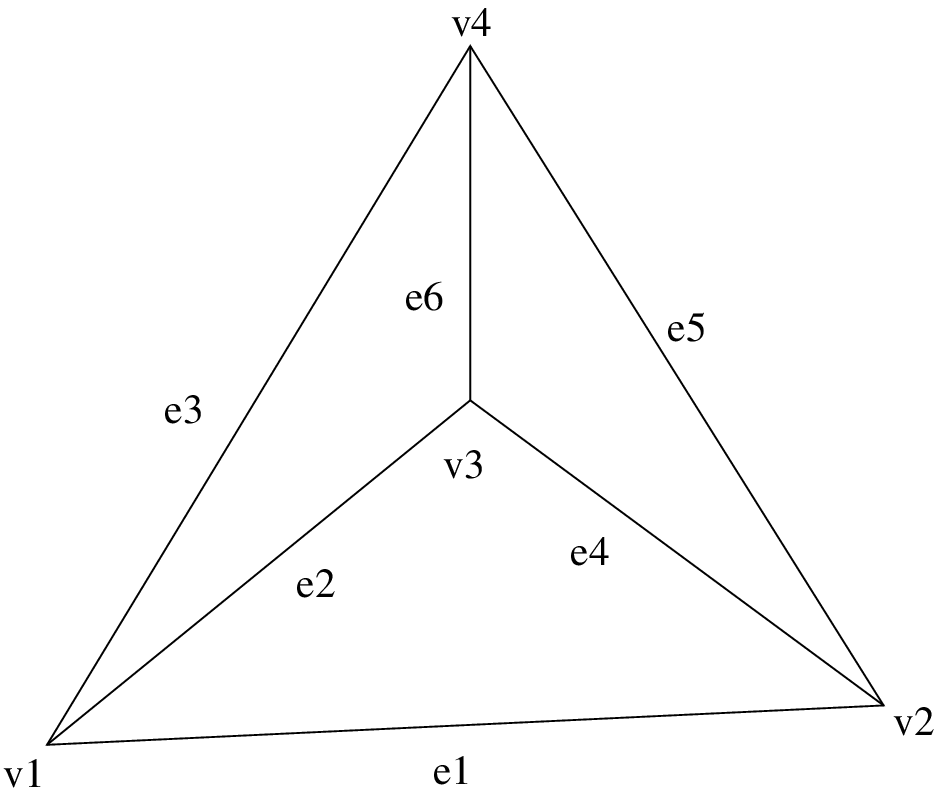}
    \caption{Undirected graph $\ul{\gamma}$.}
    \label{undirected graph}
   }
  \cmt{8}
  {\center
    \psfrag{e1}{$e_1$}
    \psfrag{e2}{$e_2$}
    \psfrag{e3}{$e_3$}
    \psfrag{e4}{$e_4$}
    \psfrag{e5}{$e_5$}
    \psfrag{e6}{$e_6$}
    
    \psfrag{v1}{$v_1$}
    \psfrag{v2}{$v_2$}
    \psfrag{v3}{$v_3$}
    \psfrag{v4}{$v_4$}
    
    \psfrag{direction}{\blue\Huge $\circlearrowleft$}
    \psfrag{orientation}{\blue $\{1,2,5,6\}$}
       
    \includegraphics[width=5cm]{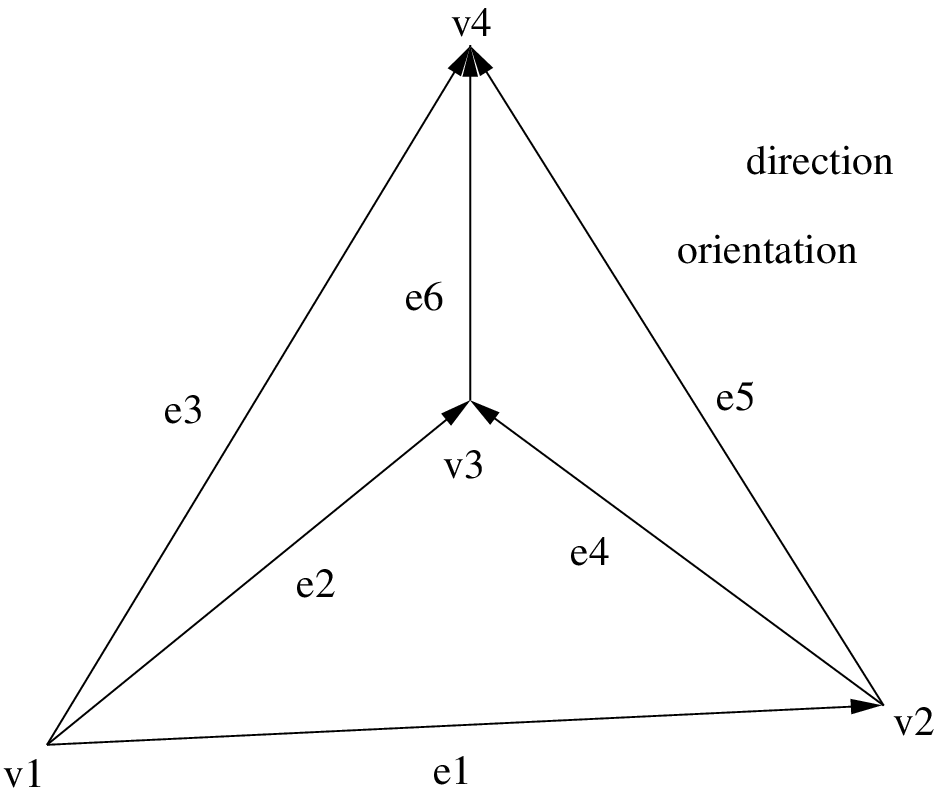}
    \caption{Directed graph $\gamma$.}
    \label{directed graph}
   }
  
\end{figure}

Consider the graph $\ul{\gamma}$ of figure \ref{undirected graph} with edge set $E=E(\ul{\gamma})=\{e_1,\ldots,e_6\}$ and vertex set $V(\ul{\gamma})=\{v_1,\ldots,v_4\}$. We notice that $\ul{\gamma}$ contains simple (undirected) cycles $\ul{X}$, for example $\{1,2,5,6\}$. The set of all such cycles is denoted by $\ul{\MC{C}}$. Then $\ul{\MC{C}}$ defines the circuits of a matroid $\ul{\MC{M}}=(E,\ul{\MC{C}})$.

Now consider the directed graph $\gamma$ of figure \ref{directed graph} now with a set $E(\gamma)=\{e_1,\ldots,e_6\}$ of oriented edges and vertex set $V(\gamma)=\{v_1,\ldots,v_4\}$. Consider the cycle  $\ul{X}=\{1,2,5,6\}$. If we introduce an anti-clockwise orientation for this cycle as indicated by $\blue\circlearrowleft$ in figure \ref{directed graph}, then it contains positive elements $X^+=\{1,5\}$ and negative elements $X^-=\{2,6\}$. Hence $X=(X^+,X^-)$ is a signed circuit or in the wording of definition \ref{Signed Subset Def}. $X=(X^+,X^-)$ is a signed set with support $\ul{X}=X^+\cup X^-$ . The set of all signed circuits obtained in this way from the oriented graph $\gamma$ is denoted by $\MC{C}$.  
This defines the oriented matroid $\MC{M}=(E,\MC{C})$.
We sometimes refer to an oriented matroid which arises from a directed graph
in this way as a \emph{graphic oriented matroid}.

We may define the extended signature $\sgn_X$  introduced in definition \ref{Signed Subset II} with respect to the chosen orientation $\circlearrowleft$, to write the signed circuit $X$ as an element of $\{+,0,-\}^{|E(\gamma)|}$:   $X=\{\sgn_X(e_1),\ldots,\sgn_X(e_6)\}=\{-,+,0,0,-,+\}$ or shorter (denoting $e_k$ as simply $k$)  $X=1\ol{2}5\ol{6}$. In this notation the set $\MC{C}$ can be written as a matrix, where each circuit is a row or the negative of a row \cite{Bjorner1999}:
\[
   \begin{array}{ccc}
    \MC{C}=&\left(
      \begin{array}{cccccc}
         +&-&0&+&0&0\\
         +&0&-&0&+&0 \\
         0&+&-&0&0&+ \\
         0&0&0&-&+&- \\
         +&-&0&0&+&- \\
         +&0&-&+&0&+ \\
         0&+&-&-&+&0 \\
      \end{array}
   \right)
   &
   \begin{array}{l}
      \leftarrow C_1=1\ol{2}4\\
      \leftarrow C_2=1\ol{3}5\\
      \leftarrow C_3=2\ol{3}6\\
      \leftarrow C_4=\ol{4}5\ol{6}\\
      \leftarrow C_5=1\ol{2}5\ol{6}\\
      \leftarrow C_6=1\ol{3}46\\
      \leftarrow C_7=2\ol{3}\ol{4}5\\
   \end{array}
   \end{array}
\]
It is instructive to check that $\MC{C}=\{\pm C_1,\ldots,\pm C_7\}$ indeed fulfills the oriented matroid circuit axioms of definition \ref{Oriented matroid circuit axioms}. In particular $(C2)$ holds. 
\\\\
The bases $\MC{B}$ for $\ul{\MC{M}}$, $\MC{M}$ are given by the set of spanning trees $T_{(\gamma)}$ of $\gamma$ \cite{White1986}. 
For a connected graph (or component) $\gamma$ the spanning tree $T_{(\gamma)}$ is given by a collection of  edges which uniquely connects any two vertices $v_1,v_2$ of $\gamma$ by a path. If $\gamma$
contains $|V(\gamma)|$ vertices, then $T_{(\gamma)}$ consists of $|V(\gamma)|-1$ edges. Accordingly the rank of $\MC{M}_\gamma$ is $|V(\gamma)|-1$.

If $\gamma$  is not connected but consists of $N$ connected components $\gamma=\{\gamma_1,\gamma_2,\ldots,\gamma_N\}$, then $\MC{B}$ is given by the collection of trees  $T_{(\gamma)}:=\{T_{(\gamma_1)},T_{(\gamma_2)},\ldots,T_{(\gamma_N)}\}$, which is also called the forest \cite{Diestel2005} of $\gamma$.

Hence in our example for $\gamma$ in figure \ref{directed graph} we find that $\MC{M}_\gamma$ has rank 3. By {\it (i)} in definition \ref{Circuits, Hyperplanes and Cocircuits of a matroid in terms of its bases intro} of bases as maximal subsets of $E$ containing no circuit we find that all triples except\footnote{The fact that we have 3-element circuits indicates that $\M_\gamma$ is non-uniform.} $\ul{C}_1,\ul{C}_2,\ul{C}_3,\ul{C}_4$ of $E(\gamma)=\{1,2,3,4,5,6\}$ are bases.

In order to compute the according chirotope, we use the signed basis graph construction (\ref{Sign of SBG}) of section \ref{Signed Basis Graphs}: As the chirotope is only given up to an overall sign, 
we choose \fbox{$\chi_{\MC{B}}(1,2,3)=-1$} for the beginning and successively apply (\ref{Sign of SBG}), 
$\chi_{\MC{B}}(B')=\chi_{\MC{B}}(B)\cdot (-1)^{1+i+j}  \cdot\sgn_C(b_i)\cdot\sgn_C(b'_j) $~~:

\barr{|c|c||c|c|c|c|c||c|c|c||c|}\hline
 B\!=\!(b_1,b_2,b_3) & B'\!=\!(b_1',b_2',b_3') & B\Delta B'\!=\!\{b_i,b_j'\} & i & j & B\cup B' & \supseteq\ul{C} & \sgn_C(b_i)&\sgn_C(b_j')&\chi_{\MC{B}}(B)&\chi_{\MC{B}}(B')
 \\\hline\hline
 (1,2,3) & (1,2,5) &\{3,5\}& 3&3 &\{1,2,3,5\}&{\ul{C}}_2 & -1& +1&-1 &-1 
 \\\hline
 (1,2,3) & (1,2,6) &\{3,6\}& 3&3 &\{1,2,3,6\}&{\ul{C}}_3 & -1& +1&-1 &-1 
 \\\hline
 (1,2,3) & (1,3,4) &\{2,4\}& 2&3 &\{1,2,3,4\}&{\ul{C}}_1 & -1& +1&-1 &+1 
 \\\hline
 (1,2,3) & (1,3,6) &\{2,6\}& 2&3 &\{1,2,3,6\}&{\ul{C}}_3 & +1& +1&-1 &-1 
 \\\hline
 (1,2,3) & (2,3,4) &\{1,4\}& 1&3 &\{1,2,3,4\}&{\ul{C}}_1 & +1& +1&-1 &+1 
 \\\hline
 (1,2,3) & (2,3,5) &\{1,5\}& 1&3 &\{1,2,3,5\}&{\ul{C}}_2 & +1& +1&-1 &+1 
 \\\hline
 (1,3,4) & (1,4,5) &\{3,5\}& 2&3 &\{1,3,4,5\}&{\ul{C}}_2 & -1& +1&+1 &-1 
 \\\hline
 (1,2,6) & (1,4,6) &\{2,4\}& 2&3 &\{1,2,4,6\}&{\ul{C}}_1 & -1& +1&-1 &-1 
 \\\hline
 (1,4,6) & (1,5,6) &\{4,5\}& 2&2 &\{1,4,5,6\}&{\ul{C}}_4 & -1& +1&-1 &-1 
 \\\hline
 (2,3,5) & (2,4,5) &\{3,4\}& 2&2 &\{2,3,4,5\}&{\ul{C}}_7 & -1& -1&+1 &-1 
 \\\hline
 (2,4,5) & (2,4,6) &\{5,6\}& 3&3 &\{2,4,5,6\}&{\ul{C}}_4 & +1& -1&-1 &-1 
 \\\hline
 (2,4,5) & (2,5,6) &\{4,6\}& 2&3 &\{2,4,5,6\}&{\ul{C}}_4 & -1& -1&-1 &-1 
 \\\hline
 (1,3,4) & (3,4,5) &\{1,5\}& 1&3 &\{1,3,4,5\}&{\ul{C}}_2 & +1& +1&+1 &-1 
 \\\hline
 (1,3,4) & (3,4,6) &\{1,6\}& 1&3 &\{1,3,4,6\}&{\ul{C}}_6 & +1& +1&+1 &-1 
 \\\hline
 (3,4,5) & (3,5,6) &\{4,6\}& 2&3 &\{3,4,5,6\}&{\ul{C}}_4 & -1& -1&-1 &-1 
 \NN\hline
\earr
Notice that in order to compute $\chi_{\MC{B}}(B)$ one can choose {\it any} $B'\in\MC{B}$ which differs from $B$ in one element and the according signed circuit contained in $B\cup B'$. For instance for $B=(123)$ one could take $B'=(125)$ instead with $B\Delta B'=\{3,5\}$, $\leadsto i=3$, $j=3$. Then one simply uses the according circuit $C_2\in\C$ with support $\ul{C}\subseteq B\cup B'=\{1235\}$ and computes equivalently $\chi_{\MC{B}}(123)=(-1)^{1+3+3}\cdot\sgn_{C_2}(3)\cdot\sgn_{C_2}(5)\cdot\chi_{\MC{B}}(125)
     =(-1)\cdot(-1)\cdot(+1)\cdot(-1)
     =-1$ .

Our findings are be summarized in table \ref{applied SBGM} , where we write down all triples contained in $E$ and decide whether they are a basis element (${\green \bullet}$) or not (${\red   \bullet}$). The initial choice for $\chi_{\MC{B}}(1,2,3)=-1$ is marked by a blue frame.

\begin{table}[htb!]
\renewcommand{\arraystretch}{1} 
\barr{rcccrcccrcccrcc}
\text{Basis} & \text{Triple} & \chi_{\MC{B}}&& 
\text{Basis} & \text{Triple} & \chi_{\MC{B}}&&
\text{Basis} & \text{Triple} & \chi_{\MC{B}}&&
\text{Basis} & \text{Triple} & \chi_{\MC{B}} \\

{\green \bullet}&123&{\blue\fbox{$-1$}}&   &&&&&&\\
{\red   \bullet}&124&0&   &&&&&\\
{\green \bullet}&125&-1&   &&&&&\\ 
{\green \bullet}&126&-1&   &&&&&\\
{\green \bullet}&134&+1&&   {\green \bullet}&234&+1&&   &\\
{\red   \bullet}&135&0&&   {\green \bullet}&235&+1&& &\\
{\green \bullet}&136&-1&&   {\red   \bullet}&236&0&& &\\
{\green \bullet}&145&-1&&   {\green \bullet}&245&-1&&  {\green \bullet}&345   &-1\\
{\green \bullet}&146&-1&&   {\green \bullet}&246&-1&&  {\green \bullet}&346&-1\\
{\green \bullet}&156&-1&&   {\green \bullet}&256&-1&&  {\green \bullet}&356&-1
                                              &&  {\red   \bullet}&456&0\\
 
\earr
\caption{Chirotope data of the graphic oriented matroid encoded in figure \ref{directed graph}, computed by the signed basis graph method.}
\label{applied SBGM}
\end{table}

We will now compute the basis chirotope $\chi_{\MC{B}^*}$ of the dual basis as defined in definition \ref{Dual Oriented Matroid}.
\[
    \MC{B}^*:=\{E\backslash B~:~B\in\MC{B}\}
\]
 
with its chirotope

\[
   \chi_{\MC{B}^*}(B^*)=\chi_{\MC{B}}(B)\cdot\sgn\big((B^*,B)\big)
                       =:\chi_{\MC{B}}(B)\cdot\sgn\big(\pi(b_1^*,b_2^*,b_3^*,b_1,b_2,b_3)\big)
\]
where we have written $B=(b_1,b_2,b_3)$ and $B^*=(b_1^*,b_2^*,b_3^*)$ and the defined lexicographically sorted ground set is $E=(1,2,3,4,5,6)$. As before $\sgn\big(\pi(\cdot)\big)$ denotes the parity of the permutation $\pi$ in order to bring the set in the argument into lexicographic order. The result is given in table \ref{Dual chirotope computed}.

\begin{table}[hbt!]
\barr{cc}
\begin{array}{c|c||c|c|c}

B & \chi_{\MC{B}}(B) & B^* &\sgn((B^*,B))&\chi_{\MC{B}^*}(B^*)
\\\hline&&&&\\
123&-1 &456 & -1 & +1
\\
125&-1 &346 & -1 & +1
\\
126&-1 &345 & +1 & -1
\\&&&&\\
134&+1 &256 & -1 & -1
\\
136&-1 &245 & -1 & +1
\\
145&-1 &236 & -1 & +1
\\
146&-1 &235 & +1 & -1
\\
156&-1 &234 & -1 & +1
\\&&&&\\
234&+1 &156 & +1 & +1
\\
235&+1 &146 & -1 & -1
\\&&&&\\
245&-1 &136 & +1 & -1
\\
246&-1 &135 & -1 & +1
\\
256&-1 &134 & +1 & -1
\\&&&&\\
345&-1 &126 & -1 & +1
\\
346&-1 &125 & +1 & -1
\\
356&-1 &124 & -1 & +1
\end{array}
&~~~

\begin{array}{c|c||c|c|c}
\multicolumn{5}{c}{{\red \text{non-bases}}}\\
B & \chi_{\MC{B}}(B) & B^* &\chi_{\MC{B}^*}(B^*)
\\\hline&&&&\\
124& 0 &356  & 0
\\
135 & 0 &246    &  0 
\\
236&0& 145& 0
\\
456 & 0 & 123 &0   
\end{array}

\nonumber
\earr
\caption{Dual chirotope data for the oriented matroid given by figure \ref{directed graph}.}
\label{Dual chirotope computed}
\end{table}

Using $\chi_{\MC{B}^*}$  from table \ref{Dual chirotope computed} we can now compute the signed cocircuits $\C^*$ as signed circuits of the dual matroid $\M^*=(E,\MC{B}^*)$. As $E=(1,2,3,4,5,6)$ consists of 6 elements we know that $\M^*$ has rank $r^*=6-r=3$. That is, all possible cocircuits
 should be supported within subsets having at most $4$ elements. Now consider all  $\left(6\atop 4\right)=15$ four-element subsets of $E$.
\[\begin{array}{cccccccccc}
   1234&1235&1236& 1245  & 1246  &1256& 1345  &1346& 1356  & 1456  
   \\
                                     &&&&&&2345&2346 &  2356  & 2456  
   \\                                                    &&&&&&&&&  3456                                   
\end{array}\]
We simply compute the circuit signature induced by $\chi_{\MC{B}}^*$ for every tuple, using again the signed basis graph construction (\ref{Sign of SBG}) in its dualized version (dual bases and cocircuits as circuits of the dual matroid). 
 
From the dual non-bases given in table \ref{Dual chirotope computed} we already see that
\[\begin{array}{cccccccccc}
   \ul{C}_1^*&:=&\{1,2,3\}&\subset&\{1,2,3,4\}, \{1,2,3,5\},\{1,2,3,6\}\\
   \ul{C}_2^*&:=&\{1,4,5\}&\subset&\{1,2,4,5\}, \{1,3,4,5\},\{1,4,5,6\}\\
   \ul{C}_3^*&:=&\{2,4,6\}&\subset&\{1,2,4,6\}, \{2,3,4,6\},\{2,4,5,6\}\\
   \ul{C}_4^*&:=&\{3,5,6\}&\subset&\{1,3,5,6\}, \{2,3,5,6\},\{3,4,5,6\}\\
\end{array}\]
Hence we have to additionally analyze the cocircuits $C_5^*,C_6^*,C_7^*$ whose support is equal\footnote{As these 4-element subsets are extensions of dual bases by one element, and we already have computed all cocircuits with fewer than 4 elements, the remaining cocircuits must be supported on 4 elements.} to $\{1,2,5,6\}$, $\{1,3,4,6\}$, $\{2,3,4,5\}$ respectively.
As an example, let us compute the signed sets for  $\pm C_1^*$. First we again choose $\sgn_{C_1^*}(1)=-1$. In order to compute the signs for the elements $2,3\in \ul{C}_1^*$  we have to choose pairs $B^*,(B^*)'$ of dual bases, such that $\ul{C}_1^*\subseteq \big(B^*\cup(B^*)'\big)$ and
$B^*\Delta (B^*)'=\{1,2\},\{1,3\}$, for example we may choose 
$B^*=(1,3,6),(B^*)'=(2,3,6)$ and $B^*=(1,2,6),(B^*)'=(2,3,6)$. Then we have, using (\ref{Sign of SBG}) and the dual chirotope data of table \ref{Dual chirotope computed}:
\ba
   \sgn_{C^*_1}(2)
   =\sgn_{C^*_1}(1)\cdot (-1)^{1+1+1}\cdot\chi_{\MC{B}^*}(1,3,6)\cdot\chi_{\MC{B}^*}(2,3,6)
   =(-1)\cdot(-1)\cdot (-1)\cdot (+1)
   =-1
   \NN
   \sgn_{C^*_1}(3)
   =\sgn_{C^*_1}(1)\cdot (-1)^{1+1+2}\cdot\chi_{\MC{B}^*}(1,2,6)\cdot\chi_{\MC{B}^*}(2,3,6)
   =(-1)\cdot(+1)\cdot (+1)\cdot (+1)
   =-1
   \nonumber
\ea
It follows that we can 
write in our compact notation $C_1^*=\ol{1}\ol{2}\ol{3}$.
One can check that $C_1^*\in\C^*$ is orthogonal to all $C_k\in\C$ as by definition \ref{Def Cocircuit Signature}. We get:
\barr{l|c|cc|cc}\label{C_1 is ON to all C}
  &\ul{C_1^*}\cap \ul{C_k}=\{a,b\} & \sgn_{C_k}(a)&\sgn_{C_k}(b)
  &\sgn_{C_1^*}(a)&\sgn_{C_1^*}(b)
  \\\hline&&&&&\\
  C_1=1\ol{2}4       & \{1,2\}  & + & - & - & - \\
  C_2=1\ol{3}5       & \{1,3\}  & + & - & - & - \\
  C_3=2\ol{3}6       & \{2,3\}  & + & - & - & - \\
  C_4=\ol{4}5\ol{6}       & \emptyset&   &   &   &   \\
  C_5=1\ol{2}5\ol{6} & \{1,2\}  & + & - & - & - \\
  C_6=1\ol{3}46      & \{1,3\}  & + & - & - & - \\
  C_7=2\ol{3}\ol{4}5 & \{2,3\}  & + & - & - & - \\
\earr
and see that the orthogonality condition is fulfilled. 
Similarly one can proceed with the remaining cocircuits, where we use as before  $B^*\Delta (B^*)':=(B^*\cup (B^*)')\bs(B^*\cap (B^*)')=\{b_i,b'_j\}$, $B^*=(b_1,b_2,b_3), (B^*)'=(b'_1,b'_2,b'_3)$ and $\ul{C}^*\subseteq \big\{B^*\cup (B^*)'\big\}$:
\barr{|c||c|c|c||c|c|c|c|c||c|}
\hline
\ul{C}^* &\text{choose}&\{b_i,b'_j\}&(b_1,b_2,b_3)& (b'_1,b'_2,b'_3) &i&j
& \chi_{\MC{B}^*}(B^*)&\chi_{\MC{B}^*}((B^*)')
&\sgn_{C^*}(b_j)
\\\hline\hline
\{1,4,5\} & \sgn_{C^*}(1)=-1 & \{1,4\} &(1,2,5)&(2,4,5)&1&2&-1&+1&+1
\\
                            && \{1,5\} &(1,2,4)&(2,4,5)&1&2&+1&+1&+1
\\\hline
\{2,4,6\} & \sgn_{C^*}(2)=-1 & \{2,4\} &(2,3,6)&(3,4,6)&1&2&+1&+1&-1
\\
                            && \{2,6\} &(2,3,4)&(3,4,6)&1&3&+1&+1&+1
\\\hline
\{3,5,6\} & \sgn_{C^*}(3)=-1 & \{3,5\} &(2,3,6)&(2,5,6)&2&2&+1&-1&-1
\\
                            && \{3,6\} &(3,4,5)&(3,4,6)&1&3&-1&+1&-1
 \\\hline
\{1,2,5,6\} & \sgn_{C^*}(1)=-1 & \{1,2\} &(1,2,5)&(2,5,6)&1&1&+1&-1&-1
\\
                              && \{1,5\} &(1,2,6)&(2,5,6)&1&2&+1&-1&+1
\\
                              && \{1,6\} &(1,2,5)&(2,5,6)&1&3&-1&-1&+1
 \\\hline
\{1,3,4,6\} & \sgn_{C^*}(1)=-1 & \{1,3\} &(1,4,6)&(3,4,6)&1&1&-1&+1&-1
\\
                              && \{1,4\} &(1,3,6)&(3,4,6)&1&2&-1&+1&+1
\\
                              && \{1,6\} &(1,3,4)&(3,4,6)&1&3&-1&+1&-1
 \\\hline
\{2,3,4,5\} & \sgn_{C^*}(2)=-1 & \{2,3\} &(2,3,4)&(3,4,5)&1&1&+1&-1&-1
\\
                              && \{2,4\} &(2,3,5)&(3,4,5)&1&2&-1&-1&-1
\\
                              && \{2,5\} &(2,3,4)&(3,4,5)&1&3&+1&-1&-1
\NN\hline                       
\earr
In this way we end up with a set of 14 cocircuits \cite{Bjorner1999} p.\ 9, given by the following signed sets and their negatives:
\barr{l}
  C_1^*=\ol{123}       \\
  C_2^*=\ol{1}45         \\
  C_3^*=\ol{24}6      \\
  C_4^*=\ol{356}       \\
  C_5^*=\ol{12}56 \\
  C_6^*=\ol{13}4\ol{6}     \\
  C_7^*=\ol{2345}
\earr
Certainly it is also possible to directly use the fact that cocircuits are orthogonal to all circuits. Writing 
$\sgn_{C_k}(a),\sgn_{C_k}(b)$ as in ({\ref{C_1 is ON to all C}) for every pair $\{a,b\}$ contained in the set $\C$ of all circuits, one can compare these data for all elements of the cocircuits and obtain their partition into positive and negative parts.

\subsection{\label{Oriented Matroids from Vector Configurations}Oriented Matroids from Vector Configurations}
Consider a finite set $E=\{{\bf v}_1,\ldots,{\bf v}_n\}$ of $n\ge r$ non-zero vectors, spanning the $r$-dimensional vector space $\mb{R}^r$. We can express the minimal linear dependencies among the elements of $E$ by the solutions to

\[
   \sum_{i=1}^n \lambda_i {\bf v}_i=0~~~~~~\text{with}~~\lambda_i\in\mb{R}.
\]
A solution is given by  elements of the form $\vec{\lambda}=(\lambda_1,\ldots,\lambda_n)\in\mb{R}^n$ having at least two non-zero components. For each such solution $\vec{\lambda}$ we define the set $\ul{C}=\{i~:~\lambda_i\ne 0\}$, called a circuit. The set of all $\ul{C}$ then builds up the set of all circuits $\ul{\MC{C}}$ of the (unoriented) matroid $\ul{\MC{M}}=(E,\ul{\MC{C}})$. By construction $\ul{\MC{M}}$ has rank $r$.

In order to obtain an oriented matroid, consider for each solution $\vec{\lambda}$ the signed set $C=(C^+,C^-)$ with $C^+=\{i~:~\lambda_i>0\}$ and $C^-=\{i~:~\lambda_i<0\}$, giving a signed circuit. We denote the set of all $C$ by $\MC{C}$. This defines an  oriented matroid $\MC{M}=(E,\MC{C})$, also called a \emph{vector oriented matroid}.

We may also introduce the extended signature $\sgn_C$ for all $C\in\MC{C}$. Then the signed circuit $C$ can be written as an element in $\{+,0,-\}^n$ as $C=(\sgn_C(\lambda_1),\ldots,\sgn_C(\lambda_n))$.
Notice that the linear dependencies encoded in each circuit $\ul{C}$ are only given up to an overall scalar. Hence each element $\ul{C}\in \ul{\MC{C}}$ gives rise to two elements $C=(C^+,C^-)$ and $-C=(C^-,C^+)$ in $\MC{C}$.

It is also straightforward to see that the set $\MC{B}$ of bases of $\MC{M}=(E,\MC{C})$ is given by all subsets $B$ of $E$ which contain $r$ linearly independent vectors. The natural notion of a basis orientation $\chi_{\MC{B}}(B)$, $B=(b_1,\ldots,b_r)$, can be written in terms of an $(r\times r)$-matrix with column vectors $({\bf v}_{b_1},\ldots, {\bf v}_{b_r})$ 
by 
\be\label{det sign equals chirotope in vector case}
   \chi_{\MC{B}}(B)= \chi_{\MC B}(b_1,\ldots, b_r)
                   =\sgn\big(\det({\bf v}_{b_1},\ldots, {\bf v}_{b_r})\big)
                   =:\sgn\big( [b_1,\ldots, b_r]\big)
                   ~~.
\ee
Here and in what follows we write $[b_1,\ldots, b_r]$ to abbreviate determinant expressions  like $\det({\bf v}_{b_1},\ldots, {\bf v}_{b_r})$. It may be checked that this notion of $\chi_{\MC B}$ is consistent with definition \ref{Oriented Matroid from Oriented Bases of a Set} by the determinant properties.

Moreover the set $\C^*$ of signed cocircuits of $\MC{M}=(E,\MC{C})$ gets the following geometric interpretation: Consider an $(r-1)$-subset $\{h_1,\ldots, h_{r-1}\}$ of $E$ such that $H=\{{\bf v}_{h_1},\ldots, {\bf v}_{h_{r-1}}\}$ spans an $(r-1)$-dimensional hyperplane which cuts $R^n$ into two half spaces $H^+,H^-$. Equip $H$ with an orientation and choose the labels $H^+,H^-$ such that $H^+$ is the positive and $H^-$ the negative half space with respect to the chosen orientation. Then a signed cocircuit $C^*$ of $\MC{M}$ with $E\backslash C^*=H$ is given by $C^*=({C^*}^+,{C^*}^-)$ where ${C^*}^\pm$ is the set of vectors in $E$ contained in the according half space $H^\pm$.  By construction  $\C^*$ contains   $\pm C^*$, since the orientation of $H$ can be chosen arbitrarily without loss of generality. 

\paragraph{Grassmann Pl\"ucker identities and determinants \cite{Bjorner1999}.} Given any two sorted sets\footnote{Not necessarily $B,B'\in\MC{B}$.} $B,B'\subset E$ where $B=(b_1,\ldots,b_r)$, $B'=(b'_1,\ldots,b'_r)$, the product $[b_1,\ldots,b_r]\cdot [b'_1,\ldots,b'_r]$ can be expressed as
\be\label{Generalized Laplace Expansion}
  [b_1,\ldots,b_r]\cdot [b'_1,\ldots,b'_r]
     =\sum_{k=1}^r~[b'_k,b_2,\ldots,b_r]\cdot [b_1',\ldots,b'_{k-1},b_1,b'_{k+1},\ldots,b'_r] ~~,
\ee
one special example being the Laplace expansion for 
determinants\footnote{  Let $B,B'\in\MC{B}$ and $(b_1,\ldots,b_r)$ be the coordinate vectors, that is $b_1=(1,0,\ldots,0)^t$ , $b_2=(0,1,\ldots,0)^t$, \ldots, $b_r=(0,0,\ldots,1)^t$ ($~^t$ stands for transposed). Let $(b'_1,\ldots,b'_r)$ be another basis.
Certainly, $[b_1,\ldots,b_r]$ corresponds to the determinant of the unit $(r\times r)$-matrix. Inserting this, (\ref{Generalized Laplace Expansion}) corresponds to the Laplace expansion of the determinant of the matrix whose column vectors are given by $(b'_1,\ldots,b'_r)$.}.
In general expressions like (\ref{Generalized Laplace Expansion}) are called {\it Grassmann Pl\"ucker relations for determinants}. 
Notice that the difference between the left hand side and right hand side of (\ref{Generalized Laplace Expansion}) gives an antisymmetric linear expression in the $(r+1)$ arguments $b_1,b'_1,b'_2,\ldots,b'_r$, that is an $(r+1)$-form on $\mb{R}^r$. This vanishes by construction  \cite{Bjorner1999}.  
\\\\
Obviously (\ref{Generalized Laplace Expansion}) implies 
$[b_1,\ldots,b_r]\cdot [b'_1,\ldots,b'_r]\ge 0$ if 
$[b'_k,b_2,\ldots,b_r]\cdot [b_1',\ldots,b'_{k-1},b_1,b'_{k+1},\ldots,b'_r]\ge 0$ holds for all $k=1\ldots r$. With the identification of signs of determinants and chirotopes given in (\ref{det sign equals chirotope in vector case}), this gives rise to 

\begin{Definition}[Grassmann Pl\"ucker relations for chirotopes \cite{Bjorner1999}.]
   \label{Grassmann Plucker relations for chirotopes}
   Let the oriented matroid $\M=(E,{\MC B})$ be given with the chirotope $\chi_{\MC B}$ giving a basis orientation of ${\MC B}$. Let  $B,B'\subset E$ be given with $B=(b_1,\ldots,b_r)$, $B'=(b'_1,\ldots,b'_r)$. Then\\
   
   \begin{tabular}{lll}
    (B2) &\cmt{14}{   For all $B,B'$ such that  
    $\chi_{\MC B}(b'_k,b_2,\ldots,b_r)\cdot \chi_{\MC B}(b_1',\ldots,b'_{k-1},b_1,b'_{k+1},\ldots,b'_r)\ge 0$~~$\forall\,k=1\ldots r$
    it holds that
    $\chi_{\MC B}(b_1,\ldots,b_r)\cdot \chi_{\MC B}(b'_1,\ldots,b'_r)\ge 0$ }
    
    \\\\&or equivalently 
    \\\\  
    (B2') &\cmt{14}{For all $B,B'$ such that $\chi_{\MC B}(b_1,\ldots,b_r)\cdot \chi_{\MC B}(b'_1,\ldots,b'_r)\ne 0$ $\exists$ $k\in\{1,\ldots,r\}$ such that
    $\chi_{\MC B}(b'_k,b_2,\ldots,b_r)\cdot \chi_{\MC B}(b_1',\ldots,b'_{k-1},b_1,b'_{k+1},\ldots,b'_r)= \chi_{\MC B}(b_1,\ldots,b_r)\cdot \chi_{\MC B}(b'_1,\ldots,b'_r)$.}  
   \end{tabular}
\end{Definition}
The equivalence of $(B2)$ and $(B2')$ can be seen as follows \cite{Bjorner1999}: For $r=1$ it is trivially fulfilled.
For $r\ge 2$ set $\epsilon_k:=\chi_{\MC B}(b'_k,b_2,\ldots,b_r)\cdot \chi_{\MC B}(b_1',\ldots,b'_{k-1},b_1,b'_{k+1},\ldots,b'_r)$. Observe that under the permutation $b_1'\leftrightarrow b_2'$ the sign tuple $(\epsilon_1,\epsilon_2,\ldots,\epsilon_r)$ changes into $(-\epsilon_2,-\epsilon_1,\ldots,-\epsilon_r)$ as well as 
$\chi_{\MC B}(b_1,\ldots,b_r)\cdot \chi_{\MC B}(b'_1,b_2',\ldots,b'_r)=-\chi_{\MC B}(b_1,\ldots,b_r)\cdot \chi_{\MC B}(b'_2,b_1',\ldots,b'_r)$.
 
 Now assume $(B2)$ holds and $\chi_{\MC B}(b_1,\ldots,b_r)\cdot \chi_{\MC B}(b'_1,\ldots,b'_r)> 0$. Then there must be a $k$ such that $\epsilon_k>0$, as otherwise (\ref{Generalized Laplace Expansion}) would give a contradiction. By permuting $b_1'\leftrightarrow b_2'$ $\chi_{\MC B}(b_1,\ldots,b_r)\cdot \chi_{\MC B}(b'_2,b'_1,\ldots,b'_r)< 0$ and as all $\epsilon_k$ change their sign also $\epsilon_k<0$. Hence $(B2')$ holds.
 
 Conversely assume $(B2')$ holds. Then we know that whenever $\chi_{\MC B}(b_1,\ldots,b_r)\cdot \chi_{\MC B}(b'_1,\ldots,b'_r)\ne 0$ there is a $k\in\{1,\ldots,r\}$ such that $\epsilon_k=\chi_{\MC B}(b_1,\ldots,b_r)\cdot \chi_{\MC B}(b'_1,\ldots,b'_r)\ne 0$. Now if $\epsilon_k\ge 0~~\forall k$ then it must hold that  $\chi_{\MC B}(b_1,\ldots,b_r)\cdot \chi_{\MC B}(b'_1,\ldots,b'_r)\ge 0$. By applying the permutation argument again it also holds for $\epsilon_k\le 0~~\forall k$ that $\chi_{\MC B}(b_1,\ldots,b_r)\cdot \chi_{\MC B}(b'_1,\ldots,b'_r)< 0$. The case $\chi_{\MC B}(b_1,\ldots,b_r)\cdot \chi_{\MC B}(b'_1,\ldots,b'_r)= 0$ is contained in both argumentations.
\\\\
The question arising here is whether $(B2)$, $(B2')$ arise only in case of a vector realization of the oriented matroid $\M=(E,{\MC B})$. The answer is given in the following 

\begin{Lemma}[Grassmann-Pl\"ucker Relations and Basis Orientation \cite{Bjorner1999}.] Let the oriented matroid $\M$ be given in terms of the ground set $E$ and its set of cocircuits $\C^*$, $\M=(E,C^*)$ with basis orientation $\chi_{\MC B}$. Then its chirotope $\chi_{\MC B}$ satisfies $(B2')$.
\end{Lemma}
To see this \cite{Bjorner1999}, consider  $B,B'\subset E$ with $B=(b_1,\ldots,b_r)$, $B'=(b'_1,\ldots,b'_r)$, such that \linebreak 
$\epsilon:=\chi_{\MC B}(b_1,\ldots,b_r)\cdot\chi_{\MC B}(b'_1,\ldots,b'_r)\in\{-1,+1\}$. If $b_1\in B'$ then $(B2')$ is trivially satisfied. 
\\
If $b_1\notin B'$ consider the basic circuit $C=(b_1,B')$ and the basic cocircuit $C^*=(b_1,B)$ defined as in definition \ref{Basic (fundamental) (co-)circuits}. By construction $b_1\in \ul{C}\cap\ul{C}^*$. By orthogonality of $C,C^*$ there is a $k\in\{1,\ldots,r\}$ such that $\sgn_C(b_1)\sgn_C(b'_k)=-\sgn_{C^*}(b_1)\sgn_{C^*}(b'_k)$. Now $\M$ is an oriented matroid and hence $(PV^*)$ and $(PV)$ are satisfied by $\chi_{\MC B}$. Therefore it holds that
\ba
  \chi_{\MC B}(b'_1,\ldots,b'_{k-1},b_1,b'_{k+1},\ldots,b'_r)
  &=& -\sgn_C(b_1)\cdot\sgn_C(b'_k)\cdot
       \chi_{\MC B}(b'_1,\ldots,b'_{k-1},b'_k,b'_{k+1},\ldots,b'_r)
  \NN
  \chi_{\MC B}(b'_k,b_2,\ldots,b_r)
  &=&~~ \sgn_{C^*}(b_1)\cdot\sgn_{C^*}(b'_k)\cdot\chi_{\MC B}(b_1,b_2,\ldots,b_r)
  \NN
  \NN
  \leadsto~~~\epsilon=\chi_{\MC B}(b'_1,\ldots,b'_r)\cdot \chi_{\MC B}(b_1,\ldots,b_r)
  &=&    \chi_{\MC B}(b'_k,b_2,\ldots,b_r)\cdot 
  \chi_{\MC B}(b'_1,\ldots,b'_{k-1},b_1,b'_{k+1},\ldots,b'_r)
  \nonumber
\ea
However the existence of $k\in\{1,\ldots,r\}$ such that the last line holds is precisely the statement of $(B2')$. 
\\\\
We see that $(B2),(B2')$ of definition \ref{Grassmann Plucker relations for chirotopes} is not a consequence of the realization of $\M$ as a vector configuration. Rather it is a
general consequence from the combinatorial definition  
\ref{Oriented Matroid from Oriented Bases of a Set}.

\paragraph{As an example \cite{Bjorner1999},} consider the vectors $E=({\bf v}_1,\ldots,{\bf v}_6)$ in $\mb{R}^3$, given by the columns of the matrix.
\[
   A=\left(
         \begin{array}{cccccc}
              1&1&1&0&0&0\\
              0&1&1&1&1&0\\
              0&0&1&0&1&1
         \end{array}
    \right)
\]
From $A$ we can read off the set of minimal linear dependencies among the vectors in $E$. For instance we have ${{\bf{v}}_1-{\bf{v}}_2+{\bf{v}}_5-{\bf{v}}_6}=0$, that is $\vec{\lambda}=(1,-1,0,0,1,-1)$ and $X=\{+,-,0,0,+,-\}$. This can be compared to the graph example in the previous section. Here we also obtain the same list of signed circuits as in the graph example above. The column vectors of the matrix $A$ contain the set $\MC{B}$ of bases of the oriented matroid $\MC{M}$ obtained in the previous section.

\subsection{\label{Oriented Matroids from Hyperplane and Sphere Arrangements}Oriented Matroids from Hyperplane and Sphere Arrangements  }
There are two additional realizations of oriented matroids, related to vector configurations, which we introduce for completeness \cite{Bjorner1999,Goodman2004}.

Consider the rank $r$ vector oriented matroid $\MC{M}=(E,\MC{C})$ as defined at the beginning of section \ref{Oriented Matroids from Vector Configurations}. Each element ${\bf v}_i$ of ground set $E=\{{\bf v}_1,\ldots,{\bf v}_n\}$ of $\M$ can be used to define an $(r-1)$-dimensional hyperplane $H_i$ through the origin in $\mb{R}^r$ as $H_i=\{{\bf x}\in\mb{R}^r~:~\big<{\bf v}_i\,,\,{\bf x}\big>=0\}$, where $\big<\cdot,\cdot\big>$ denotes the usual Euclidean inner product on $\mb{R}^r$. That is, ${\bf v}_i$ is the normal vector to $H_i$.  
Obviously the orientation of ${\bf v}_i$ can be used to define an orientation of $H_i$. 
Then we may define the positive side of $H_i$ by $H_i^+:=\{{\bf x}\in\mb{R}^r~:~\big<{\bf v}_i\,,\,{\bf x}\big> > 0\}$ and the negative side $H_i^-:=\{{\bf x}\in\mb{R}^r~:~\big<{\bf v}_i\,,\,{\bf x}\big> < 0\}$.

In this way $E$ corresponds to an oriented arrangement $\MC{A}=\{H_1,\ldots,H_n\}$ of $n$ $(r-1)$-dimensional hyperplanes. 
In turn $\MC{A}$ gives rise to a decomposition of $\mb{R}^r$ into $r$-dimensional cells $\MC{W}$. The interior of each such cell is uniquely described by its relative position with respect to the hyperplanes contained in  $\MC{A}$. For each $H_i\in\MC{A}$ we can say if $\MC{W}$ is on the positive or negative side of $H_i$. That is $\MC{W}$ is uniquely described by an element of $\{+1,-1\}^n$.  

Obviously the oriented vector matroid $\M=(E,\MC{B})$ can be equivalently encoded in the arrangement $\MC{A}$. To specify $\M$ we need to specify the unique cell $\MC{W}\equiv (+1,\ldots,+1)$ situated on the positive side of every hyperplane.
\\\\

We may alternatively consider the intersection of $\MC{A}$ with the unit sphere $S^{r-1}=\{{\bf x}\in\mb{R}^r~:~\|{\bf x}\|=1\}$ around the origin of $\mb{R}^r$.
This gives rise to an arrangement $\MC{S}=\{S_1,\ldots,S_n\}$ of $n$ unit $(r-2)$-spheres $S_i$ on $S^{r-1}$ defined by $S_i:=S^{r-1}\cap H_i$. Equivalently we define $S_i^{\pm}:=S^{r-1}\cap H_i^{\pm}$. This is depicted in figure \ref{split sphere a}.
The sphere arrangement $\MC{S}=\{S_1,\ldots,S_n\}$ together with the specification of a positive and negative side $S_i^\pm$ for every $S_i\in\MC{S}$ is called a signed sphere arrangement.

\begin{figure}[hbt!]
\center
 \psfrag{Seplus}{$S_i^+$}
 \psfrag{Seminus}{$S_i^-$}
 \psfrag{e}{$S_i$}
 \psfrag{vi}{$\blue \frac{1}{\|{\bf v}_i\|}{\bf v}_i$}
 \includegraphics[width=5cm]{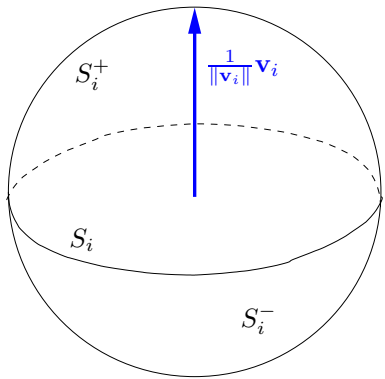}
 \caption{$S^{r-1}$ split into two open hemispheres $S_i^+$, $S_i^-$ by an
   $r-2$-sphere $S_i$, induced by the element ${\bf v}_i$ of the ground set
   $E$.}
 \label{split sphere a} 
\end{figure}
Now recall the rank $r$ vector oriented matroid $\M=(E,\C)$ from the beginning of section \ref{Oriented Matroids from Vector Configurations}. For a signed circuit $C\in\C$ it holds that\footnote{\label{Remark on different notation}Notice that the definition of $H_i^\pm$ is slightly different in \cite{Bjorner1999} and \cite{Goodman2004}. In the former, $H_i^\pm$ contains $H_i$
and is thus closed, whereas in the latter $H_i$ is defined as an open half space of $\mb{R}^r$. We regard the notation of \cite{Goodman2004} as more convenient, as it avoids difficulties in assigning a sign to points $x\in H_i\subset \mb{R}^{r}$,
 and separates $H_i$ from $H_i^\pm$.
In general a circuit $C\in\C$ corresponds to a minimal system of closed hemispheres that cover the whole unit sphere $S^{r-1}$.}
\[
   \displaystyle\bigcup_{{\bf v}_i\in\ul{C}} \ol{S_i^{\sgn_C({\bf v}_i)}}=S^{r-1} \text{~~where~~~~}  S_i^{\sgn_C({\bf v}_i)}= S_i^{\pm} \text{~~~~and~~}\ol{S_i^\pm} \text{~~denotes the closure~ }\ol{S_i^\pm}=S_i^\pm\cup S_i
\]
We will see in section \ref{Realizability of Oriented Matroids} how sphere arrangements can be used for deciding the so-called realizability problem of oriented matroids.

\subsection{\label{Realizability of Oriented Matroids}Realizability of Oriented Matroids}
In sections \ref{Oriented Matroids from Vector Configurations} and \ref{Oriented Matroids from Hyperplane and Sphere Arrangements} we have outlined how oriented matroids can arise from different geometric setups corresponding to vector configurations. This section will be concerned with the opposite point of view:
Given an oriented rank $r$ matroid $\M=(E,\MC{B})$ defined as in section \ref{Oriented Matroids}, when can it be represented as a vector configuration? This question is called the realizability problem for oriented matroids. It is of particular relevance if one wants to compute all possible vector oriented matroids of rank $r$ constructable over an abstract ground set $E$. For instance one would like to compute all classes of diffeomorphic embeddings of a graph vertex as discussed in section \ref{Local Properties: Signfactors Characterizing the Volume Spectrum}. 

The main statement \cite{Bjorner1999,Goodman2004} here is that reorientation\footnote{See section \ref{Constructions with Oriented Matroids} for the definition of reorientation.} equivalence classes ($\equiv$ equivalence classes under relabelling of the elements of the ground set $E$ and a possible change of orientation for each element $e\in E$) of oriented matroids correspond to signed arrangements of modified arrangements of spheres called signed pseudosphere arrangements on $S^{r-1}\subset\mb{R}^r$.

Consider an $(r-2)$-sphere $S_i$ as depicted in figure \ref{split sphere a}. A subset\footnote{Notice that our notation distinguishes between pseudospheres and spheres, in contrast to \cite{Bjorner1999}.} $\mathscr{S}_i$ of $S^{r-1}$ is called a \emph{pseudosphere} if $\mathscr{S}=h(S_i)$ for some homeomorphism $h:~S^{r-1}\rightarrow S^{r-1}$. Using $h$ we can also define the open subspaces $\mathscr{S}_i^\pm:=h(S_i^\pm)$, as depicted in figure \ref{split pseudo sphere}.
The arrangement $\mathscr{S}=\{\mathscr{S}_1,\ldots,\mathscr{S}_1\}$ of pseudospheres $\mathscr{S}_i\subset S^{r-1}\subset\mb{R}^r$ together with a fixed choice of positive side $\mathscr{S}_i^+$ and negative side $\mathscr{S}_i^-$ for each $\mathscr{S}_i$ is called a signed pseudosphere arrangement. 
\begin{figure}[hbt!]
\center
    \psfrag{Seplus}{$\mathscr{S}_i^+$}
    \psfrag{Seminus}{$\mathscr{S}_i^-$}
    \psfrag{e}{$\mathscr{S}_i$}

    \includegraphics[width=5cm]{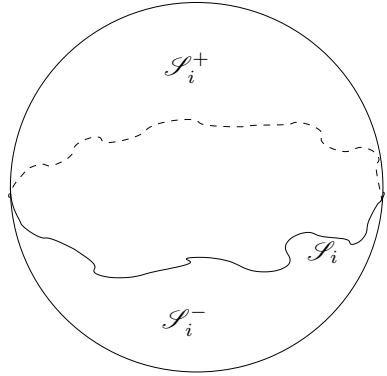}
    \caption{$S^{r-1}$ split into two half spaces $\mathscr{S}_i^+$, $\mathscr{S}_i^-$ by an $(r-2)$-pseudosphere $\mathscr{S}_i$ homeomorphic to the $(r-2)$-sphere $S_i$ induced by the matroid element ${\bf v}_i$ as shown in figure \ref{split sphere a}.}
    \label{split pseudo sphere}
  
\end{figure}

Now conversely let a signed pseudosphere arrangement $\mathscr{S}=\{\mathscr{S}_1,\ldots,\mathscr{S}_n\}\subset S^{r-1}\subset \mb{R}^r$ be given. Choose an arbitrary $n$-element set $E=\{1,\ldots,n\}$ as a ground set and associate to every $i\in E$ a $\mathscr{S}_i\in\mathscr{S}$. Define $\C(\MC{\mathscr{S}})$ as the set of sign vectors $C\in\{+,-,0\}^{|E|}$ with support $\ul{C}=\{e\in E : \sgn_C(e)\ne 0\}$, 
which satisfy\footnote{Again notice the difference in notation between \cite{Bjorner1999} and \cite{Goodman2004} as described in footnote \ref{Remark on different notation}. As in the case of spheres $S_i$ we understand the $\mathscr{S}_i^\pm$ as open subpaces in the sense of \cite{Goodman2004}. Moreover we understand $C\in \{+,-,0\}^{|E|}$ as coming from an extended signature map, assigning a sign to every $e\in E$ in the sense of definition \ref{Signed Subset II}.}
\begin{itemize}
   \item[\it(i)] $\displaystyle\bigcup_{e\in\ul{C}} \ol{\mathscr{S}_e^{\sgn_C(e)}}=S^{r-1}$ where  $\mathscr{S}_e^{\sgn_C(e)}:= \mathscr{S}_e^{\pm}$
   $\text{~~~~and~~}\ol{\mathscr{S}_e^\pm} \text{~~denotes the closure~ }\ol{\mathscr{S}_e^\pm}=\mathscr{S}_e^\pm\cup \mathscr{S}_e$
   
   \item[\it (ii)] The support $\ul{C}$ is minimal with respect to {\it (i)}, that is $\ul{C}$ contains only minimal subsets of $E$ such that the union of the closed half spaces $\ol{\mathscr{S}_e^\pm}$ covers $S^{r-1}$.
\end{itemize}

Then we get the following

\begin{Theorem}[Topological Representation Theorem.] 
The following conditions are equivalent \cite{Bjorner1999}:
   \begin{itemize}
      \item[(1)] If $\mathscr{S}=(\mathscr{S}_e)_{e\in E}$ is a signed arrangement of pseudeospheres in $S^{r-1}$, then $\C(\mathscr{S})$ is the family of circuits of a rank $r$ simple oriented matroid on $E$.
      \item[(2)] If $(E,\C)$ is a rank $r$ simple oriented matroid, then there exists a signed arrangement of pseudospheres $\mathscr{S}$ in $S^{r-1}$ such that $\C=\C(\mathscr{S})$.
       
       \item[(3)] $\C(\mathscr{S})=\C(\mathscr{S}')$ for two signed arrangements $\mathscr{S}$ and $\mathscr{S}'$ in $S^{r-1}$ iff $\mathscr{S}'=h(\mathscr{S})$ for some self-homeomorphism $h$ of $S^{r-1}$.
   \end{itemize}
 
\end{Theorem}

It follows that there is a one-to-one correspondence between (equivalence classes of) pseudosphere arrangements in $S^{r-1}$ and (reorientation classes of) simple rank $r$ oriented matroids. That is, a pseudosphere arrangement specifies an oriented matroid only up to reorientation and permutation (relabeling of the ground set).

Moreover the stretchability of the arrangement (existence of a self-homeomorphism on $S^{r-1}$ transforming each pseudosphere $\mathscr{S}_i$ into a proper sphere $S_i$ ) is equivalent to the realizability of the oriented matroid as a vector configuration. Note that the problem of deciding stretchability can be encoded in a system of equalities and inequalities of polynomial expressions. Finding a solution to this system is equivalent to deciding the realizability problem for an oriented matroid given by a pseudosphere arrangement \cite{Bjorner1999}. For rank $r=3$ it turns out that from $n\ge 9$ there exist oriented matroids whose corresponding pseudosphere arrangements cannot be stretched and which are thus not realizable \cite{Richter-Gebert1992}. For rank $r=3$ results for such a counting are shown in table \ref{Sign factor combinatorics table}.

\paragraph{Remark.} Notice that the term realizability is used in two different contexts throughout this work. In the mathematics literature such as \cite{Bjorner1999} realizability of a given matroid aims at answering the question whether a given set $\C$ of oriented circuits of a matroid can be realized by a vector configuration. 
In the context of \cite{BrunnemannRideout2008,BrunnemannRideout2008a}  the term realizability is directed towards the question whether a given set $\vec{\epsilon}$ of sign factors in the definition of the volume operator $\hat{V}_v$ at a vertex $v$ in the vertex set of a graph corresponds to the chirotope of an oriented matroid of rank 3 which can be realized (in the mathematical sense) as a vector configuration.
Either question of realizability not only decides whether a given family of oriented sets respectively sign factors represents an oriented matroid, but it also can be used to count the number of realizable oriented matroids of a given rank $r$ over an $n$-element set $E$.

\subsection{\label{Constructions with Oriented Matroids}Operations on Oriented Matroids}
Finally we would like to define certain operations which can be performed on oriented matroids. Let again $\M=(E,\C)=(E,\MC{B})$ be an oriented rank $r$ matroid over the ground set $E$. Then the following operations are defined \cite{Bjorner1999}.

\paragraph{Reorientation. }According to definition \ref{Signed Subset Def} one can define a reorientation $\lu{-F~}C=:\WT{C}$ of a signed set $C$ on $F\subseteq E$ by  
   $(\WT{C})^\pm=(C^\pm\backslash F)\cup (C^\mp\cap F)$, that is
   \begin{itemize}
     \item[(a)] $\sgn_{\WT{C}}(e):=(-1)^{|F\cap\{e\}|}\cdot \sgn_C(e)$ ~~~~~~~~~and
      
     \item[(b)] $\lu{-F~}\chi_{\MC{B}}(e_1,\ldots,e_r)=\WT{\chi}_{\MC{B}}(e_1,\ldots,e_r):=\chi_{\MC{B}}(e_1,\ldots,e_r)\cdot (-1)^{|F\cap \{e_1,\ldots,e_r\}|} $.  
   \end{itemize}

 \paragraph{Mutation.} Let $\M$ furthermore be a uniform\footnote{See section \ref{Further Definitions} for the definition.} oriented matroid. Then every $r$-tuple $B=(b_1,\ldots,b_r)$ is a basis. An $r$-tuple $\ol{B}$ 
yields a mutation of $\M$ if the mapping $\lu{\ol{B}~}\chi_{\MC{B}}$ given by
 \be
    \lu{\ol{B}~}\chi_{\MC{B}}(B):=\left\{\begin{array}{rl}
                                           -\chi_{\MC{B}}(B) & \text{if} ~\ol{B}=B\\\\
                                            \chi_{\MC{B}}(B) & \text{otherwise}
                                        \end{array}\right.
 \ee
 defines another oriented matroid, that is  $\lu{\ol{B}~}\chi_{\MC{B}}$ satisfies the Grassmann Pl\"ucker relations {\it (B2) / (B2')} of definition \ref{Grassmann Plucker relations for chirotopes}.

\paragraph{Deletion.} If a subset $A\subset E$ is removed from $E$ then the remaining matroid $\M\backslash A$ over the ground set $E\bs A$ is defined by its set of signed circuits
\be
   \C\bs A:=\{C\in\C~:~C\cap A=\emptyset\} \;.
\ee
Suppose $\M\backslash A$ has rank $s<r$. Choose an arbitrary basis $B\in\MC{B}$. Now take the $(r-s)$ tuple $B\cap A=:(a_1,\ldots,a_{r-s})\in A$. Then the chirotope of $\chi_{\MC{B}\bs A}$ can be constructed (up to an overall sign) as follows:  
\ba\label{chirotope after deletion}
  \chi_{B\bs A}:~~~~ (E\bs A)^s&\rightarrow& \{-1,0,+1\}\NN
                 (e_1,\ldots,e_s) &\mapsto& \chi_{\MC{B}}(e_1,\ldots,e_s,a_1,\ldots,a_{r-s}) 
\ea

By construction $\chi_{B}\bs A$ has all properties of a chirotope. Moreover the above definition specifies $\chi_{B\bs A}$ uniquely up to a sign \cite{Bjorner1999}, independent of the choice of $B\in\MC{B}$. 
\\~\\
{\it Example.} As an example consider a ground set $E=(e_1,e_2,e_3)$ of 3 vectors $e_1,e_2,e_3$ in $\mb{R}^2$. Let a family $\MC{B}$ of bases be given by $\MC{B}=(B_1,B_2)$ with $B_1=(e_1,e_3)$, $B_2=(e_2,e_3)$, assume $e_1,e_2$ to be co-linear, such that $\ul{\C}=\{\ul{C}\}$ with $\ul{C}=\{e_1,e_2\}$. The resulting oriented matroid $\M=(E,\MC{B})$ then has rank $r=2$. Now consider $A=\{e_3\}$. Clearly $\M\bs A$ has rank $s=1$ with bases $\MC{B}\bs A=\{e_1,e_2\}$. in order to construct $\chi_{\MC{B}\bs A}$, choose $B_2\in\MC{B}$ with $B_2\cap A=(e_3)=:(a_1)$. Applying (\ref{chirotope after deletion}) then gives $\chi_{\MC{B}\bs A}(e_1)=\chi_{\MC{B}}(e_1,a_1)$ and $\chi_{\MC{B}\bs A}(e_2)=\chi_{\MC{B}}(e_2,a_1)$.

\section{\label{Oriented Matroids in Loop Quantum Gravity}Oriented Matroids in Loop Quantum Gravity}
Having outlined the construction of LQG in section \ref{LQG Combinatorics}, and oriented matroids in the previous section, we will now demonstrate how oriented matroids naturally enter into LQG. As we have seen, both the local geometric properties of a graph $\gamma$ as well as its global topological (connectedness) properties can be described by oriented matroids. 
\subsection{\label{Local Properties: Signfactors Characterizing the Volume Spectrum}Local Properties: Signfactors Characterizing the Volume Spectrum }
As described in section \ref{LQG Combinatorics}, the measure $\mu_0$ on the kinematical Hilbert space $\Ho$ is sensitive only to  $\mathrm{supp}(l(\gamma))$ \footnote{$l(\gamma)$ is defined in section \ref{LQG Combinatorics}.} of the graph $\gamma$ underlying a spin network function \cite{Bahr2007} in $\Ho$, but not to other  properties of $\gamma$ such as relative orientations of edges at the vertices. 

In contrast to $\mu_0$, the flux and a composite thereof,  
namely the volume operator \cite{Ashtekar1998}, {\it are} sensitive to relative orientations of edges. These relations are preserved by analytic diffeomorphisms. Even more, this sensitivity is crucial for a consistent formulation of the theory \cite{Giesel2006a} and the constraint operators \cite{Thiemann1996}. 
This is based on the fact that the spectrum of the volume operator due to \cite{Ashtekar1998} is characterized by signature factors resulting from the embedding of $\gamma$ in $\Sigma$ as shown in \cite{BrunnemannRideout2008,BrunnemannRideout2008a}. 
They encode the relative orientation of all tangent vectors of $N_v$ edges intersecting at a vertex $v$ of $\gamma$. The classification of all possible signature factor combinations turns out to be intimately related to a central question of oriented matroid theory \cite{Bjorner1999,Ziegler1998}, namely whether a given oriented matroid of rank $3$ can be realized as a configuration of $N_v\ge 3$ vectors in $\mathbbm{R}^3$.

\subsubsection{Construction of the Volume Operator}

Within LQG, the operator corresponding to the volume of a region $R$ in three
dimensional Riemannian space plays a prominent role for the implementation of
the scalar (Hamilton) constraint operator on the quantum level.
In  \cite{BrunnemannRideout2008,BrunnemannRideout2008a}, the spectral properties of the volume operator according to \cite{Ashtekar1998,Thiemann1998} were analyzed on $\Hgauss$, the gauge invariant subspace of the kinematical Hilbert space $\Ho$ of LQG. Starting from the classical volume expression rewritten in terms of Ashtekar variables\footnote{ Here $\epsilon_{abc},\epsilon^{ijk}$ denote the antisymmetric symbol ($\epsilon_{123}=1=-\epsilon_{213}$ etc.) and we use Einstein's sum convention. Moreover $\det q(x)$ denotes the determinant of the spatial metric on the Cauchy surface $\Sigma$.  $E^a_i(x)$ is an $su(2)$-valued vector density, called a densitized triad. It holds that $E^a_i(x)E^b_j(x)\delta^{ij}=\det q(x) q^{ab}(x)$. Here $\delta^{ij}$ is the Cartan-Killing metric of $su(2)$, which is just the Euclidean metric. Moreover $q^{ab}(x)$ are the components of the inverse spatial metric in the tangent space $T_x\Sigma$ at the point $x\in\Sigma$, with respect to a given basis (e.g.\ a coordinate basis $\partial_a$).}
\ba\label{Classical Volume Expression}
   V(R)&=&\int_R~d^3x~\sqrt{\det q(x)}
   =\int_R~d^3x~\sqrt{\Big|\frac{1}{3!} \epsilon^{ijk}\epsilon_{abc}E^a_i(x)E^b_j(x)E^c_k(x) \Big|}
\ea
one can work out the action of $V(R)$ on a cylindrical function $f_\gamma: G^{n}\ni(h_{e_1},\ldots,h_{e_n})\rightarrow f_\gamma(h_{e_1},\ldots,h_{e_n})$, which has support on the $n$ copies of $G=SU(2)$, each labelled by one of the edges $(e_1,\ldots,e_n)$ contained in the edge set $E(\gamma)$ of a graph $\gamma$ embedded into three dimensional Riemannian space.   
Notice that the graph $\gamma$ underlying the cylindrical function $f_\gamma$ is taken to be adapted to the flux operator\footnote{Let $S\subset\Sigma$ be an orientable two-dimensional surface in $\Sigma$. Then the flux $E_i(S)$ is defined as the densitized triad $E^c_i(x)$ integrated over the 2-surface $S$, that is  $E_i(S)=\int_{S}\,du\,dv\,\epsilon_{abc}\frac{\partial x^a(u,v)}{\partial u}\frac{\partial x^b(u,v)}{\partial v} E^c_i(x(u,v)) $ where $x:(u,v)\mapsto x(u,v)$ denotes the embedding of the surface $S$ into $\Sigma$ parametrized by the pair $(u,v)$.} $E_i(S)$. That is, the elements in $E(\gamma)$, which intersect $S$, are subdivided, such that their subdivisions either end or start at $S$. With these assumptions the action of $E(S)$ on $f_\gamma$ is explicitly given as follows \cite{Thiemann2001}:
\ba\label{general flux action}
  \big[E_i(S)f_\gamma\big](h_{e_1},\ldots,h_{e_n})
  &=&
  \frac{1}{2}\sum_{e\in E(\gamma)}\sum_{AB} \epsilon(e,S)
  ~\Big(\delta_{e\cap S,b(e)}\frac{\tau_i}{2}h_e + \delta_{e\cap S,f(e)}h_e \frac{\tau_i}{2}\Big)_{AB}
  \frac{\partial f_\gamma(h_{e_1},\ldots,h_{e_n})}{\partial (h_e)_{AB}}
  \NN
  &=&
  \frac{1}{4}\sum_{e\in E(\gamma)} \epsilon(e,S)
  ~\Big[\big(\delta_{e\cap S,b(e)}R^i_e + \delta_{e\cap S,f(e)}L^i_e\big)f_\gamma\Big]
 \big(h_{e_1},\ldots,h_{e_n}\big)~~.
\ea
Here we denote by $[R_e^i f_\gamma](h_{e_1},\ldots,h_{e_n}):= \frac{d}{dt}f_\gamma\big(h_{e_1},\ldots,\mb{e}^{t\, \tau_i}h_e,\ldots,h_{e_n}\big)\big|_{t=0}$ and respectively by \linebreak $[L_e^i f_\gamma](h_{e_1},\ldots,h_{e_n}):= \frac{d}{dt}f_\gamma\big(h_{e_1},\ldots,h_e\mb{e}^{t\,\tau_i},\ldots,h_{e_n}\big)\big|_{t=0}$  the action of right / left invariant vector fields on the copy of $SU(2)$ labeled by $e\in E(\gamma)$. By $\tau_i$ we represent a basis of the Lie algebra $su(2)$, given by $\tau_i=-\mb{i}~\sigma_i$ with $\sigma_i$ being the according Pauli matrix
\footnote{
\[
\begin{array}{ccccc}
      \sigma_1=\left(\begin{array}{cc}
                     0 & 1 \\
		     1 & 0
                 \end{array}\right)
    &&\sigma_2=\left(\begin{array}{cc}
                     0 & -\mb{i}~ \\
		     \mb{i} & 0
                 \end{array}\right)		 
    &&\sigma_3=\left(\begin{array}{cc}
                     1 & 0 \\
		      0 & -1
                 \end{array}\right)	~~~~~~~\mbox{with}
~~~~~\big[\sigma_i,\sigma_j \big]=2\mb{i}\,\epsilon^{ijk}\sigma_k	 
  \end{array}
\]
}. Moreover $\delta_{e\cap S,b(e)}=1$ if the intersection $e\cap S$ is the (b)eginning point of $e$, that is $e$ is outgoing from $S$ and zero otherwise. Accordingly $\delta_{e\cap S,f(e)}=1$ if the intersection $e\cap S$ is the (f)inal point of $e$, that is $e$ is incoming to $S$ and zero otherwise. Also in (\ref{general flux action}) each $e\in E(\gamma)$ labels a particular copy of $SU(2)$.
The sum over $A,B$ can be taken in defining a representation of $SU(2)$, but in principle any representation could be chosen\footnote{This is a quantization ambiguity.}. The orientation factor $\epsilon(e,S)$ indicates the relative orientation of the edge tangent at the intersection point and the chosen surface orientation $\vec{n}_S$ of $S$.  

\begin{figure}[h!]
\center
\cmt{3}
  {\center
    \psfrag{nS}{$\red\vec{n}_S$}
    \psfrag{S}{$S$}
    \psfrag{x}{$x$}
    \psfrag{de}{$\blue\dot{e}_1(x)$}
    \psfrag{e}{$e_1$}
    \psfrag{ep}{$\epsilon(S,e_1)\!=\!+1$~~$x\!=\!b(e_1)$}
    
    \includegraphics[width=3cm]{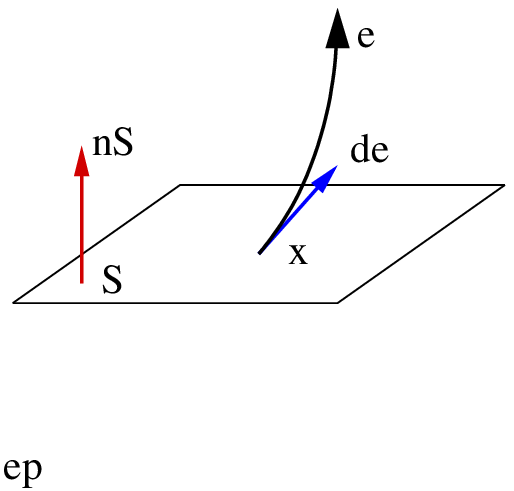}
   }~~~~~~~~
\cmt{3}
  {\center
    \psfrag{nS}{$\red\vec{n}_S$}
    \psfrag{S}{$S$}
    \psfrag{x}{$x$}
    \psfrag{de}{$\blue\dot{e}_2(x)$}
    \psfrag{e}{$e_2$}
    \psfrag{ep}{$\epsilon(S,e_2)\!=\!-1$~~$x\!=\!f(e_2)$}
    
    \includegraphics[width=3cm]{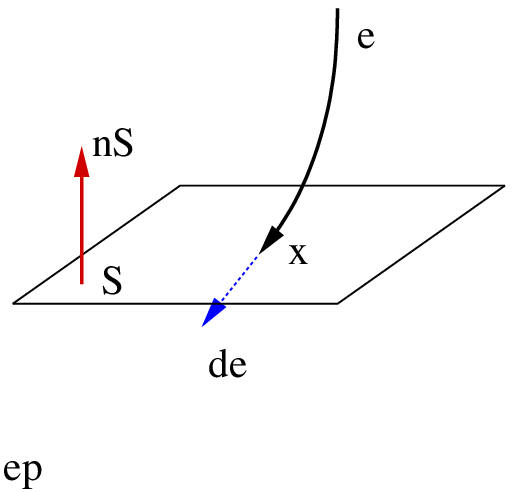}
   }~~~~~~~~
\cmt{3}
  {\center
    \psfrag{nS}{$\red\vec{n}_S$}
    \psfrag{S}{$S$}
    \psfrag{x}{$x$}
    \psfrag{de}{$\blue\dot{e}_3(x)$}
    \psfrag{e}{$e_3$}
    \psfrag{ep}{$\epsilon(S,e_3)\!=\!+1$~~$x\!=\!f(e_3)$}
    
    \includegraphics[width=3cm]{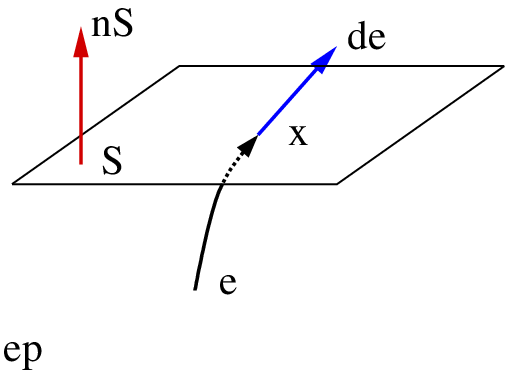}
   }~~~~~~~~
\cmt{3}
  {\center
    \psfrag{nS}{$\red\vec{n}_S$}
    \psfrag{S}{$S$}
    \psfrag{x}{$x$}
    \psfrag{de}{$\blue\dot{e}_4(x)$}
    \psfrag{e}{$e_4$}
    \psfrag{ep}{$\epsilon(S,e_4)\!=\!-1$~~$x\!=\!b(e_4)$}
    
    \includegraphics[width=3cm]{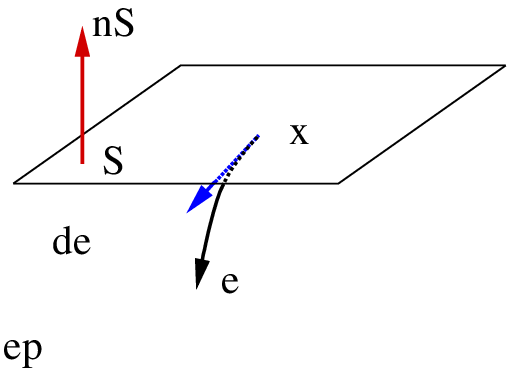}
   }
   \caption{The four distinct configurations of the surface $S$ and an edge $e$. The intersection point of $S$ and $e$ is denoted by $x$. If $S\cap e=e$ or $e\cap S=\emptyset$ then $\epsilon(S,e)=0$.}
    \label{flux1}
\end{figure}
Moreover if we set  $\WT{f}_\gamma(\ldots, h,\ldots):=f_\gamma(\ldots, h^{-1}, \ldots)$ then it holds that
\be\label{left/right-switch}
   [R^i_e \WT{f}_\gamma](\ldots, h_e, \ldots)
   = \frac{d}{dt}f_\gamma\big(\ldots, (\mb{e}^{t\,\tau_i}h_e)^{-1}, \ldots\big)\big|_{t=0}
   = \frac{d}{dt}f_\gamma\big(\ldots, h_e^{-1}\,\mb{e}^{-t\,\tau_i}, \ldots\big)\big|_{t=0}
   = -[L^i_e f_\gamma](\ldots, h_e^{-1}, \ldots)
\ee
by the exchange property of left / right invariant vector fields.
Now consider a cylindrical function $f_{e_1}(h_{e_1})$, where $e_1$ is given as in the first case of figure  \ref{flux1}. Moreover $e_1^{-1}=e_2$, $l(e_1)=l(e_2)$ where $e_2$ is given as in the second case of figure  \ref{flux1}. Then   
\ba\label{special flux action}
  \big[E_i(S)f_{e_1}\big](h_{e_1})
  &=&
  \frac{1}{4} 
  ~\big[R^i_{e_1} f_{e_1}\big]
 \big(h_{e_1}\big)
 =
  \frac{1}{4} 
  ~\big[L^i_{e_1} f_{e_1}\big]
 \big(h_{e_1}^{-1}\big)
 =\frac{1}{4} 
  ~\big[L^i_{e_1} f_{e_1}\big]
 \big(h_{e_1^{-1}}\big)
 =\frac{1}{4} 
  ~\big[L^i_{e_2} f_{e_2}\big]
 \big(h_{e_2}\big)
\ea
where we have used the transformation property of the edge holonomy under edge reorientation $e\rightarrow e^{-1}$, that is ${h_e}^{-1}=h_{e^{-1}}$. Notice that $\epsilon(S,e_1)=-\epsilon(S,{e_1}^{-1})=  -\epsilon(S,e_2)$ compensates the minus sign between the action of $R^i_{e_1}$ and $L^i_{e_1}$. A similar statement holds for case 3 and 4 in figure \ref{flux1}. That is, the action (\ref{special flux action}) of the flux operator on a cylindrical function $f_\gamma$ only depends on the support $l(\gamma)$, and not on the actual direction of edges contained in $E(\gamma)$. 

In contrast to this, situation 1 and 3 in figure  \ref{flux1} have identical relative orientations $\epsilon(S,e_1)=\epsilon(S,e_3)=1$, but 
\ba\label{special flux action 2}
  \big[E_i(S)f_{e_1}\big](h_{e_1}) =  \frac{1}{4} \big[R^i_{e_1} f_{e_1}\big] \big(h_{e_1}\big) 
  &\text{~~~and~~~}&
  \big[E_i(S)f_{e_3}\big](h_{e_3}) =  \frac{1}{4} \big[L^i_{e_3} f_{e_3}\big] \big(h_{e_3}\big)
\ea
because $x=b(e_1)=f(e_3)$. Strictly speaking,  (\ref{special flux action 2}) describes the action of the flux on two {\it different} cylindrical functions $f_{e_1},f_{e_3}$, which have different supports $l(e_1)\ne l(e_3)$.
\\~\\
Now we are going to discuss the implication of this on the volume operator. 
In its final form, 
the action of the volume operator on a cylindrical function $f_\gamma$ is given by \cite{Ashtekar1998,Thiemann1998}:
\ba
  \hat{V}(R)_{\gamma}~f_{\gamma}(\cdot)&=&\int\limits_R d^3x\widehat{\sqrt{det(q(x))_{\gamma}}}~f_{\gamma}(\cdot)
                     =\int\limits_R d^3x~ \hat{V}(x)_{\gamma}~f_{\gamma}(\cdot)
\ea
where the ``~$\widehat{~~}$~'' symbolizes the operator corresponding to the classical expression, $R\subseteq\Sigma$ denotes a region in $\Sigma$, and
\be
  \label{nullte} \hat{V}(x)_{\gamma}=\ell_P^3 \sum_{v\in V(\gamma)}\delta^3(x,v)~\hat{V}_{v,\gamma}~~.
\ee
Here $\ell_P$ denotes the Planck length and $\delta^3(x,v)$ is Dirac's delta distribution. The action of $\hat{V}(R)_\gamma$ decomposes into a local action at all vertices $v$ in the vertex set $V(\gamma)$ with valence $N_v$
\ba  
  \label{erste} \hat{V}_{v,\gamma} &=&\sqrt{\Big| Z 
                   \sum_{I<J<K\le N_v} 
		   \epsilon (I, J, K)~ \hat{q}_{IJK}\Big|}\\
\label{definition qIJK}
   \hat{q}_{IJK}&=&\sum_{i,j,k=1}^3 \varepsilon_{ijk}\, R_I^i\,R_J^j\,R_K^k\propto \Big[(R_{IJ})^2,(R_{JK})^2\Big] ~~~\text{with}~~(R_{IJ})^2=\sum_{k=1}^3 (R^k_I+R^k_J)^2~~.
\ea
Here $Z$ is an overall regularization constant which does not change the spectral properties modulo a possible rescaling. We will set $Z=1$ in numerical computations\footnote{Notice however that in \cite{Giesel2006a} it was found that $Z=\frac{1}{48}$, which is relevant if concrete numerical values are required as e.g.\ in \cite{Flori2008}.}.
At each vertex $v$ one has to sum over all possible ordered triples $(e_I,e_J,e_K)\equiv (I,J,K)$, $I<J<K\le N_v$ of outgoing edges at  $v$. Here  $\epsilon(IJK)=\mbox{sgn}\big(\det{(\dot{e}_I(v),\dot{e}_J(v),\dot{e}_K(v))}\big)$ denotes the sign of the determinant of the tangents of the three edges $e_I,e_J,e_K$ intersecting at $v$. In the sequel we will denote by $\vec{\epsilon}:=\{\epsilon(I,J,K)\}_{I<J<K\le N_v}$ the set of $\epsilon$-signfactors for all triples $I<J<K\le N_v$.

Notice that the graph $\gamma$ is adapted to $\hat{V}(R)_{\gamma}$ such that all vertices $v\in V(\gamma)$ with $N_v\ge 3$ are the beginning points of edges. This can be achieved by cutting each edge $e\in E(\gamma)$ into two pieces $e', e''$ such that $e$ is given as a composition $e=e'\circ (e'')^{-1}$ with $b(e)=b(e')$, $f(e)= b(e'')$. This introduces a two valent vertex $v'=f(e')=f(e'')$ which does not contribute to the volume, as the action of $ \hat{V}(R)_{\gamma}$ on two valent vertices is trivial. This adaption makes it possible to rewrite the action of $\hat{V}(R)_{\gamma}$ entirely in terms of right invariant vector fields $R^k_I$.

 Moreover, if gauge invariance is imposed then the volume operator can be rewritten as

\ba\label{Volume definition gauge invariant 3}
   \hat{V}_{v,\gamma}&=:& \sqrt{\Big|Z\cdot 
          \sum\limits_{I<J<K<N_v}\sigma(I,J,K)~\hat{q}_{IJK} \Big|}        
\ea
where
\be\label{sigmas}
\sigma(I,J,K):= \epsilon(I,J,K)-\epsilon(I,J,N)+\epsilon(I,K,N)-\epsilon(J,K,N)~~.
\ee 

We will accordingly denote by  $\vec{\sigma}:=\{\sigma(I,J,K)\}_{I<J<K< N_v}$ the set of $\sigma$-signfactors for all triples $I<J<K< N_v$.

\subsubsection{Sign Factors as Chirotopes}
\label{chirotopes-sec}

In \cite{BrunnemannRideout2008,BrunnemannRideout2008a} it was shown that the spectral properties of the volume operator depend strongly on the sign factors $\epsilon(I,J,K)$, respectively $\sigma(I,J,K)$.  In particular the presence of a smallest non-zero eigenvalue (also referred to as a `volume gap') depends crucially on this underlying sign configuration.
In \cite{BrunnemannRideout2008,BrunnemannRideout2008a} the set of all possible sign factors for an $N_v$ valent vertex $v$ was computed by a Monte-Carlo method, which sprinkles $N_v$ points on the unit sphere and collects the occurring distinct sign factors in a table. With this method the sign configurations resulting from $N_v$ vectors, without coplanar triples, were determined up to $N_v=7$. 
However, the combinatorics of the sign configurations were not fully understood at that point.
\\~\\
This situation has now improved, as the problem can be viewed in the broader context of oriented matroids. 
The set of $\vec{\epsilon}$ sign factors  resulting from a configuration  of $N_v$ vectors (without coplanar triples) can be understood as a chirotope of a (uniform) 
oriented matroid of rank 3 over $N_v$ elements, represented as a set of vectors in $\mathbb{R}^3$.

As already pointed out in \cite{BrunnemannRideout2008a}, the choice of a particular edge labeling has no significance, and thus the
volume spectrum should be invariant under the permutation of edge labels.  We thus wish
to group our eigenvalue data into equivalence classes, defined by the action of permutations on the edge labels.  For the case of edge spins, the action of a permutation is straightforward.  
Such permutations, however, act in a non-trivial way upon the chirotope.  
For example, if we exchange edge labels $I\leftrightarrow J$, then $\epsilon(I,J,K)\mapsto  \epsilon(J,I,K)=-\epsilon(I,J,K)=:\epsilon'(I,J,K)$, 
$\epsilon(I,L,M)\mapsto  \epsilon(J,L,M)=:\epsilon'(I,L,M)$ and 
$\epsilon(J,L,M)\mapsto  \epsilon(I,L,M)=:\epsilon'(J,L,M)$. 
We thus identify chirotopes which can be transformed
into each other by a permutation into equivalence classes, each
identified by a canonical
representative.\footnote{We represent each chirotope by a string of ${N_v
    \choose 3}$ bits, and choose the smallest representative of each class,
  where the bit string is interpreted directly as a non-negative integer.
  Here `-' is 0, `+' 1, and the triples start at the low end with 123 at $2^0$,
  124 at $2^1$, etc.}

For each such representative, we compute the sigma configuration, from
(\ref{sigmas}).  For the 6- and 7-vertex, there are a small number of
duplicates --- multiple permutation equivalence classes of chirotopes which
give rise to the same $\vec{\sigma}$-configurations.

~\\
It turns out that a more convenient symmetry, from the perspective of the
mathematics literature \cite{Richter-Gebert1992} (c.f.\ section \ref{Realizability of Oriented Matroids}), is to regard
chirotopes as invariant under a reorientation of any subset of the edge
vectors $\dot{e}_I$, in addition to any permutation of the edge labels.
A classification into such reorientation equivalence classes is thus well studied, 
and is shown in the rightmost column of table \ref{Sign factor combinatorics table}.

Using the according data \cite{Kortenkamp}, we are now able to exactly reproduce\footnote{
One can compute the permutation equivalence classes from the reorientation equivalences classes as follows.  
Choose a representative $\vec{\epsilon}_{\text{Reor}}$ of each reorientation equivalence class, and apply all $2^{N_v}$ reorientations to it.  For each such reorientation apply each of the $N_v!$ permutations of the elements, and record the canonical representative of the resulting chirotope.  The set of such canonical representatives gives 
the set of permutation equivalence classes of chirotopes.}
the permutation equivalence classes of chirotopes, that were previously computed in \cite{BrunnemannRideout2008, BrunnemannRideout2008a} by a Monte Carlo sprinkling process for $N_v\le 8$. This confirms the remarkable applicability of the Monte Carlo method to this problem. Having the oriented matroid data at hand, we are now sure that we had detected all uniform oriented matroids of rank 3 over a ground set of up to $N_v=8$ elements.\footnote{
To be completely correct we were not able to perform the sprinkling of the 8-vertex in \cite{BrunnemannRideout2008, BrunnemannRideout2008a}, because the enormous number of chirotopes overwhelmed the system memory of any machines we had available.  We performed the sprinkling with $N_v=8$ more recently, by computing the canonical representative of each chirotope as it arose, thus having to save only 28 287 chirotopes in place of more than $10^8$.}

In addition, one can write down an analytic expression which gives a lower
bound to the total number $\#\vec{\epsilon}\,(N_v)$ of realizable
$\vec{\epsilon}$ sign factors for $N_v$ vectors 
\cite{Barbero2008}.
For $N_v\ge 3$, it is
\be\label{Sign Factor formula}
   \#\vec{\epsilon}\,(N_v) \geq \frac{(N_v-2)!}{4^{N_v-2}}\prod_{s=1}^{N_v-2}(s^3-2s^2+7s+2) \;.
\ee
Note that this expression is able to exactly reproduce
$\#\vec{\epsilon}\,(N_v)$ for $N_v$ up to 6.  A derivation of this bound will appear in \cite{BrunnemannRideout2010}.

\begin{table}[hbt!]
\center 
\begin{tabular}{|c|r|r|r|r|r|r|}  
\hline 
$N_v$ 
& \cmt{1.5}{\# triples} 
& \cmt{2.5}{$\#\vec{\epsilon}\,(N_v)$\\ (lower bound by (\ref{Sign Factor formula}) )} 
& \cmt{2.5}{$\#\vec{\epsilon}\,(N_v)$ \\ (sprinkling)} 
& \cmt{2}{\#$\vec{\epsilon}$\\ permutation equivalence classes\\}
& \cmt{1.6}{\# $\vec{\sigma}$\\ confi-gurations}
& \cmt{2.5}{\# realizable reorientation equiv. classes}
\\[5mm]\hline
3  &   1 &               2&                  2 &        1 & 1   &     1      \\
4  &   4 &              16&                 16 &        3 & 3   &     1      \\
5  &  10 &             384&                384 &        4 & 4   &     1      \\
6  &  20 &          23~808&             23~808 &       41 & 39  &     4      \\
7  &  35 &       3~333~120&          3~486~720 &      706 & 673 &    11      \\
8  &  56 &     939~939~840&  $\ge$ 747~735~880 &  28~ 287 &&   135        \\
9  &  84 & 486~888~837~120&                  ? &        ? && 4~381 
\\\hline 
\end{tabular}
\caption{Sign factor combinatorics for 3--9-valent non-coplanar vertices embedded in $\mb{R}^3$.} 
\label{Sign factor combinatorics table}
\end{table}

As can be seen from (\ref{erste}), the volume operator evaluated at a vertex $v$ is not only sensitive to the local edge tangent vector configuration encoded in the $\epsilon(I,J,K)$ factors, but also on 
whether $v$ is the beginning or final point of the $N_v$ edges intersecting at $v$.
To see this, notice that the reorientation of an edge, say $e_L$, outgoing from $v$, flips the edge tangent $\dot{e}_L(v)$, and hence inverts all signs $\{\epsilon(IJL)\}$ containing the label $L$. However, at the same time, the action of $[R^i_{e_L}f_\gamma](\ldots,h_{e_L},\ldots)$ changes to $[L^i_{e_L}f_\gamma](\ldots,h_{e_L}^{-1},\ldots)
=-[R^i_{e_L}\WT{f}_\gamma](\ldots,h_{e_L},\ldots)$ according to (\ref{left/right-switch}). This compensates the sign inversion of all  $\epsilon(IJL)$ similarly to (\ref{special flux action}). This situation is depicted in figure \ref{vertex1} and figure \ref{vertex2}. There $e_3$ is reoriented, and hence the orientation of the edge tangent $\dot{e}_3(v_2)$ is flipped compared to $\dot{e}_3(v_1)$. Although the configuration of sign factors in figures \ref{vertex1} and \ref{vertex2} differs accordingly, the volume spectrum at $v_1$ and $v_2$ is identical.

To rephrase this in terms of oriented matroids: If an element in the graphic matroid constructed from $\gamma$ is reoriented, this induces a reorientation of the according element in the local vector oriented matroid at $v$.
Because of the described sign compensation, 
the volume spectrum is thus invariant under reorientations of the graphic oriented matroid of $\gamma$.

In contrast to this,  figure \ref{vertex3} seems to have the same sign configuration $\vec{\epsilon}$ induced from the edge tangents at $v_3$ as we find at $v_2$. 
However, the support (tracing of $e_3$) of the underlying graph $\gamma$ is different in figure \ref{vertex2} and figure \ref{vertex3}. Hence by (\ref{special flux action 2}) the signs $\vec{\epsilon}$ of (\ref{erste}) containing $e_3$ are effectively opposite at $v_2$ and $v_3$, and it follows that the volume spectra at $v_2$ and $v_3$ are different. This is equivalent to stating that $ \hat{V}(R)$ is evaluated on spin network functions supported on non-diffeomorphic graphs.

It follows that the spectrum of the volume operator is invariant under relabellings (permutation) of edges adjacent to $v$.\footnote{In \cite{Thiemann1998,Brunnemann2006a,BrunnemannRideout2008a} the details for computing the matrix elements $\hat{q}_{IJK}$ in (\ref{Volume definition gauge invariant 3}) of the volume operator with respect to a basis of gauge invariant spin network functions is given. This basis is formulated in terms of so-called recoupling schemes. Permuting edge labels amounts to changing the recoupling order, which is a unitary basis transformation. This can be seen from the fact that the transformation matrix between two recoupling schemes can be written in terms of Clebsch-Gordan coefficients which are unitary. Details will be given in \cite{BrunnemannRideout2010}. }
It is furthermore invariant under reorientations of the adjacent edges, which induce reorientations on the vector oriented matroid encoded by the $\vec{\epsilon}$ factors (figures \ref{vertex1} and \ref{vertex2}). However it is {\it not invariant} under an isolated reorientation of the oriented matroid encoded in the $\vec{\epsilon}$ factors alone (figure \ref{vertex3}). Hence different volume spectra at $v$ are characterized by permutation equivalence classes of $\vec{\epsilon}$ configurations.

\begin{figure}[hbt!]
\center
\cmt{5}
  {\center
    \psfrag{e1}{$e_1$}
    \psfrag{e2}{$e_2$}
    \psfrag{e3}{$e_3$}
    \psfrag{e4}{$e_4$}
    \psfrag{e5}{$e_5$}
    \psfrag{e6}{$e_6$}
    
    \psfrag{e1d}{$\blue\dot{e}_1$}
    \psfrag{e2d}{$\blue\dot{e}_2$}
    \psfrag{e3d}{$\blue\dot{e}_3$}
    \psfrag{e4d}{$\blue\dot{e}_4$}
    \psfrag{e5d}{$\blue\dot{e}_5$}
    \psfrag{e6d}{$\blue\dot{e}_6$}
    
    \psfrag{v}{$v_1$} 
       
    \includegraphics[width=4cm]{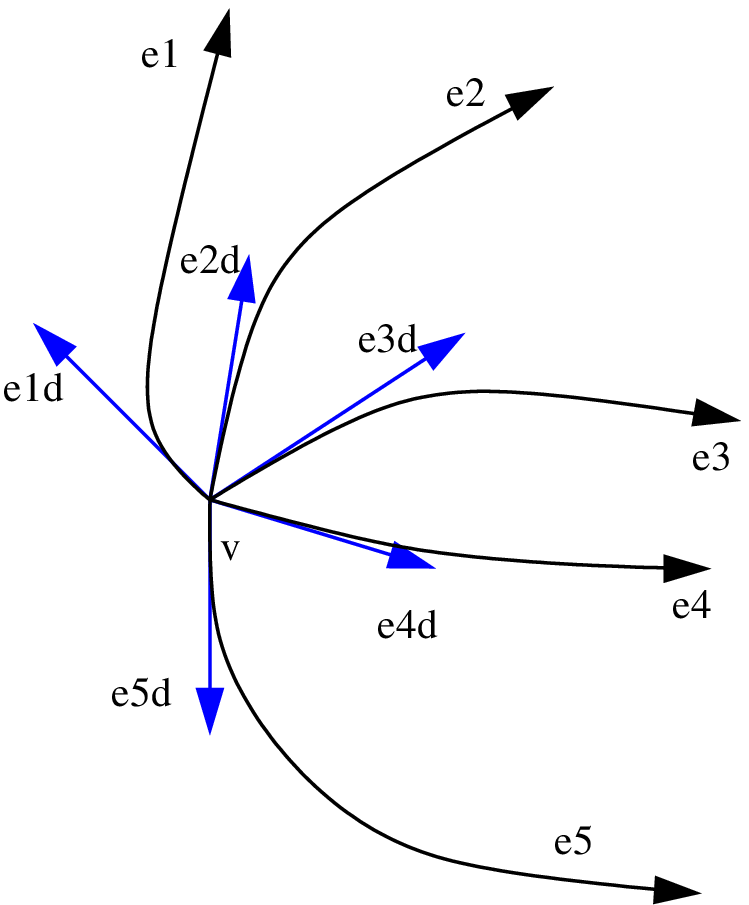}
    \caption{Vertex $v_1$ all edges outgoing.}
    \label{vertex1}
   }~~~~~
  \cmt{5}
  {\center
    \psfrag{e1}{$e_1$}
    \psfrag{e2}{$e_2$}
    \psfrag{e3}{$e_3$}
    \psfrag{e4}{$e_4$}
    \psfrag{e5}{$e_5$}
    \psfrag{e6}{$e_6$}
    
    \psfrag{e1d}{$\blue\dot{e}_1$}
    \psfrag{e2d}{$\blue\dot{e}_2$}
    \psfrag{e3d}{$\blue\dot{e}_3$}
    \psfrag{e4d}{$\blue\dot{e}_4$}
    \psfrag{e5d}{$\blue\dot{e}_5$}
    \psfrag{e6d}{$\blue\dot{e}_6$}
    
    \psfrag{v}{$v_2$} 
       
    \includegraphics[width=4cm]{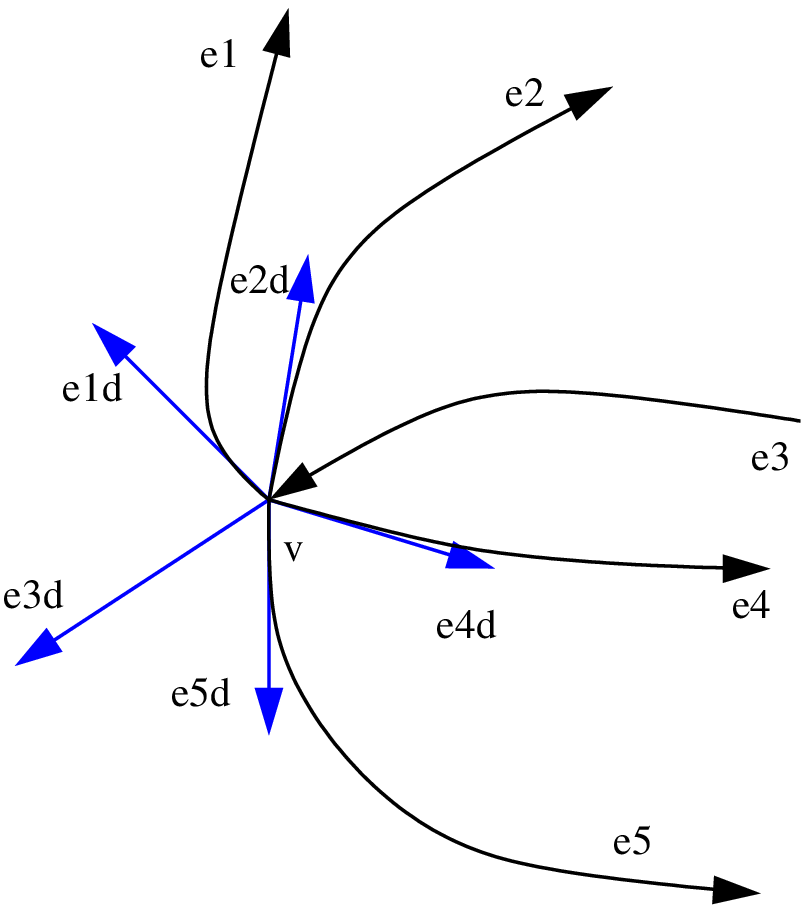}
    \caption{Vertex $v_2$ generated from $v_1$ by orientation reversal of edge $e_3$.}
    \label{vertex2}
   }~~~~~
   \cmt{5}
  {\center
    \psfrag{e1}{$e_1$}
    \psfrag{e2}{$e_2$}
    \psfrag{e3}{$e_3$}
    \psfrag{e4}{$e_4$}
    \psfrag{e5}{$e_5$}
    \psfrag{e6}{$e_6$}
    
    \psfrag{e1d}{$\blue\dot{e}_1$}
    \psfrag{e2d}{$\blue\dot{e}_2$}
    \psfrag{e3d}{~\vspace{2mm}$\blue\dot{e}_3$}
    \psfrag{e4d}{$\blue\dot{e}_4$}
    \psfrag{e5d}{$\blue\dot{e}_5$}
    \psfrag{e6d}{$\blue\dot{e}_6$}
    
    \psfrag{v}{$v_3$} 
       
    \includegraphics[width=5cm]{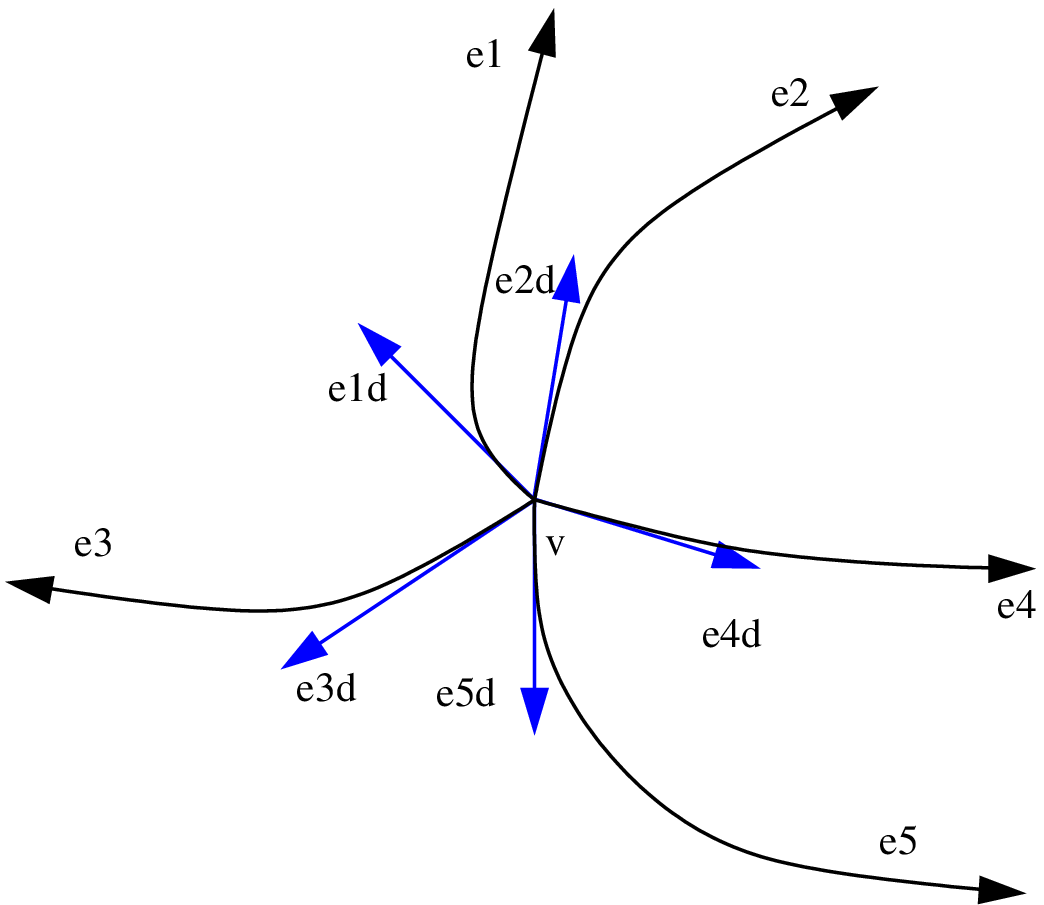}
    \caption{Vertex $v_3$ with identical edge tangent orientation as $v_2$ but with all edges outgoing.}
    \label{vertex3}
   }
\end{figure}

A more general discussion, also including non-uniform oriented matroids (with
coplanar edges), using methods developed in \cite{Finschi2002}, will be
presented in a forthcoming paper \cite{BrunnemannRideout2010}.

\subsection{\label{Gauge Invariance and Circuits on Digraphs}Global Properties: Gauge Invariance and Circuits on Digraphs}

We have seen in section \ref{Oriented Matroids from Oriented Graphs} that connectedness properties of (directed) graphs can be described by (oriented) matroids.
Although originally introduced for planar graphs \cite{White1986}, the definition of graphic oriented matroids 
in terms of its signed circuits depends only on the topology of the graph (adjacency of vertices, orientation of edges), 
and not on any embedding.
Therefore the framework of graphical oriented matroids 
is general enough to be applied to any abstract graph, and in particular to 
the graphs underlying (gauge invariant) spin network functions in LQG\footnote{The possibility of winding numbers between edges certainly needs to be further investigated. See also \cite{Wan2007,Bilson-Thompson2008} and section \ref{Discussion}.}.

Given the fact that the construction of LQG as outlined in section \ref{LQG Combinatorics} is based on partially ordered sets of graphs, it is natural to ask if we can reformulate this construction in terms of oriented matroids. This seems feasible, because it makes sense to speak of the restriction of an (oriented) matroid, if its ground set $E$ is restricted. This can be directly seen from the deletion property in section \ref{Constructions with Oriented Matroids}. Hence, similarly to the partial order among the label set $\MC{L}$ of section \ref{LQG Combinatorics}, the according graphic (oriented) matroids will be equipped with a partial order given by set inclusion of their ground sets. 
This will be further investigated in \cite{Brunnemann2010}. By the definition of the graphic oriented matroid in terms of circuits it seems obvious 
that the oriented matroid formalism should be applied at the gauge invariant level $\Hgauss$. See also section \ref{Discussion} for a discussion.

In order to demonstrate the possibly new and interesting features of such a reformulation, we would like to give the following example on $\Hgauss$:

If the Gauss constraint, and hence gauge invariance, is imposed on spin network functions, then the fundamental group of its underlying graph $\gamma$
becomes 
relevant\footnote{Consider a cylindrical function $f_\gamma(h_{e_1},\ldots,h_{e_N})$ defined over the digraph $\gamma$ with edge set $E(\gamma)=\{e_1,\ldots,e_N\}$ and vertex set $V(\gamma)=\{b(e_K),f(e_K)\}_{e_K\in E(\gamma)}$. In the context of LQG, gauge invariance of $f_\gamma$ denotes invariance under generalized gauge transformations $\ol{\MC{G}}$, which consist of all mappings $g: \Sigma\ni x\mapsto g(x)\in SU(2)$ (no continuity assumption). Applying $g\in \ol{\MC{G}}$, a holonomy $h_{e_K}$ transforms as $h_{e_K}\stackrel{g}{\mapsto} h_{e_K}^{(g)}:=g\big(b(e_K)\big)h_{e_K}g\big(f(e_K)\big)^{-1}$.}.
Taking into account the properties of the measure $\mu_0$ as described in section  \ref{LQG Combinatorics}, a gauge invariant spin network function $f_\gamma$ over $\gamma$ is equivalent to a gauge variant spin network function over another graph $\WT{\gamma}\subset\gamma$, the remainder of $\gamma$ after the retraction to a spanning tree $T_{(\gamma)}$ of edges\footnote{To see this, choose a spanning tree $T_{(\gamma)}=\{e_1,\ldots,e_t\}\subseteq E(\gamma)$ of $\gamma$, with cardinality $t$. Denote $\WT{E}:=E\bs T_{(\gamma)}=\{\WT{e}_1,\ldots,\WT{e}_{N-t}\}$. Consider the cylindrical function  $f_\gamma(h_{\WT{e}_1},\ldots,h_{\WT{e}_{N-t}},h_{e_1},\ldots,h_{e_t})$ as before. Given {\it any} $(h_{\WT{e}_1},\ldots,h_{\WT{e}_{N-t}},h_{e_1},\ldots,h_{e_t})\in SU(2)^N$, we can choose a $g\in\ol{\MC{G}}$, such that $h_{e_K}^{(g)}=\mb{1}_{SU(2)}$ for {\it every} $e_K\in T_{(\gamma)}$. This implies that the gauge invariant information of $f_\gamma$ can be encoded in $f_{\WT\gamma}$, where $\WT{E}=E(\WT\gamma)$.}  
\cite{Freidel2003}.
 This also holds if additionally invariance under graph automorphisms is imposed \cite{Bahr2007}. 
The circuit structure of a directed graph can in turn be used  to define an oriented matroid \cite{Bjorner1999,Ziegler1998}.
By definition, the oriented matroid $\M_\gamma$ constructed from $\gamma$ has the set of all spanning trees of $\gamma$ as its basic family $\MC{B}(\M_\gamma)$.  Then each possible edge set  $E(\WT{\gamma})$ of $\WT{\gamma}$ obtained by removal of a $B\in\MC{B}(\M_\gamma)$ corresponds to an element $B^*$ of the dual basis $\MC{B}^*$ by definition \ref{Dual Oriented Matroid}. Hence we can reformulate the findings of \cite{Freidel2003} by saying that gauge invariant information of $f_\gamma$ can be encoded in cylindrical functions $f_{\WT{\gamma}}$, which have support only on $\MC{B}^*$, the family of bases of the dual oriented matroid $\M^*_\gamma$.

\section{\label{New Volume Data}Spectrum of the Volume Operator Revisited}

In \cite{BrunnemannRideout2008,BrunnemannRideout2008a} an extensive numerical
analysis of the spectrum of the volume operator was presented.  For valences
$N_v = 4-7$, and spins up to some $\jmax(N_v)$, for every chirotope which
arose from the Monte Carlo sprinkling discussed in section
\ref{chirotopes-sec}, we presented graphs and histograms which detail various
aspects of the collection of eigenvalues of the volume operator for vertices
within this range of parameters.

Casting the analysis into the context of oriented matroids now allows us to
better organize the collection of eigenvalues, in particular to take
advantage of the permutation symmetry of the operator, as touched upon in
section \ref{chirotopes-sec}.  Instead of computing the volume spectrum for
all realizable chirotopes, we can restrict attention to the canonical
representatives of each permutation equivalence class.  Since the volume
operator depends upon these chirotopes through the sigma values of
(\ref{sigmas}) (a sigma configuration $\vec{\sigma}$), we compute this for
each canonical representative of a permutation equivalence class.

When counting eigenvalues to form histograms, we follow the same practice as
in \cite{BrunnemannRideout2008,BrunnemannRideout2008a}, regarding each
eigenvalue from each chirotope as distinct.  Thus we associate a redundancy
with each permutation equivalence class of chirotopes, equal to the number of
members in that class.  We then add these redundancies for each chirotope
which yields the same sigma configuration, to get a redundancy for each sigma
configuration.  The numbers of eigenvalues reported in histograms are then
weighted by (half of)\footnote{
This half stems
  back to the overall sign symmetry of chirotopes.  In our code we consider
  only chirotopes which assign +1 to the basis 123.
However it is interesting
  to note that a chirotope $C$ will in general \emph{not} lie in the same
  permutation equivalence class as its negative $-C$.  However we have
  carefully checked that nevertheless the true redundancies as defined here
  are exactly double of those that arise by ignoring all chirotopes with basis
  123 = -1, in spite of the separation of $C$ from $-C$ by permutations.}
these redundancies.

~\\
The analysis in \cite{BrunnemannRideout2008,BrunnemannRideout2008a} considers
only `sorted' spin values $j_1\le\ldots\le j_{N_v}$, because any other
assignment of spins to the edges is related to one of these by a
permutation.  However, this same permutation will also transform the
chirotope $\vec{\epsilon}$ to another $\vec{\epsilon}'$.  Thus the separation
of the eigenvalues into those arising from a given chirotope $\vec{\epsilon}$, while at the
same time considering only sorted spin values $j_1\le\ldots\le j_{N_v}$, is `misleading',
in the sense that the eigenvalues which arise from the same chirotope 
$\vec{\epsilon}$, but with non-sorted spin values $j_1, \ldots, j_{N_v}$, are
found instead under a different chirotope $\vec{\epsilon}\,'$ which is related
to $\vec{\epsilon}$ by a permutation which brings the spins into a sorted ordering $j_1\le\ldots\le j_{N_v}$.

A more `correct' presentation of the eigenvalue data would combine all
eigenvalues which arise from vertices which are related to each other by a
permutation.  We achieve this by working only with the canonical
representative of each permutation equivalence class of the chirotopes, but
allow \emph{all} values of spins on the edges $j_1, \ldots, j_{N_v}$, not
just those which are sorted $j_1\le\ldots\le j_{N_v}$.

~\\
The resulting spectra are presented below. 
Remarkably it appears that the property found in
\cite{BrunnemannRideout2008,BrunnemannRideout2008a}, that different
$\vec{\sigma}$-configurations have a different behavior of the smallest
non-zero eigenvalue (increases, decreases, or remains constant as $\jmax$ is increased) is
preserved among the permutation equivalence classes for the 5-vertex. 
There we find precisely four permutation equivalence classes and the corresponding four different spectral
properties as $\jmax$ is increased: identically zero spectrum,
increasing/decreasing smallest non-zero eigenvalue, and constant smallest
non-zero eigenvalue.  
At the six vertex we also observe all four such behaviors, however at the
7-vertex the increasing smallest eigenvalue sequences seem to
vanish. This may  affect the notion of a semiclassical limit for the volume operator as discussed in section \ref{Discussion}.

~\\
Our improved understanding of the action of permutations on the whole spin
network vertex, including edge spins and orientation of the embedded edge
tangents (as encoded in the chirotope) together, has revealed some errors in
our previous analysis of
\cite{BrunnemannRideout2008,BrunnemannRideout2008a}.

One such error is
in the handling of the overall sign symmetry of chirotopes (that if $\chi_\MC{B}$ is a valid chirotope (equivalently $\vec{\epsilon}$ a valid sign configuration), then so $-\chi_\MC{B}$ (equivalently $-\vec{\epsilon}$).
There we made a
mistake in mapping this symmetry to the sigma configurations, and thus some
of the sigma configurations we used in our analysis were incorrect.  This
affects some of the counting of sigma configurations which yield the
different behaviors of the smallest non-zero eigenvalues, and some of the
eigenvalues themselves.  However none of the qualitative results are
affected.

We were also able to utilize the permutation symmetry of the volume operator
as a cross check on our code, which at the level of sigma configurations is
extremely subtle.  In using this we
discovered a bug which affected all our results on the 7-vertex.  The
corrected eigenvalues are shown in section \ref{7v-sec}.

In addition there was a bug in our code with respect to the handling of the
sigma configuration for the cubic 6-vertex.  Corrected eigenvalues are shown
in section \ref{cubic-sec}.

~\\
The results from the computations shown below come from running a modified
version of the code developed for references
\cite{BrunnemannRideout2008,BrunnemannRideout2008a}, which is written as a
`thorn' within the Cactus high performance computing framework \cite{cactus},
on the `whale' cluster of the SHARCNET grid in Southern Ontario.  The
computations for the 7-vertex ran efficiently on 225 cores.

\subsection{Gauge Invariant 5-Vertex}

In figure \ref{5v_sci_hist} we show histograms of the three non-trivial
permutation equivalence classes of the 5-vertex, for $\jmax$ up to $\frac{25}{2}$.  Each
is distinct yet shows similar behavior.  
Note the small lip at zero for the purple $\vec{\sigma} = (2, 2, 4, 0)$
histogram.  This results from a non-negligible fraction of the eigenvalues
descending toward zero at larger spins, c.f.\ figure \ref{5v_min-fig}.  All
histograms are drawn as described in \cite{BrunnemannRideout2008a}.
\begin{figure}[htbp]
\center
\psfrag{frequency}{\#eigenvalues}
\psfrag{eigenvalue}{volume in $\ell_{\text{Planck}}^3$}
\includegraphics{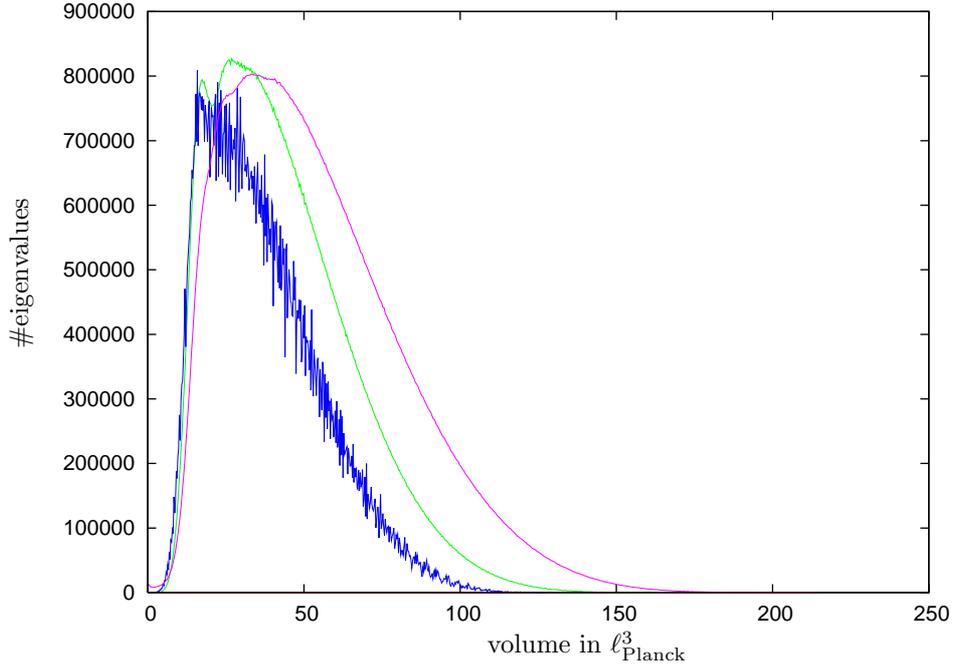}
\caption{Histograms for each sigma configuration $\vec{\sigma}$ at the
  5-vertex, up to $\jmax=25/2$.  From bottom to top at volume 100, the blue is for $\vec{\sigma} =
  (\sigma_{123},\sigma_{124},\sigma_{134},\sigma_{234}) = (2, 0, 0, 0)$, the
  green for $\vec{\sigma} = (2, 2, 2, 0)$, and the purple for $\vec{\sigma} =
  (2, 2, 4, 0)$.  Each histogram has 512 bins.}
\label{5v_sci_hist}
\end{figure}

Figure \ref{5v_jmax_hist} depicts overall histograms for the 5-vertex, at
various values of $\jmax$ up to 25/2.  Note that the lip at zero is not
really visible at this scale.
\begin{figure}[htbp]
\center
\psfrag{frequency}{\#eigenvalues}
\psfrag{eigenvalue}{volume in $\ell_{\text{Planck}}^3$}
\includegraphics{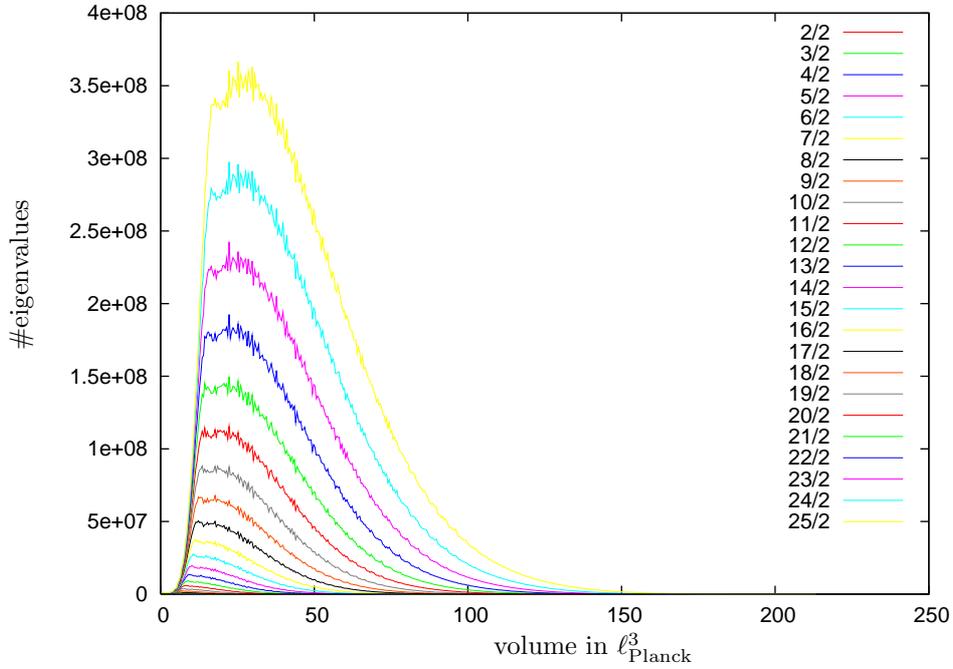}
\caption{Histograms for the overall generic 5-vertex, up to every $\jmax \leq
  25/2$.  (By `generic' we mean excluding co-planar edges, which in a sense
  form a set of measure zero.)  Each histogram has 512 bins.}
\label{5v_jmax_hist}
\end{figure}
Figure \ref{5v_jmax_hist-zoom} zooms in on the small eigenvalue region.
There we can discern the lip just beginning to show at $\jmax=11$.
\begin{figure}[htbp]
\center
\psfrag{frequency}{\#eigenvalues}
\psfrag{eigenvalue}{volume in $\ell_{\text{Planck}}^3$}
\includegraphics{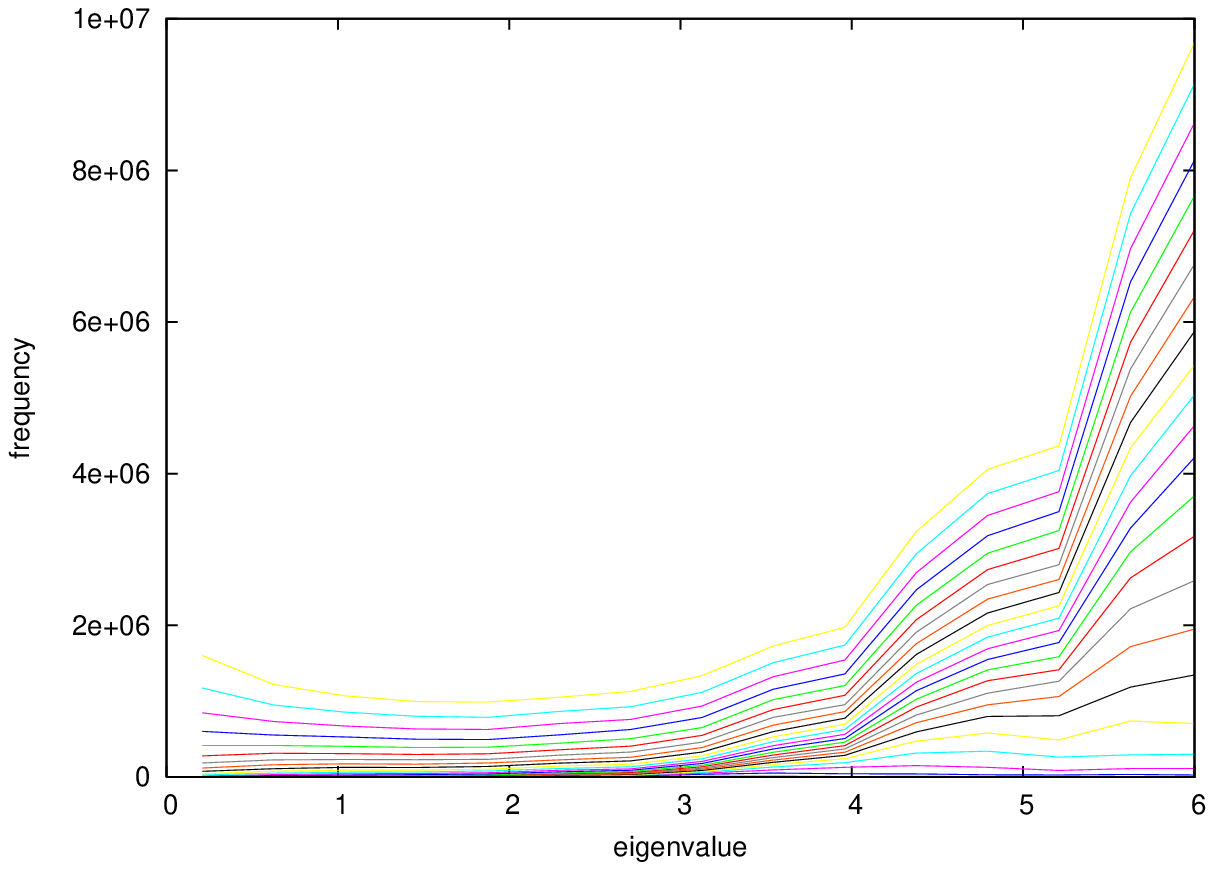}
\caption{Histograms for the overall 5-vertex, up to every $\jmax \leq 25/2$,
  zoomed to the small eigenvalue region.}
\label{5v_jmax_hist-zoom}
\end{figure}

Figures \ref{5v_min-fig} and \ref{5v_max-fig} show the smallest and largest
non-zero eigenvalues of the 5-vertex, for each $\vec{\sigma}$ and $\jmax$.
For these $\vec{\sigma}$-index 0 corresponds to $\vec{\sigma} = (2, 2, 2,
0)$, 1 to $\vec{\sigma} = (2, 0, 0, 0)$, and 2 to $\vec{\sigma} = (2, 2, 4,
0)$.  Here we see the three behaviors of sigma configurations with respect
to their smallest non-zero eigenvalues, as mentioned in section
\ref{New Volume Data}, with $\vec{\sigma}$-index 0 increasing, 1 constant,
and 2 decreasing with $\jmax$.  The largest eigenvalues behave similarly for
each non-trivial $\vec{\sigma}$.\footnote{There is a fourth $\vec{\sigma}$ at
  the 5-vertex, which is all zeros, and hence leads to only zero
  eigenvalues.  There will be such a $\vec{\sigma}$ for every valence.  Zero eigenvalues are suppressed on all plots.}
\begin{figure}[htbp]
 \begin{minipage}[t]{9cm}
    \psfrag{min_eval}{$\lambda_{\text{min}}$}
    \psfrag{jmax}{$2\,\jmax$}
    \psfrag{sci}{$\vec{\sigma}$-index}
    \includegraphics[width=9cm]{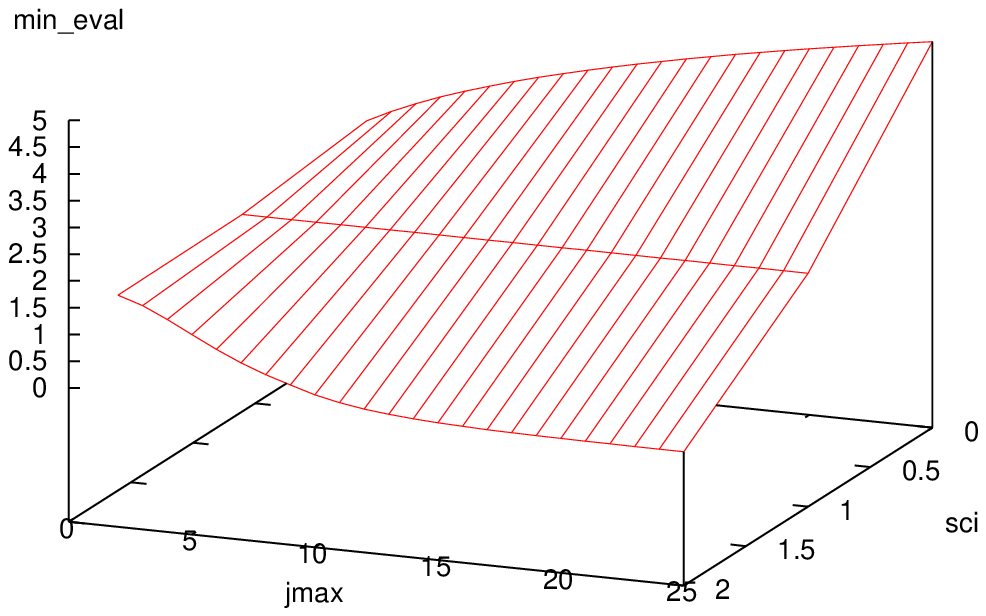}
    \caption{ \label{5v_min-fig}
      Smallest non-zero eigenvalues $\lambda_{\text{min}}$ at the 5-vertex.}
\end{minipage}
~~~~
 \begin{minipage}[t]{9cm}
    \psfrag{max_eval}{$\lambda_{\text{max}}$}
    \psfrag{jmax}{$2\,\jmax$}
    \psfrag{sci}{$\vec{\sigma}$-index}
    \includegraphics[width=9cm]{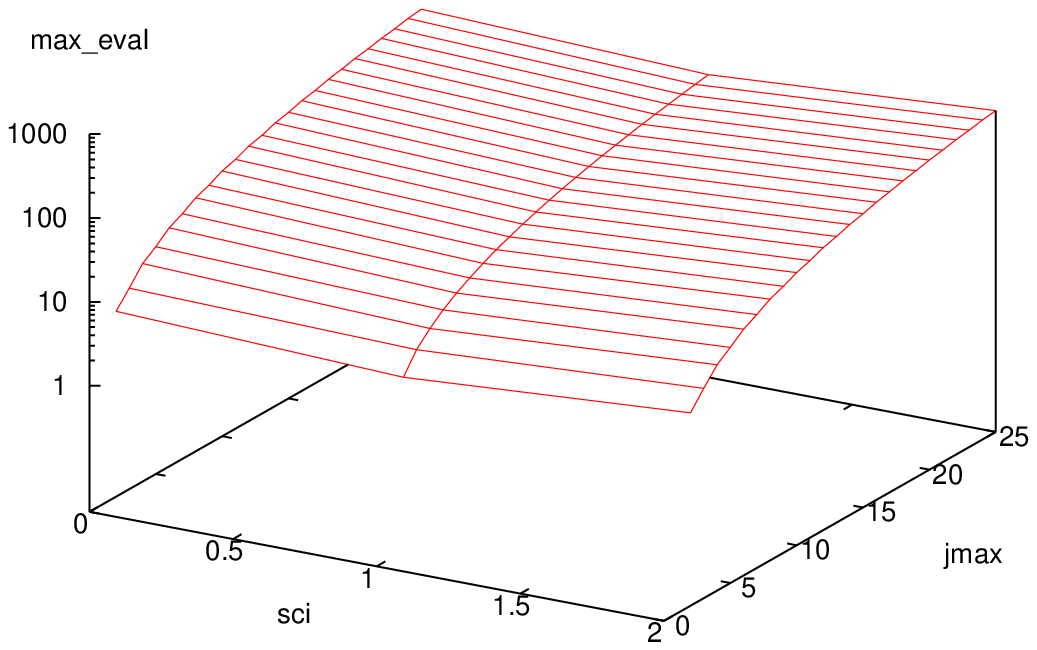}
    \caption{ \label{5v_max-fig}
      Largest eigenvalues $\lambda_{\text{max}}$ of the 5-vertex.}
 \end{minipage}
\end{figure}

\subsection{Gauge Invariant 6-Vertex}

Histograms for the 6-vertex are displayed in figures \ref{6v_sci_hist},
\ref{6v_sci_hist-zoom}, and \ref{6v_jmax_hist}, for $\jmax$ up to 13/2.  Here
there are 39 distinct sigma configurations.  
The beginnings of a lip at zero in the spectrum are just visible for several
sigma configurations $\vec{\sigma}$, as seen in figure
\ref{6v_sci_hist-zoom}.

\begin{figure}[htbp]
\center
\psfrag{frequency}{\#eigenvalues}
\psfrag{eigenvalue}{volume in $\ell_{\text{Planck}}^3$}
\includegraphics{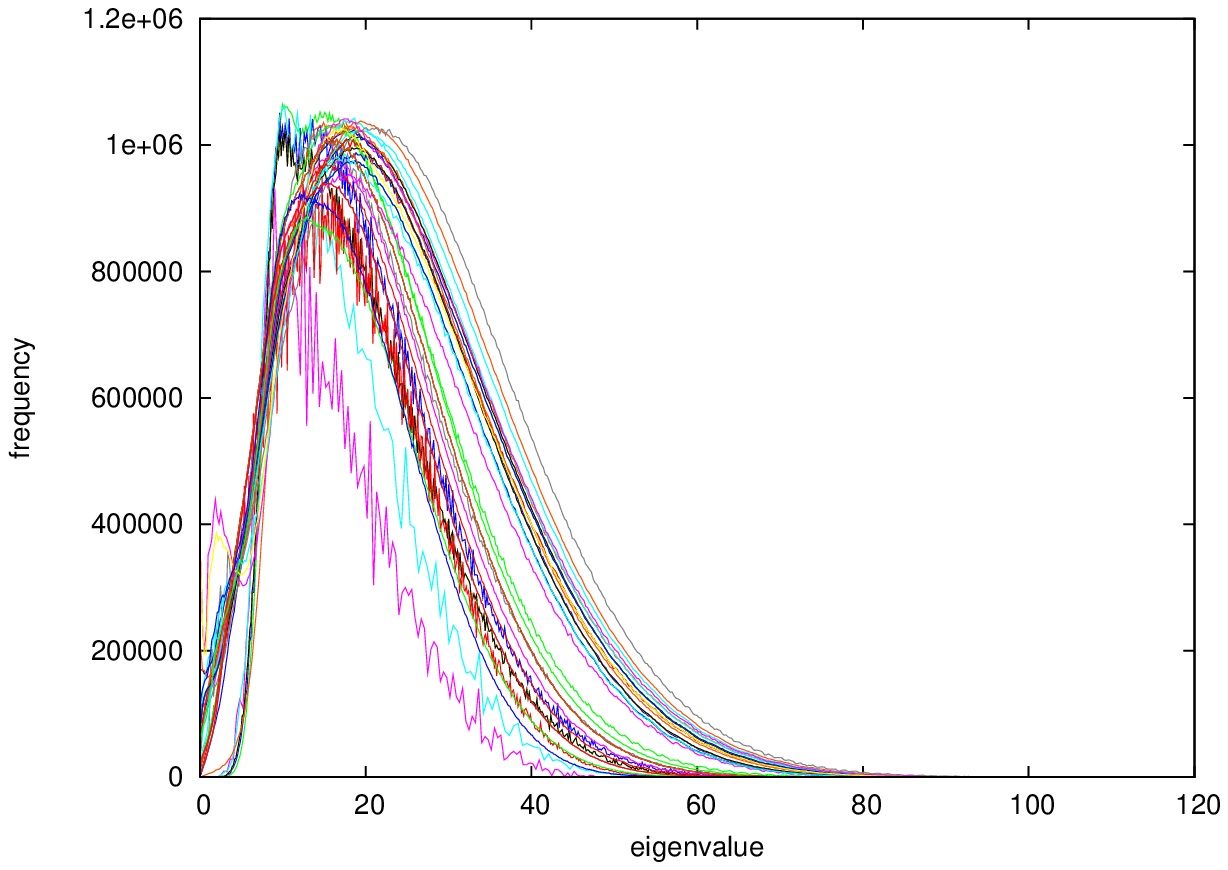}
\caption{Histograms for each sigma configuration $\vec{\sigma}$ at the
  6-vertex, up to $\jmax=13/2$.  Two histograms have 128 bins, all others
  have 512.}
\label{6v_sci_hist}
\end{figure}

\begin{figure}[htbp]
\center
\psfrag{frequency}{\#eigenvalues}
\psfrag{eigenvalue}{volume in $\ell_{\text{Planck}}^3$}
\includegraphics{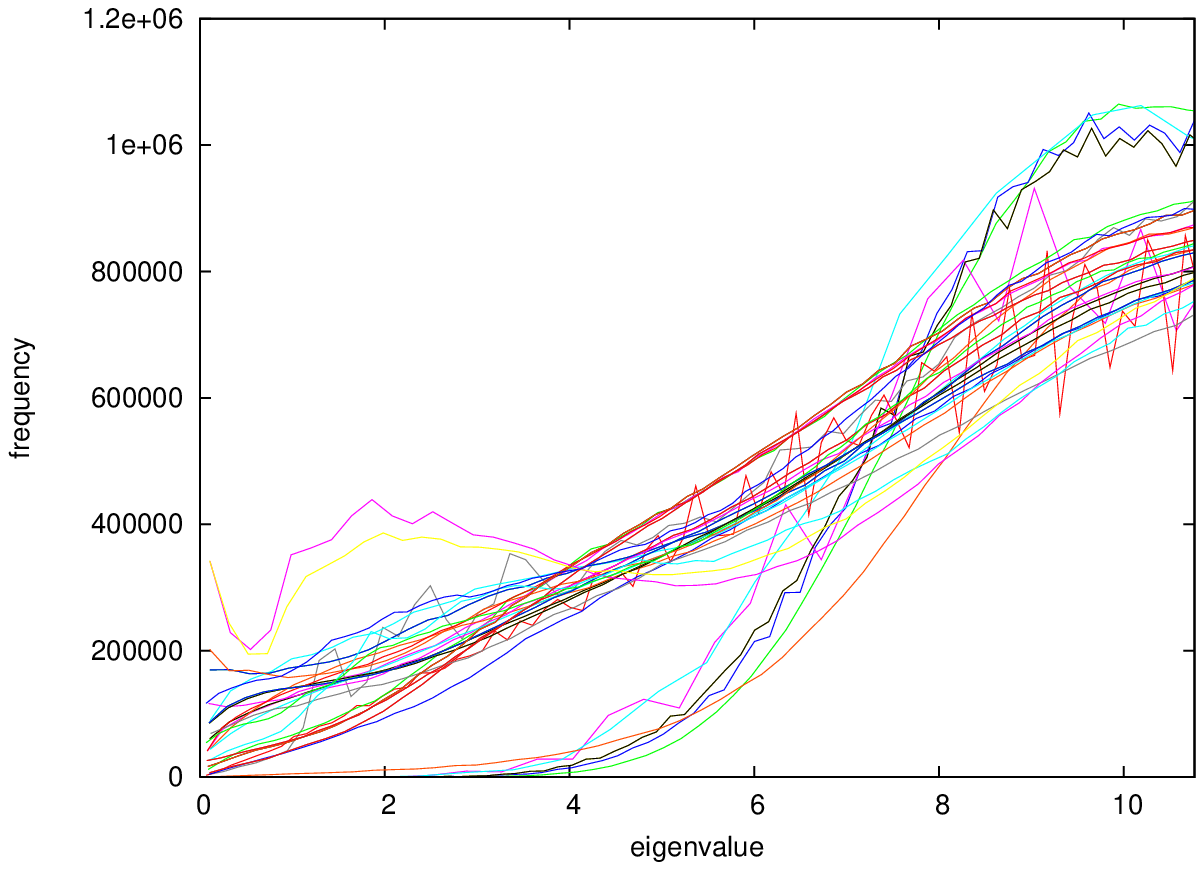}
\caption{Histograms for each sigma configuration $\vec{\sigma}$ at the
  6-vertex, up to $\jmax=13/2$, zoomed to the small eigenvalue region.}
\label{6v_sci_hist-zoom}
\end{figure}

\begin{figure}[htbp]
\center
\psfrag{frequency}{\#eigenvalues}
\psfrag{eigenvalue}{volume in $\ell_{\text{Planck}}^3$}
\includegraphics{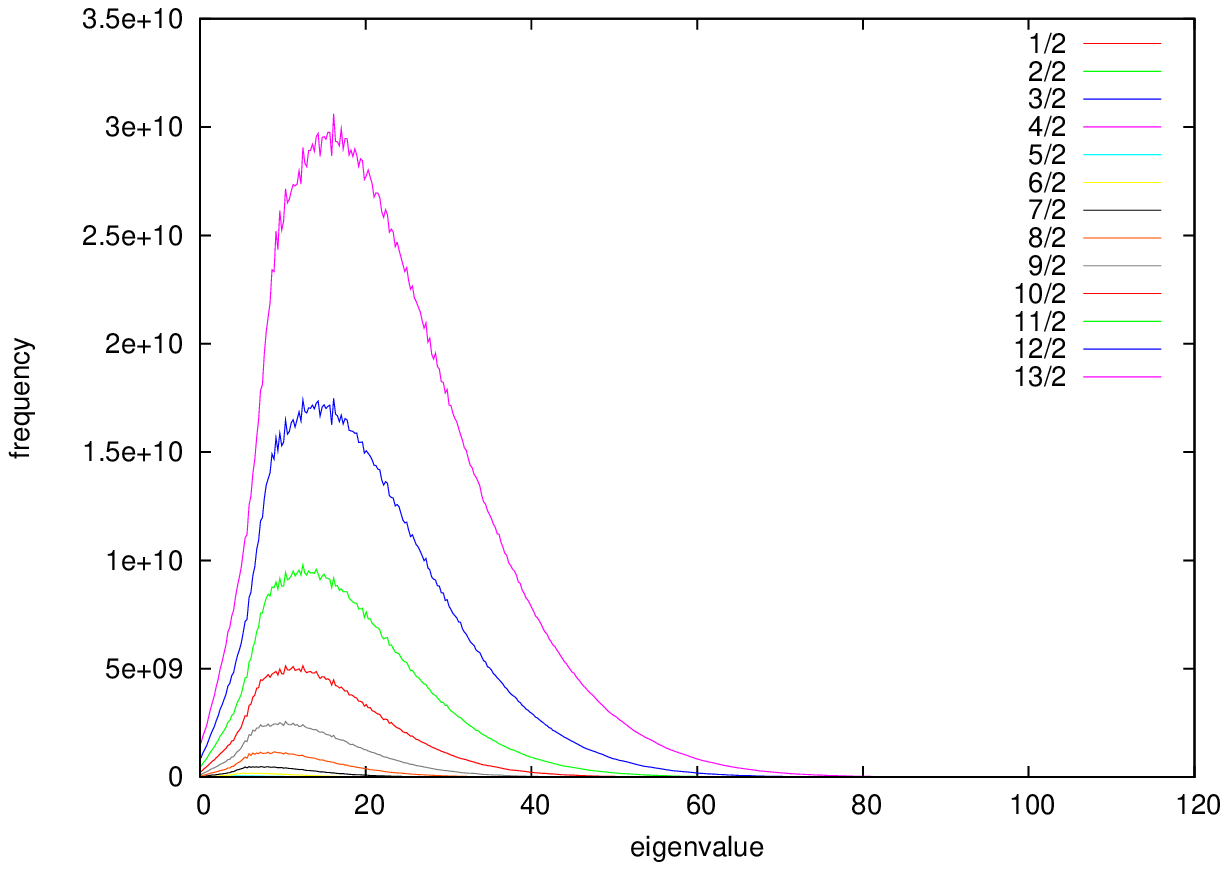}
\caption{Histograms for the overall (generic) 6-vertex, up to every $\jmax \leq 13/2$.  All have 512 bins.}
\label{6v_jmax_hist}
\end{figure}

Figures \ref{6v_min-fig} and \ref{6v_max-fig} depict the smallest and largest
non-zero eigenvalues for each $\vec{\sigma}$, as a function of maximum spin.
Most of the sigma configurations yield smallest non-zero eigenvalues which
decrease with $\jmax$, however a number of them remain constant (such as the rather degenerate $\vec{\sigma} = (2 0 0 0 0 0 0 0 0 0)$, in which only four of the six spins play any role beyond their effect in determining the recoupling basis), and one
($\vec{\sigma} = (0 2 2 2 2 2 2 2 2 0)$)
leads to increasing smallest non-zero eigenvalues.  It can also be noted, in
particular in figure \ref{6v_min-fig}, that in general there are several
sigma configurations which yield the same exact sequence of smallest non-zero
eigenvalues.
The largest eigenvalues all appear to increase by 
a power law, as expected from (\ref{definition qIJK}).
\begin{figure}[htbp]
 \begin{minipage}[t]{8.5cm}
    \psfrag{min_eval}{$\lambda_{\text{min}}$}
    \psfrag{jmax}{$2\,\jmax$}
    \psfrag{sci}{$\vec{\sigma}$-index}
    \includegraphics[width=9cm]{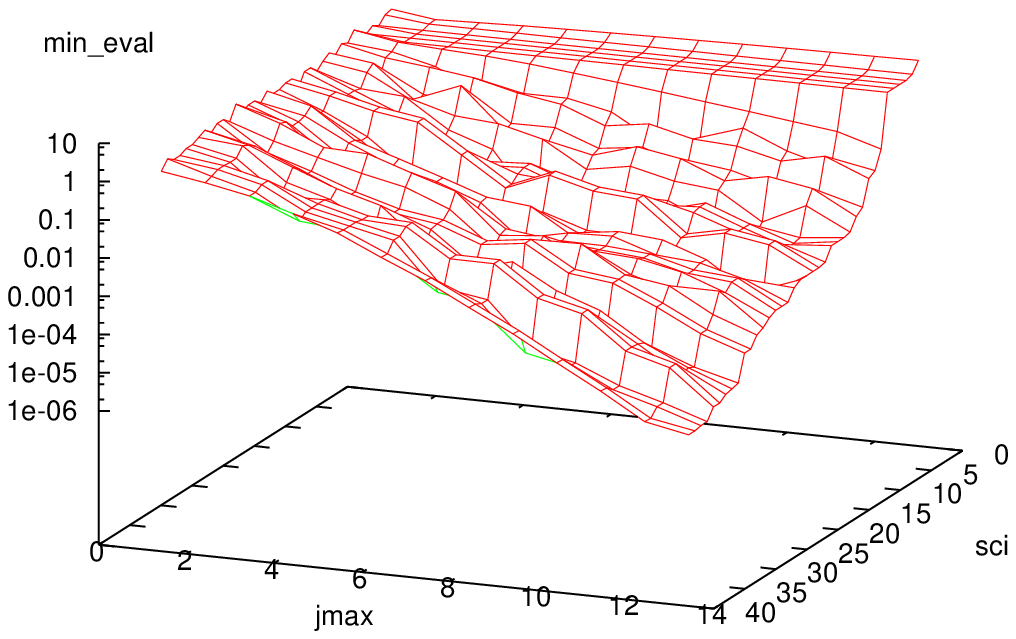}
    \caption{ \label{6v_min-fig}
      Smallest non-zero eigenvalues $\lambda_{\text{min}}$ at the 6-vertex.}
\end{minipage}
~~~~
 \begin{minipage}[t]{8.5cm}
    \psfrag{max_eval}{$\lambda_{\text{max}}$}
    \psfrag{jmax}{$2\,\jmax$}
    \psfrag{sci}{$\vec{\sigma}$-index}
    \includegraphics[width=9cm]{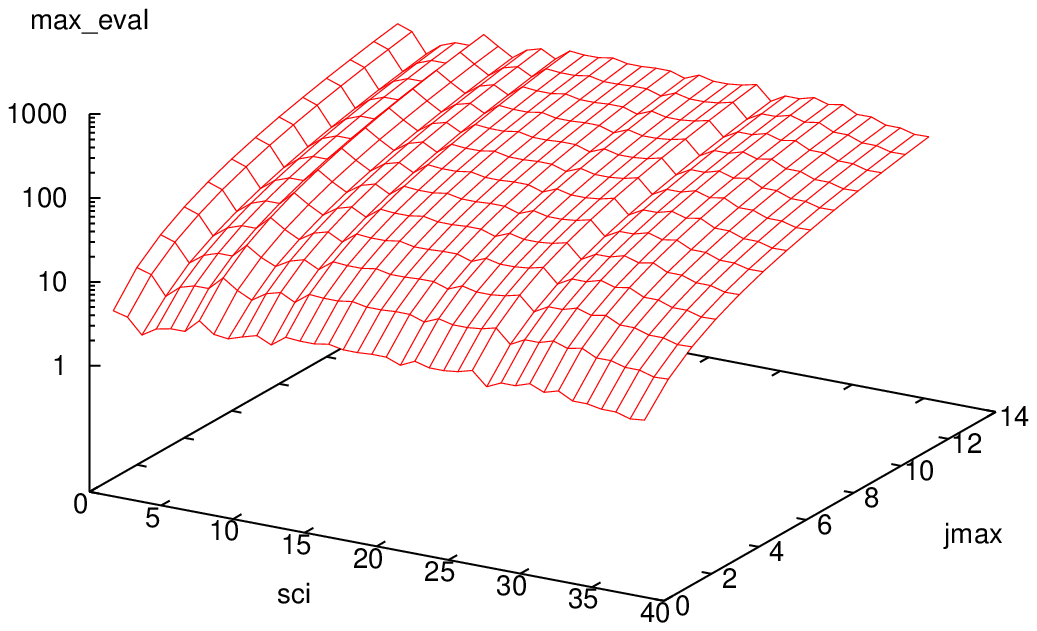}
    \caption{ \label{6v_max-fig}
      Largest eigenvalues $\lambda_{\text{max}}$ of the  6-vertex.}
 \end{minipage}
\end{figure}

\subsection{Cubic 6-Vertex}
\label{cubic-sec}

The cubic 6-vertex arises from a `tilation' of $\mb{R}^3$ by cubes.  I.e.\ it
is the network of lines formed by restricting the Cartesian coordinates $x, y
\in \mb{Z}$, with $z\in \mb{R}$, and likewise for $x, z \in \mb{Z}$, $y\in
\mb{R}$, and $y, z \in \mb{Z}$, $x\in \mb{R}$.  The edge tangents at such a
vertex are of course coplanar --- this is the one exception to our rule 
of excluding coplanar edges.  The resulting chirotope and sigma configuration
are detailed in \cite{BrunnemannRideout2008a}.

In figure \ref{cubic_hist-fig} we present histograms for the volume
eigenvalues of the cubic 6-vertex, for maximam spins $\jmax$ up to 5.
\begin{figure}[htbp]
\center
\psfrag{frequency}{\#eigenvalues}
\psfrag{eigenvalue}{volume in $\ell_{\text{Planck}}^3$}
\includegraphics{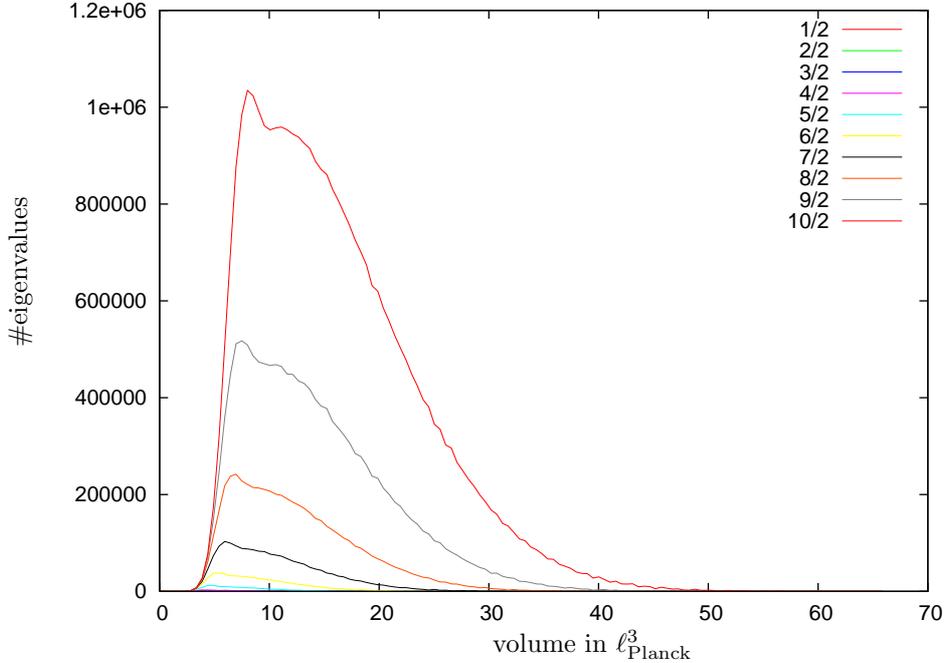}
\caption{Histograms of volume eigenvalues for the cubic 6-vertex, one for
  each value of $\jmax$ up to 5.}
\label{cubic_hist-fig}
\end{figure}
Figure \ref{cubic_minmax-fig} displays the minimum and maximum non-zero
eigenvalues, as a function of $\jmax$.  One can see that the cubic 6-vertex
is one of the rare ones for which the smallest non-zero eigenvalue increases
with spin.
\begin{figure}[htbp]
\center
    \psfrag{min_eval}{$\lambda_{\text{min}}$}
    \psfrag{max_eval}{$\lambda_{\text{max}}$}
    \psfrag{max_spin}{$2\,\jmax$}
    \psfrag{eval}{extremal eigenvalue}
    \includegraphics{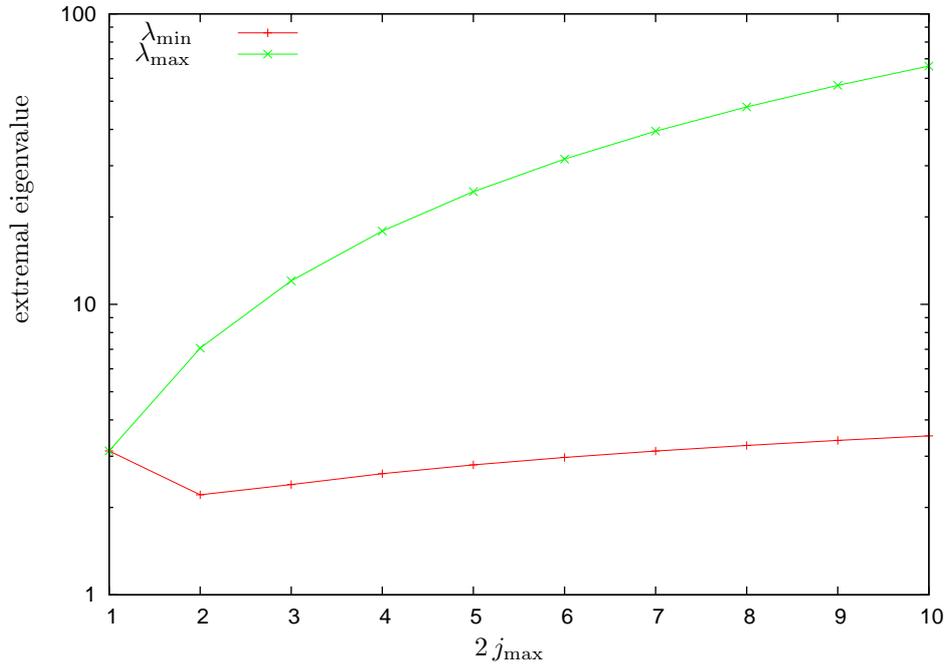}
    \caption{Smallest and largest non-zero volume eigenvalues of the cubic
      6-vertex, as a function of $\jmax \leq 5$.}
    \label{cubic_minmax-fig}
\end{figure}

\subsection{Gauge Invariant 7-Vertex}
\label{7v-sec}

Histograms for the gauge invariant 7-vertex are shown in figures
\ref{7v_sci_hist}, \ref{7v_sci_hist-zoom}, and \ref{7v_jmax_hist}, for $\jmax$ up to 7/2.  There are 673
permutation equivalence classes of sigma configurations for the 7-vertex.
Here there seem to be roughly two sorts of sigma configurations --- those
that lead to extremely jagged histograms in figure \ref{7v_sci_hist} (we
believe that these come from degenerate $\vec\sigma$, which have many zeros,
and thus can be effectively of lower valence), and the
others which all have more-or-less the same smooth shape.  $\jmax=7/2$ is 
too small to see much evidence for a lip in the spectrum near zero volume,
however there seems to be one sigma configuration which is already heading in
this direction in figure \ref{7v_sci_hist-zoom}.

\begin{figure}[htbp]
\center
\psfrag{frequency}{\#eigenvalues}
\psfrag{eigenvalue}{volume in $\ell_{\text{Planck}}^3$}
\includegraphics{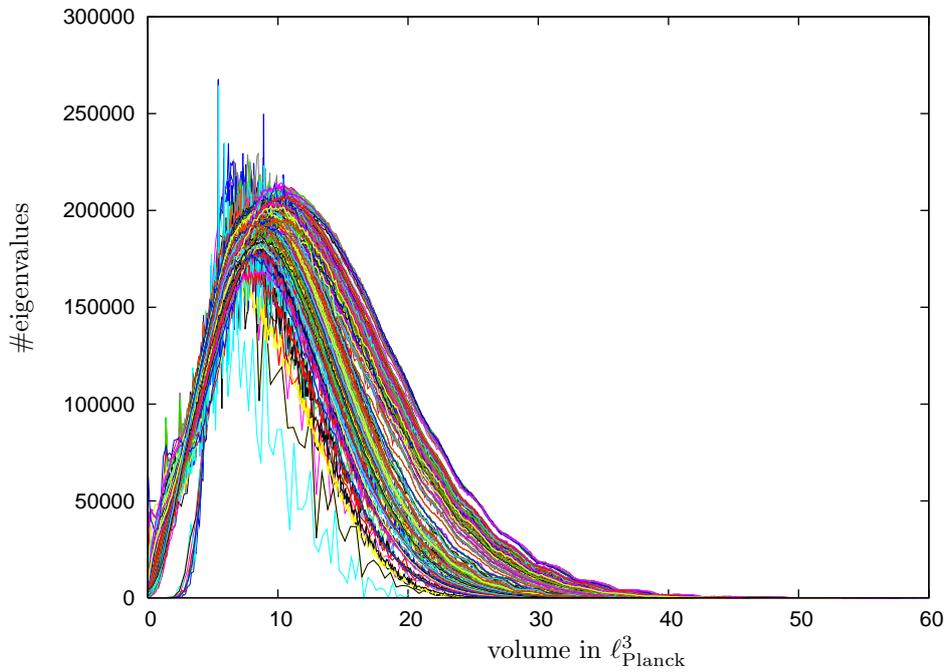}
\caption{Histograms for each sigma configuration $\vec{\sigma}$ at the
  7-vertex, up to $\jmax=7/2$.}
\label{7v_sci_hist}
\end{figure}

\begin{figure}[htbp]
\center
\psfrag{frequency}{\#eigenvalues}
\psfrag{eigenvalue}{volume in $\ell_{\text{Planck}}^3$}
\includegraphics{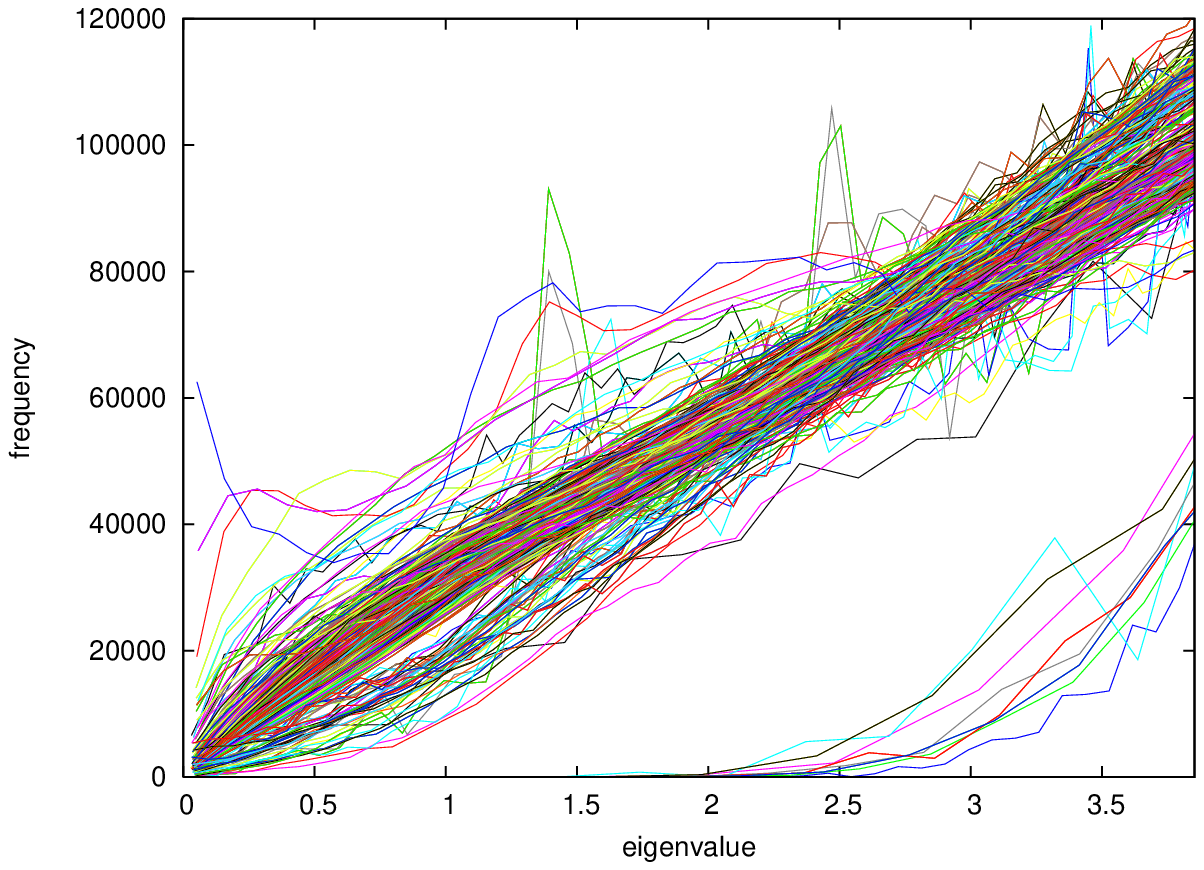}
\caption{Histograms for each sigma configuration $\vec{\sigma}$ at the
  7-vertex, up to $\jmax=7/2$, zoomed to the small eigenvalue region.}
\label{7v_sci_hist-zoom}
\end{figure}

\begin{figure}[htbp]
\center
\psfrag{frequency}{\#eigenvalues}
\psfrag{eigenvalue}{volume in $\ell_{\text{Planck}}^3$}
\includegraphics{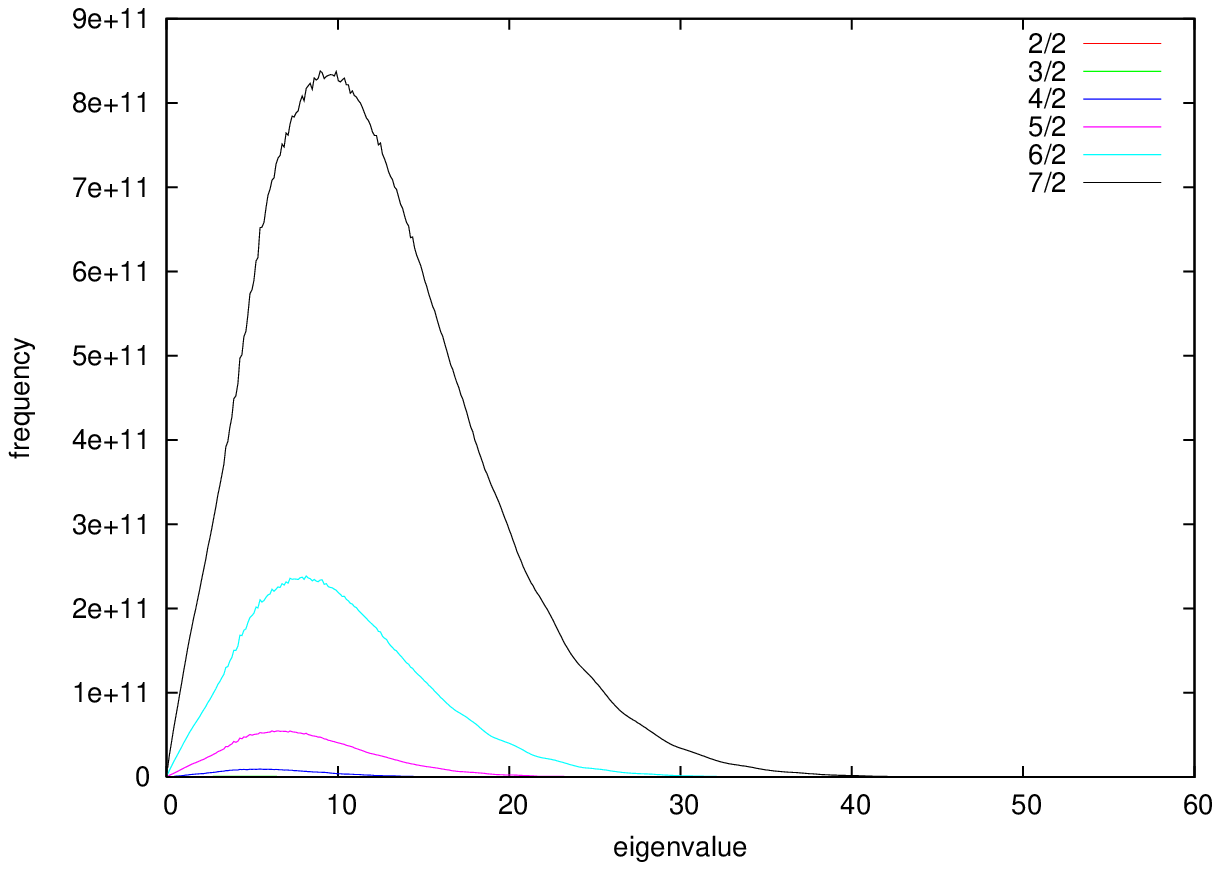}
\caption{Histograms for the overall (generic) 7-vertex, up to every $\jmax \leq 7/2$.}
\label{7v_jmax_hist}
\end{figure}

Figures \ref{7v_min_evals-fig} and \ref{7v_max_evals-fig} show the extremal
eigenvalues for each $\jmax$ and $\vec\sigma$ at the 7-vertex.  Here there
are no $\vec\sigma$ for which the smallest non-zero eigenvalue increases with
$\jmax$, though there are several for which this smallest eigenvalue is
constant, such as $\vec\sigma = (2 2 0 0 2 0 0 0 0 0 2 0 0 0 0 0 0 0 0 0)$,
which is one of the $\vec\sigma$ which yields the largest smallest non-zero
eigenvalues, and the `4-vertex-like' $\vec\sigma = (2 0 0 0 0 0 0 0 0 0 0 0 0
0 0 0 0 0 0 0)$.  $\vec\sigma = (0 0 0 0 0 0 0 0 0 2 2 0 0 0 0 0 0 0 0 0)$ is
an example of a 7-vertex sigma configuration with decreasing smallest
non-zero eigenvalue.  The largest eigenvalues behave as usual.

\begin{figure}[htbp]
 \begin{minipage}[t]{8.5cm}
    \psfrag{min_eval}{$\lambda_{\text{min}}$}
    \psfrag{jmax}{$2\,\jmax$}
    \psfrag{sci}{$\vec{\sigma}$-index}
    \includegraphics[width=9cm]{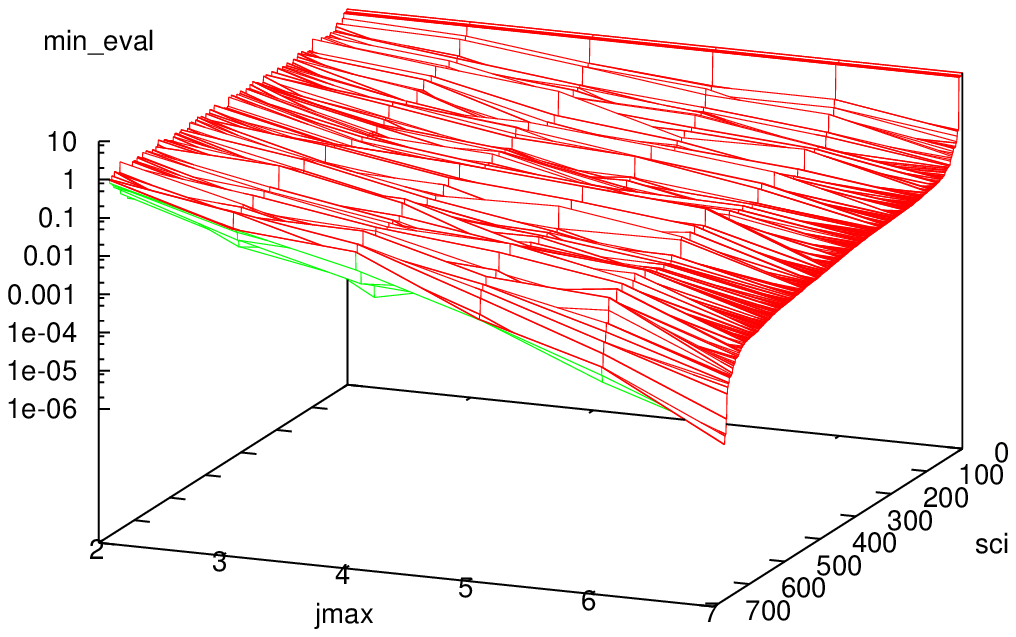}
    \caption{\label{7v_min_evals-fig} Smallest non-zero eigenvalues $\lambda_{\text{min}}$ at 7-vertex.}
\end{minipage}
~~~~
 \begin{minipage}[t]{8.5cm}
    \psfrag{max_eval}{$\lambda_{\text{max}}$}
    \psfrag{jmax}{$2\,\jmax$}
    \psfrag{sci}{$\vec{\sigma}$-index}
    \includegraphics[width=9cm]{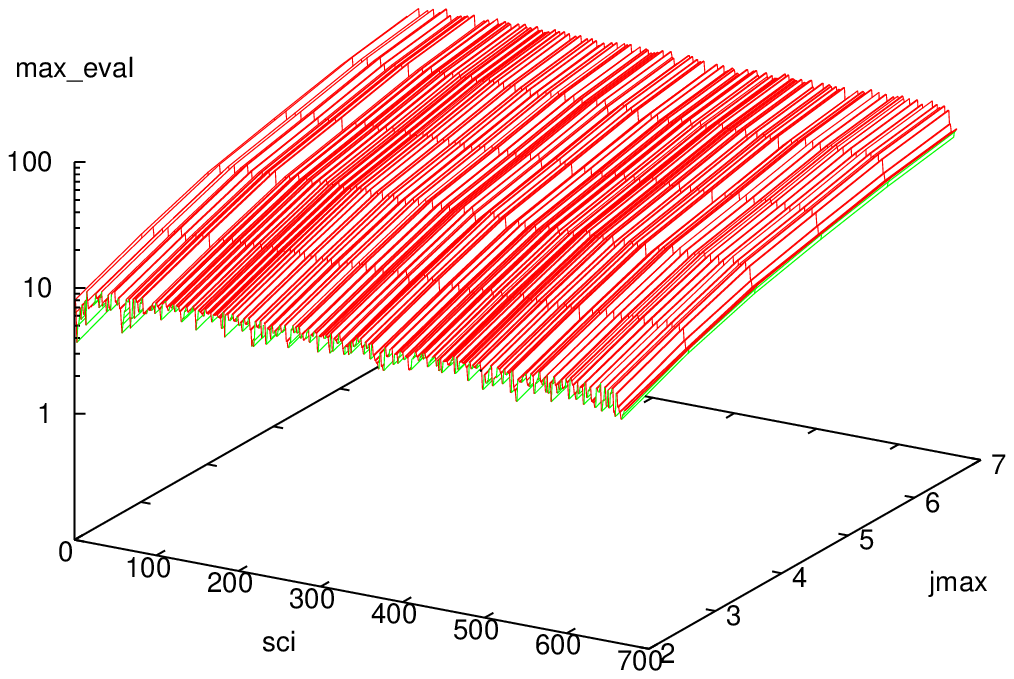}
    \caption{\label{7v_max_evals-fig} Largest eigenvalues $\lambda_{\text{max}}$ at 7-vertex.}
 \end{minipage}
\end{figure}

It is interesting to note that the sigma configuration $\vec\sigma$ in which
the only non-zero sigma is $\sigma(1,2,3) = +2$ appears to arise at every
valence, as the sigma configuration corresponding to the chirotope
$\epsilon(1,2,3)=+1$, $\epsilon(I,J,K)=-1$ for all other triples $I,J,K$.
This chirotope appears as the canonical representative of an equivalence
class at every valence, because it corresponds to the integer 1 in our
numbering scheme (and presumably because chirotopes `0' and `1' cannot be
transformed into each other by a permutation, at any valence).  It leads to a
smallest non-zero eigenvalue of $\sqrt[4]{12}$ for every $\jmax$.

\section{\label{Discussion}Discussion}

As we have demonstrated, the framework of oriented matroids can capture both local embedding properties of graphs as well as their global connectedness properties. 

Originally developed for describing planar graphs, the framework of oriented matroids is general enough to be applied to graphs embedded in three dimensional Riemannian space, as is the case for LQG. Although the example from \cite{Bjorner1999}, revisited in detail in section \ref{Oriented Matroids from Oriented Graphs}, corresponds to a planar graph and uses a global orientation, the computation of the set $\C$ of signed circuits of the according oriented matroid does not rely on this fact. 
Because the set $\C$ of signed circuits as well as the chirotope $\chi_{\MC{B}}$ are symmetric with respect to a global sign reversal\footnote{That is if $\chi_{\MC{B}}$ is a chirotope, then so is $-\chi_{\MC{B}}$. Also $\pm C\in\C$ for all signed circuits $C$. Similar statements hold for the dual oriented matroid.}, one can just take any vertex $v$ contained in a circuit $\ul{C}$ as a starting point and then follow $\ul{C}$ along 
either direction until one comes back to $v$. The relative orientation of edges to the chosen direction then determines $C^\pm$.

In the outlined treatment, we certainly neglect the global knotting of edges (e.g.\ one edge could wind around another) of graphs embedded in three dimensional space \cite{Wan2007}. If one wants to include these properties, one has to think about a possible extension of the oriented matroid framework for embedded digraphs. On the other hand, the knotting properties are not seen by any quantum operator corresponding to classical geometric quantities, unless they are interpreted as e.g.\ matter excitations as suggested e.g.\ in \cite{Wan2007,Bilson-Thompson2008}. Note that knotting also becomes irrelevant if automorphisms of graphs are considered \cite{Bahr2007}. 
Therefore we prefer to postpone considerations of knotting of edges until the connection between oriented matroids and the framework of LQG is worked out in more detail.
 
Regarding the local embedding of graph vertices, the occurrence of so-called moduli parameters was discussed in \cite{Fairbairn2004}, which seems to prevent $\Hdiff$ from having a countable label set. However the solution suggested there erases all local geometric information of the vertex embedding, encoded in sign factors.
In contrast, we think that this information is physically relevant and should be kept, as it is crucial for the implementation of the Hamilton constraint operator \cite{Thiemann1996} and a consistent formulation of LQG \cite{Giesel2006a}. A solution to this issue is suggested in \cite{Thiemann2005}, giving $\Hdiff$ a countable basis, while keeping the information on the local embedding.

The level at which the graphic oriented matroids can be introduced into the LQG formalism has to be further analyzed. 
At the level of $\Ho$, one can have graphs underlying spin network functions, which have ``open ends'', that is edges which have a 1-valent vertex as their beginning or final point. This difficulty could be dealt with by treating such ``open ends'' as effective re-tracings, however as one finally has to solve the Gauss constraint, one can directly start at the level of gauge invariant spin network functions in $\Hgauss$, because there every edge of a graph has at least a two-valent vertex as beginning and final point. 
From the combinatorics side introducing graphic oriented matroids directly at the level of $\Hdiff$ seems to be the most natural approach.

We have seen in section \ref{New Volume Data} that the spectral properties of the volume operator are closely related to the oriented matroid resulting from the local embedding of the vertices of a graph. Interestingly, it is not only sensitive to those `local embedding chirotopes',
but also to the supports of the edges at the graph, as illustrated in figures \ref{vertex1}--\ref{vertex3}. Rephrasing this, a reorientation in the graphic matroid (corresponding to re-directing an edge in the graph) causes a reorientation in the local vector oriented matroid. Hence, global and local graph properties are related in this way.

~\\
The different possibilities of embedding a vertex cause different behaviors of the lowest non-zero eigenvalues of the volume spectrum as the maximum spin $\jmax$ at a vertex is increased. We find, in agreement with \cite{BrunnemannRideout2008,BrunnemannRideout2008a}, increasing, constant, and decreasing smallest non-zero eigenvalue sequences. 
In the latter two cases even very large spins can contribute microscopically small non-zero eigenvalues, and hence one cannot say that the limit $\jmax\rightarrow\infty$ produces only large volume eigenvalues. This is different from the area operator \cite{Ashtekar1997}, for which large spins imply large eigenvalues. If one wants to keep this property one might regard this as a ``physical'' preference for vertices whose smallest non-zero eigenvalue grows with $\jmax$, such as the 4-vertex, cubic 6-vertex, or the (permutation equivalence class of) 6-vertex with $\vec{\sigma} = (0 2 2 2 2 2 2 2 2 0)$.

~\\

As a consequence we would like to 
point out that high valent vertices do not necessarily overcount the volume in a semiclassical analysis as presented in \cite{Flori2008}. That is, high valence does not automatically shift the volume spectrum towards larger eigenvalues, as one might naively expect from the sum structure in (\ref{Volume definition gauge invariant 3}). This is due to, for example,  the presence of vertex embeddings which only have a small number of non-zero signs $\sigma(I,J,K)$, as defined in (\ref{sigmas}). In particular the number of triples $(I,J,K)$ with $\sigma(I,J,K)\ne 0$ can be independent of the number of edges attached to the vertex. One example for such a vertex  embedding is given by $\vec{\sigma}=(2,0,0,\ldots,0)$, as noted in section \ref{7v-sec}, which we find to be contained in the permutation equivalence classes of 5- up to 7-valent vertices.
Such sigma configurations seem able to lead to smallest non-zero eigenvalues
which are independent of both $\jmax$ \emph{and valence} $N_v$.

Given this fact, it will be instructive to analyze if one can find embedded higher valent vertices which also resemble the correct semiclassical limit in the sense of \cite{Flori2008}.
Approximating the volume of a spatial region by semiclassical coherent state techniques introduced in \cite{Thiemann2001a} might be a subtle mixture of the embedding of the graph supporting the coherent state, as well as choosing its topology, in particular the number of vertices and their connectedness in that region.  Here the unified description of these properties in terms of oriented matroids might give us a better understanding of the possible choices one has to make in the construction of the complexifier coherent states used in \cite{Flori2008}.

\section{\label{Summary and Outlook}Summary and Outlook}

In this work we have
demonstrated a connection 
between the LQG framework and the field of oriented matroids. For this we have described the combinatorial properties of LQG in section \ref{LQG Combinatorics}, and introduced matroids and oriented matroids in section \ref{Matroids and Oriented Matroids} in an abstract way, without referring to any particular realization. 
In sections \ref{Oriented Matroids from Oriented Graphs} and \ref{Oriented Matroids from Vector Configurations} we showed how this abstract framework can be applied in order to describe global (connectedness) and local (geometry of vertex embedding) graph properties by oriented matroids. 
We have also briefly discussed the issue of realizability of oriented matroids in terms of pseudosphere arrangements in section \ref{Realizability of Oriented Matroids}, that is the question of when an abstract oriented matroid 
of rank $r$ gives rise to a configuration of vectors in $\mb{R}^r$.
The obvious connection of oriented matroids to LQG is then discussed in section \ref{Oriented Matroids in Loop Quantum Gravity}.

We see several possible benefits in this approach.
First, it provides {\it one unified  framework}  to describe local and global graph combinatorics, 
in an abstract way. That is, in this setup we do not 
need to refer to the underlying manifold  
into which graphs are embedded.  In addition the oriented matroid concept
introduces the possibility of a dual description of 
oriented graphs in terms of vector configurations, and vice versa.  

Secondly, 
it gives explicit ways to classify oriented matroids, and e.g.\ to generate
representatives of equivalence classes of vector configurations under
permutation.
We have already taken advantage of this 
in section \ref{New Volume Data}, where we revisit the results of \cite{BrunnemannRideout2008,BrunnemannRideout2008a}. Using our insights from oriented matroids we are able to drastically simplify the presentation, and to confirm the vertex combinatorics used in \cite{BrunnemannRideout2008,BrunnemannRideout2008a} by direct computation from reorientation classes of oriented matroids.\\

The present work will serve as a starting point to further explore this subject.

The constructions outlined in section \ref{Constructions with Oriented Matroids}, e.g.\ reorientation and deletion, can be used in order to construct an analogy to the projective limit of the graph poset in LQG, using only oriented matroids. Here the construction of matroids for infinite graphs \cite{Bruhn2009} will become relevant. This question will be analyzed in \cite{Brunnemann2010}.
We have seen in section \ref{Gauge Invariance and Circuits on Digraphs} gauge invariance of a cylindrical function can be decribed by the selection of a dual basis of the underlying graph. This points to a connection between oriented matroids and recoupling theory of $SU(2)$ representations. This is an interesting perspective for a possible extension of the oriented matroid framework by techniques from the graphical calculus of angular momentum, as for example given in \cite{Yutsis1962}.  
For LQG it will also be crucial to describe the action of graph changing operators (holonomies) by modifications to the graphic oriented matroid. If this turns out to be possible, then it becomes feasible to cast the action of the Hamilton \cite{Thiemann1996}, and respectively Master constraint operator \cite{Thiemann2006,Thiemann2005}, in LQG into the oriented matroid framework. The difficulty of finding eigenstates could then be re-formulated in terms of oriented matroids.

In section \ref{Oriented Matroids} an abstract notion of orthogonality between signed subsets is given. An interesting question is if this notion can be related to the inner product on $\Hdiff$, the diffeomorphism invariant  sector of LQG,  where the inner product is given by a rigging map construction \cite{Ashtekar1995,Thiemann2007}. Also, in the context of \cite{Bahr2007}, it will be instructive to see if the outstanding problem of giving a basis for the automorphism invariant sector of LQG can be tackled using an oriented matroid labelling.  

Besides these exciting conceptual perspectives, we have already seen that oriented matroids can be used in order to perform explicit computations in LQG. 
We expect the spectral analysis of the volume operator to be completed\footnote{In the sense that we can analyze all possible vertex embeddings, including coplanar triples. Of course, this is limited by finite computational resources.} in \cite{BrunnemannRideout2010}, where also non-uniform oriented matroids, that is coplanar triples of tangent vectors, will be included using techniques from \cite{Finschi2002}.
In this context it should also be possible to extend the analysis of \cite{Flori2008} to all realizable vertex embeddings.  Additionally the issue of taking the semiclassical limit, as discussed at the end of section \ref{Discussion}, can be addressed in this setup.

Moreover it is now 
possible to classify diffeomorphic graphs\footnote{Modulo knotting, see the discussion in section \ref{Discussion}.}, and to compute the number of diffeomorphism equivalence classes. For this the issue of realizability in section \ref{Realizability of Oriented Matroids} becomes important. The development of effective algorithms for computing realizable oriented matroids for larger ground sets will be crucial for this. 
In the mid-term perspective, oriented matroid techniques may provide effective techniques in order to develop a computational toolkit for performing numerical simulations in full LQG.

\section*{Acknowledgments}
\addcontentsline{toc}{section}{\numberline{}Acknowledgements}
The authors would like to thank Fernando Barbero for useful discussions and for providing the analytic counting formula (\ref{Sign Factor formula}) of generic sign configurations. Moreover we thank J\"urgen Richter-Gebert for discussions and handing out the manuscripts \cite{Bokowski1990a,Bokowski1990b} to us and additionally Ulrich Kortenkamp and Lukas Finschi for providing code for computing reorientation classes of oriented matroids.  
The work of JB has been supported by the Emmy-Noether-Programm (grant FL 622/1-1) of the Deutsche Forschungsgemeinschaft.
The work of DR was supported by the Perimeter Institute for
Theoretical Physics.  Research at the Perimeter Institute is supported by the
Government of Canada through Industry Canada and by the Province of Ontario
through the Ministry of Research \& Innovation.
JB would like to thank the Perimeter Institute for hospitality and financial support.
DR would also like to thank the Emmy-Noether-Programm (grant FL 622/1-1) of the Deutsche Forschungsgemeinschaft for financial support.
The numerical results were generated using the facilities of the
Shared Hierarchical Academic Research Computing Network
(SHARCNET:www.sharcnet.ca).
We thank the anonymous referees for several helpful suggestions.

\addcontentsline{toc}{section}{\numberline{}References}

\end{document}